\documentclass[aps,prd,twocolumn,superscriptaddress,nofootinbib]{revtex4-1}

\usepackage{latexsym}
\usepackage{amsmath}
\usepackage{amssymb}
\usepackage{amsfonts}
\usepackage{bm}

\usepackage{color}

\usepackage{supertabular} 
\usepackage{placeins}
\usepackage{epsfig}
\usepackage{graphicx}

\usepackage{color}
\definecolor{purple}{rgb}{0.5,0,0.5}
\definecolor{blue}{rgb}{0.0,0,0.9}
\definecolor{prdblue}{rgb}{0.133,0.118,0.498}
\usepackage[colorlinks=true, pdfstartview=FitV, linkcolor=prdblue, citecolor= prdblue, urlcolor=prdblue]{hyperref}

\begin{document}

% Use the \preprint command to place your local institutional report
% number in the upper righthand corner of the title page in preprint mode.
% Multiple \preprint commands are allowed.
% Use the 'preprintnumbers' class option to override journal defaults
% to display numbers if necessary
%\preprint{}

%Title of paper
\title{Bottomonium-like tetraquarks in a chiral quark model}

% repeat the \author .. \affiliation  etc. as needed
% \email, \thanks, \homepage, \altaffiliation all apply to the current
% author. Explanatory text should go in the []'s, actual e-mail
% address or url should go in the {}'s for \email and \homepage.
% Please use the appropriate macro foreach each type of information

% \affiliation command applies to all authors since the last
% \affiliation command. The \affiliation command should follow the
% other information
% \affiliation can be followed by \email, \homepage, \thanks as well.
\author{Gang Yang }
\email[]{yanggang@zjnu.edu.cn}
\affiliation{Department of Physics, Zhejiang Normal University, Jinhua 321004, China}
%\homepage[]{Your web page}
%\thanks{}
%\altaffiliation{}

\author{Jialun Ping}
\email[]{jlping@njnu.edu.cn}
\affiliation{Department of Physics and Jiangsu Key Laboratory for Numerical Simulation of Large Scale Complex Systems, Nanjing Normal University, Nanjing 210023, P. R. China}

\author{Jorge Segovia}
\email[]{jsegovia@upo.es}
\affiliation{Departamento de Sistemas F\'isicos, Qu\'imicos y Naturales, \\ Universidad Pablo de Olavide, E-41013 Sevilla, Spain}

%Collaboration name if desired (requires use of superscriptaddress
%option in \documentclass). \noaffiliation is required (may also be
%used with the \author command).
%\collaboration can be followed by \email, \homepage, \thanks as well.
%\collaboration{}
%\noaffiliation

\date{\today}

\begin{abstract}
The low-lying bottomonium-like tetraquarks $b\bar{b}q\bar{q}$ $(q=u,\,d,\,s)$ with spin-parity $J^P=0^+$, $1^+$ and $2^+$, and isospin $I=0,\,1$ or $\frac{1}{2}$, are systematically investigated within the theoretical framework of real- and complex-scaling range of a chiral quark model, which has already been successfully applied in analysis of several multiquark systems.
A complete four-body $S$-wave function, which includes meson-meson, diquark-antidiquark and K-type arrangements of quarks, along with all possible color configurations are considered.
In the $b\bar{b}q\bar{q}$ $(q=u,\,d)$ tetraquark system, we found resonance states of $B^{(*)} \bar{B}^{(*)}$, $\Upsilon\omega$, $\Upsilon\rho$ and $\eta_b \rho$ nature with all possible $I(J^P)$ quantum numbers. Their masses are generally located in the energy range $11.0-11.3$ GeV and their widths are less than $10$ MeV. In addition, extremely narrow resonances, with two-meson strong decay widths less than $1.5$ MeV, are obtained in both $b\bar{b}u\bar{s}$ and $b\bar{b}s\bar{s}$ tetraquark systems. Particularly, four radial excitations of $\Upsilon K^*$ and $\eta_b K^*$ are found at $\sim11.1$ GeV in $J^P=0^+$, $1^+$ and $2^+$ channels of $b\bar{b}u\bar{s}$ system. One $\Upsilon(1S)\phi(2S)$ resonance state is obtained at $11.28$ GeV in the $J^P=1^+$ sector.
\end{abstract}

% insert suggested PACS numbers in braces on next line
% insert suggested keywords - APS authors don't need to do this

%\maketitle must follow title, authors, abstract, \pacs, and \keywords
\maketitle

%%%%%%%%%%%%%%%%%%%%%%%%%%%%%%%%%%%%%%%%%%%%%%%%%%%%%%%%%%%%%%%%%%%%%%%%%%%%%%%%

\section{Introduction}

During the past two decades several exotic structures have been reported experimentally, and this fact undoubtedly opens crucial access in exploring the quantum field theory of the strong interaction, \emph{i.e.} Quantum Chromodynamics (QCD), and its related perturbative and non-perturbative theoretical formulations; in particular, QCD-inspired phenomenological quark models and their extensions.

Focusing on the most recent discoveries, on one hand, the first hidden-charm pentaquark $P^+_c(4380)$ was announced by the LHCb collaboration in $2015$~\cite{Aaij:2015tga}; later on, many more candidates of exotic hadrons with 5-quark content were reported, \emph{e.g.}, the $P^+_c(4312)$, $P^+_c(4337)$, $P^+_c(4440)$ and $P^+_c(4457)$~\cite{lhcb:2019pc, LHCb:2021chn}, and the hidden-charm pentaquark with strangeness $P^0_{cs}(4459)$~\cite{LHCb:2020jpq}. On the other hand, substantial data on tetraquark candidates have been accumulated in the last two years by worldwide high-energy experiments such as the LHCb, BES~III and Belle~II. Particularly, there are several open- and hidden-charm structures newly reported, such as the charm-strange tetraquarks $X_{0,1}(2900)$~\cite{LHCb:2020pxc, LHCb:2020bls}, the doubly charmed tetraquark $T^+_{cc}$~\cite{LHCb:2021vvq, LHCb:2021auc}, the charged hidden-charm tetraquarks with strangeness $Z_{cs}(3985)$~\cite{BESIII:2020qkh}, $Z_{cs}(4000)$, $Z_{cs}(4220)$, $X(4630)$ and $X(4685)$~\cite{LHCb:2021uow}, and the fully charmed tetraquark $X(6900)$~\cite{LHCb:2020bwg}.

The experimental progress within hidden-bottom sector can be dated back to $2011$, when two charged exotic candidates $Z_b(10610)$ and $Z_b(10650)$ were reported by the Belle collaboration~\cite{Belle:2011aa}, and confirmed in later studies~\cite{Belle:2014vzn, Belle:2015upu}. A wide range of phenomenological approaches have been employed in investigating their properties. Generally, a $B^{(*)} \bar{B}^*$ molecule interpretation for these $Z_b$ hadrons is favored by effective field theory~\cite{Cleven:2011gp, He:2014nya}, phenomenology models~\cite{Sun:2011uh, Yang:2017rmm, Zhang:2022hfa} and heavy quark symmetry~\cite{Nieves:2011zz}. However, either virtual state or cusp effects may be not discarded~\cite{Dias:2014pva, Guo:2016bjq, Wang:2018jlv, Ortega:2021xst}. Meanwhile, a $b\bar{b}u\bar{d}$ tetraquark has been very recently studied in lattice-regularized QCD and no significant attraction, or repulsion, is found~\cite{Sadl:2021bme}; in contrast, several attraction features of $B^{(*)}_s B^{(*)}_s$ systems have been suggested by effective field theory investigation~\cite{Dai:2022ulk}. A way of distinguishing among the mentioned structural solutions is studying the decay properties of $Z_b(10610)$ and $Z_b(10650)$ which have been analyzed in quark model approaches~\cite{Goerke:2017svb, Wang:2018pwi}.

The exciting advances in the exploration of exotic hadrons with heavy quark content have triggered many theoretical studies using a wide variety of approaches. Comprehensive reviews of the status of the field have been published over the years~\cite{Dong:2020hxe, Chen:2016qju, Chen:2016spr, Guo:2017jvc, Liu:2019zoy, Yang:2020atz, Dong:2021bvy, Chen:2021erj}. Given the fact that several hidden-charm tetraquarks with strangeness have been just experimentally reported along the last year, and they have been investigated by us within a chiral quark model formalism~\cite{Yang:2021zhe}, a natural extension to the bottomonium-like tetraquark $b\bar{b}q\bar{q}$ $(q=u,\,d,\,s)$ investigation is proposed herein.

A QCD-based chiral quark model is employed, and this formalism has already been successfully applied in the study of a wide variety of multiquark systems, \emph{e.g.} hidden- and double-charm pentaquarks~\cite{Yang:2015bmv, gy:2020dcp}, hidden-bottom pentaquarks~\cite{Yang:2018oqd}, doubly and fully heavy tetraquarks~\cite{gy:2020dht, gy:2020dhts, Yang:2021hrb}, and one-heavy-quark tetraquarks with strangeness~\cite{Yang:2021izl}. The formulation of our chiral quark model in either real- and complex-scaling method has been discussed extensively in Ref.~\cite{Yang:2020atz}, which serves also as a nice review of our latest results for multiquark systems. The complex-scaling method (CSM) allows us to distinguish bound, resonance and continuum states directly, and thus perform a complete analysis of the scattering singularities within the same formalism. Furthermore, the meson-meson, diquark-antidiquark and K-type configurations, plus their couplings, shall be considered for the $b\bar{b}q\bar{q}$ tetraquark system with all possible color configurations.

The manuscript is arranged as follows. In Sec.~\ref{sec:model} the theoretical framework is presented; we briefly discuss the complex-range method applied to a chiral quark model and the $b\bar{b}q\bar{q}$ $(q=u,\,d,\,s)$ tetraquark wave-functions. Section~\ref{sec:results} is devoted to the analysis of the obtained low-lying bottomonium-like tetraquark states with $J^P=0^+$, $1^+$ and $2^+$, and isospin $I=0,\,1$ or $\frac{1}{2}$. And, finally, a summary is presented in Sec.~\ref{sec:summary}.

%%%%%%%%%%%%%%%%%%%%%%%%%%%%%%%%%%%%%%%%%%%%%%%%%%%%%%%%%%%%%%%%%%%%%%%%%%%%%%%%

\section{Theoretical framework}
\label{sec:model}

A throughout review of the theoretical formalism used herein has been recently published in Ref.~\cite{Yang:2020atz}. We shall, however, focused on the most relevant features of the chiral quark model and the numerical method concerning the hidden-bottom tetraquarks $b\bar{b}q\bar{q}$ $(q=u,\,d,\,s)$.
 
Within the so-called complex-range studies, the relative coordinate of a two-body interaction is rotated in the complex plane by an angle $\theta$, \emph{i.e.}, $\vec{r}_{ij}\to \vec{r}_{ij} e^{i\theta}$. Therefore, the general form of the four-body Hamiltonian reads:
\begin{equation}
H(\theta) = \sum_{i=1}^{4}\left( m_i+\frac{\vec{p\,}^2_i}{2m_i}\right) - T_{\text{CM}} + \sum_{j>i=1}^{4} V(\vec{r}_{ij} e^{i\theta}) \,,
\label{eq:Hamiltonian}
\end{equation}
where $m_{i}$ and $\vec{p}_i$ are, respectively, the mass and momentum of a quark; and $T_{\text{CM}}$ is the center-of-mass kinetic energy. According to the so-called ABC theorem~\cite{JA22269, EB22280}, the complex scaled Schr\"odinger equation:
\begin{equation}\label{CSMSE}
\left[ H(\theta)-E(\theta) \right] \Psi_{JM}(\theta)=0	
\end{equation}
has (complex) eigenvalues which can be classified into three types, namely bound, resonance and scattering states. In particular, bound-states and resonances are independent of the rotated angle $\theta$, with the first ones always fixed on the real-axis (there is no imaginary part of the eigenvalue), and the second ones located above the continuum threshold with a total decay width $\Gamma=-2\,\text{Im}(E)$.

The dynamics of the $b\bar b q\bar q$ tetraquark system is driven by a two-body potential
\begin{equation}
\label{CQMV}
V(\vec{r}_{ij}) = V_{\chi}(\vec{r}_{ij}) + V_{\text{CON}}(\vec{r}_{ij}) + V_{\text{OGE}}(\vec{r}_{ij})  \,,
\end{equation}
which takes into account the most relevant features of QCD at its low energy regime: dynamical chiral symmetry breaking, confinement and the perturbative one-gluon exchange interaction. Herein, the low-lying $S$-wave positive parity $b\bar{b}q\bar{q}$ tetraquark states shall be investigated, and thus the central and spin-spin terms of the potential are the only ones needed.

One consequence of the dynamical breaking of chiral symmetry is that Goldstone boson exchange interactions appear between constituent light quarks $u$, $d$ and $s$. Therefore, the chiral interaction can be written as~\cite{Vijande:2004he}:
\begin{equation}
V_{\chi}(\vec{r}_{ij}) = V_{\pi}(\vec{r}_{ij})+ V_{\sigma}(\vec{r}_{ij}) + V_{K}(\vec{r}_{ij}) + V_{\eta}(\vec{r}_{ij}) \,,
\end{equation}
given by
\begin{align}
&
V_{\pi}\left( \vec{r}_{ij} \right) = \frac{g_{ch}^{2}}{4\pi}
\frac{m_{\pi}^2}{12m_{i}m_{j}} \frac{\Lambda_{\pi}^{2}}{\Lambda_{\pi}^{2}-m_{\pi}
^{2}}m_{\pi} \Bigg[ Y(m_{\pi}r_{ij}) \nonumber \\
&
\hspace*{1.20cm} - \frac{\Lambda_{\pi}^{3}}{m_{\pi}^{3}}
Y(\Lambda_{\pi}r_{ij}) \bigg] (\vec{\sigma}_{i}\cdot\vec{\sigma}_{j})\sum_{a=1}^{3}(\lambda_{i}^{a}
\cdot\lambda_{j}^{a}) \,, \\
& 
V_{\sigma}\left( \vec{r}_{ij} \right) = - \frac{g_{ch}^{2}}{4\pi}
\frac{\Lambda_{\sigma}^{2}}{\Lambda_{\sigma}^{2}-m_{\sigma}^{2}}m_{\sigma} \Bigg[Y(m_{\sigma}r_{ij}) \nonumber \\
&
\hspace*{1.20cm} - \frac{\Lambda_{\sigma}}{m_{\sigma}}Y(\Lambda_{\sigma}r_{ij})
\Bigg] \,,
\end{align}
\begin{align}
& 
V_{K}\left( \vec{r}_{ij} \right)= \frac{g_{ch}^{2}}{4\pi}
\frac{m_{K}^2}{12m_{i}m_{j}}\frac{\Lambda_{K}^{2}}{\Lambda_{K}^{2}-m_{K}^{2}}m_{
K} \Bigg[ Y(m_{K}r_{ij}) \nonumber \\
&
\hspace*{1.20cm} -\frac{\Lambda_{K}^{3}}{m_{K}^{3}}Y(\Lambda_{K}r_{ij}) \Bigg] (\vec{\sigma}_{i}\cdot\vec{\sigma}_{j}) \sum_{a=4}^{7}(\lambda_{i}^{a} \cdot \lambda_{j}^{a}) \,, \\
& 
V_{\eta}\left( \vec{r}_{ij} \right) = \frac{g_{ch}^{2}}{4\pi}
\frac{m_{\eta}^2}{12m_{i}m_{j}} \frac{\Lambda_{\eta}^{2}}{\Lambda_{\eta}^{2}-m_{
\eta}^{2}}m_{\eta} \Bigg[ Y(m_{\eta}r_{ij}) \nonumber \\
&
\hspace*{1.20cm} -\frac{\Lambda_{\eta}^{3}}{m_{\eta}^{3}
}Y(\Lambda_{\eta}r_{ij}) \Bigg] (\vec{\sigma}_{i}\cdot\vec{\sigma}_{j})
\Big[\cos\theta_{p} \left(\lambda_{i}^{8}\cdot\lambda_{j}^{8}
\right) \nonumber \\
&
\hspace*{1.20cm} -\sin\theta_{p} \Big] \,,
\end{align}
where $Y(x)=e^{-x}/x$ is the standard Yukawa function. The physical $\eta$ meson, instead of the octet one, is considered by introducing the angle $\theta_p$. The $\lambda^{a}$ are the SU(3) flavor Gell-Mann matrices. Taken from their experimental values, $m_{\pi}$, $m_{K}$ and $m_{\eta}$ are the masses of the SU(3) Goldstone bosons. 

The value of $m_\sigma$ is given by the partially conserved axial current relation $m_{\sigma}^{2}\simeq m_{\pi}^{2}+4m_{u,d}^{2}$~\cite{Scadron:1982eg}. Note, however, that better determinations of the mass of the $\sigma$-meson have been reported since then~\cite{Garcia-Martin:2011nna, Albaladejo:2012te} -- see also the recent review~\cite{Pelaez:2015qba}; one should simply consider the value used here as a model parameter. Finally, the chiral coupling constant, $g_{ch}$, is determined from the $\pi NN$ coupling constant through
\begin{equation}
\frac{g_{ch}^{2}}{4\pi}=\frac{9}{25}\frac{g_{\pi NN}^{2}}{4\pi} \frac{m_{u,d}^{2}}{m_{N}^2} \,,
\end{equation}
which assumes that flavor SU(3) is an exact symmetry only broken by the different mass of the strange quark.

Color confinement should be encoded in the non-Abelian character of QCD. An attractive linearly rising potential proportional to the distance between infinite-heavy quarks is the consequence of multi-gluon exchanges as demonstrated by lattice-regularized QCD~\cite{Bali:2005fu}. However, the spontaneous creation of light-quark pairs from the QCD vacuum may give rise at the same scale to a breakup of the created color flux-tube~\cite{Bali:2005fu}. These two observations can be phenomenologically described by~\cite{Segovia:2008zza}:
\begin{equation}
V_{\text{CON}}(\vec{r}_{ij})=\left[-a_{c}(1-e^{-\mu_{c}r_{ij}})+\Delta \right] 
(\lambda_{i}^{c}\cdot \lambda_{j}^{c}) \,,
\label{eq:conf}
\end{equation}
where $a_{c}$, $\mu_{c}$ and $\Delta$ are model parameters, and the SU(3) color Gell-Mann matrices are denoted as $\lambda^c$. One can see in Eq.~\eqref{eq:conf} that the potential is linear at short inter-quark distances with an effective confinement strength $\sigma = -a_{c} \, \mu_{c} \, (\lambda^{c}_{i}\cdot \lambda^{c}_{j})$, while it becomes constant at large distances, $V_{\text{thr.}} = (\Delta-a_c)(\lambda^{c}_{i}\cdot \lambda^{c}_{j})$.

Beyond the chiral symmetry breaking scale one expects the dynamics to be
governed by QCD perturbative effects. In particular, the one-gluon exchange potential, which includes the so-called coulomb and color-magnetism interactions, is the leading order contribution:
\begin{align}
&
V_{\text{OGE}}(\vec{r}_{ij}) = \frac{1}{4} \alpha_{s} (\lambda_{i}^{c}\cdot \lambda_{j}^{c}) \Bigg[\frac{1}{r_{ij}} \nonumber \\ 
&
\hspace*{1.60cm} - \frac{1}{6m_{i}m_{j}} (\vec{\sigma}_{i}\cdot\vec{\sigma}_{j}) 
\frac{e^{-r_{ij}/r_{0}(\mu_{ij})}}{r_{ij} r_{0}^{2}(\mu_{ij})} \Bigg] \,,
\end{align}
where $r_{0}(\mu_{ij})=\hat{r}_{0}/\mu_{ij}$ is a regulator which depends on the reduced mass of the $q\bar{q}$ pair, the Pauli matrices are denoted by $\vec{\sigma}$, and the contact term has been regularized as
\begin{equation}
\delta(\vec{r}_{ij}) \sim \frac{1}{4\pi r_{0}^{2}(\mu_{ij})}\frac{e^{-r_{ij} / r_{0}(\mu_{ij})}}{r_{ij}} \,.
\end{equation}

An effective scale-dependent strong coupling constant, $\alpha_s(\mu_{ij})$, provides a consistent description of mesons and baryons from light to heavy quark sectors. We use the frozen coupling constant of, for instance, Ref.~\cite{Segovia:2013wma}
\begin{equation}
\alpha_{s}(\mu_{ij})=\frac{\alpha_{0}}{\ln\left(\frac{\mu_{ij}^{2}+\mu_{0}^{2}}{\Lambda_{0}^{2}} \right)} \,,
\end{equation}
in which $\alpha_{0}$, $\mu_{0}$ and $\Lambda_{0}$ are parameters of the model.

The model parameters are listed in Table~\ref{tab:model}. They have been fixed in advance reproducing hadron~\cite{Segovia:2009zz, Segovia:2011zza, Segovia:2015dia, Segovia:2016xqb, Yang:2017xpp, Yang:2019lsg}, hadron-hadron ~\cite{Ortega:2016mms, Ortega:2016pgg, Ortega:2016hde, Ortega:2017qmg, Ortega:2018cnm, Ortega:2020uvc} and multiquark~\cite{Yang:2017rpg, Yang:2015bmv, Yang:2018oqd, Yang:2020fou, Yang:2020twg, Yang:2021izl} phenomenology. Additionally, for later concern, Table~\ref{MesonMass} lists theoretical and experimental (if available) masses of $1S$, $2S$ and $3S$ states of $q\bar{q}$, $b\bar{q}$ $(q=u,\,d,\,s)$ and $b\bar{b}$ mesons.

\begin{table}[!t]
\caption{\label{tab:model} Model parameters.}
\begin{ruledtabular}
\begin{tabular}{llr}
Quark masses     & $m_q\,(q=u,\,d)$ (MeV) & 313 \\
                 & $m_s$ (MeV) &  555 \\
                 & $m_b$ (MeV) & 5100 \\[2ex]
Goldstone bosons & $\Lambda_\pi=\Lambda_\sigma~$ (fm$^{-1}$) &   4.20 \\
                 & $\Lambda_\eta=\Lambda_K$ (fm$^{-1}$)      &   5.20 \\
                 & $g^2_{ch}/(4\pi)$                         &   0.54 \\
                 & $\theta_P(^\circ)$                        & -15 \\[2ex]
Confinement      & $a_c$ (MeV)         & 430 \\
                 & $\mu_c$ (fm$^{-1})$ & 0.70 \\
                 & $\Delta$ (MeV)      & 181.10 \\[2ex]
OGE              & $\alpha_0$              & 2.118 \\
                 & $\Lambda_0~$(fm$^{-1}$) & 0.113 \\
                 & $\mu_0~$(MeV)           & 36.976 \\
                 & $\hat{r}_0~$(MeV~fm)    & 28.170 \\
\end{tabular}
\end{ruledtabular}
\end{table}

\begin{table*}[!t]
\caption{\label{MesonMass} Theoretical and experimental (if available) masses of $nL=1S, 2S$ and $3S$ states of $q\bar{q}$, $q\bar{b}\,(q=u, d, s)$ and $b\bar{b}$ mesons.}
\begin{ruledtabular}
\begin{tabular}{lccclccc}
Meson & $nL$ & $M_{\text{The.}}$ (MeV) & $M_{\text{Exp.}}$ (MeV)  & Meson & $nL$ & $M_{\text{The.}}$ (MeV) & $M_{\text{Exp.}}$ (MeV) \\
\hline
$\pi$ & $1S$ &  $149$ & $140$  & $\eta$ & $1S$ &  $689$ & $548$ \\
    & $2S$ & $1291$ & $1300$    &  & $2S$ & $1443$ & $1295$ \\
    & $3S$ & $1749$ & $1800$    &  & $3S$ & $1888$ & - \\[2ex]
$\rho$ & $1S$ &  $772$ & $770$  & $\omega$ & $1S$ &  $696$ & $782$ \\
    & $2S$ & $1479$ & $1450$    &  & $2S$ & $1449$ & $1420$ \\
    & $3S$ & $1940$ & -    &  & $3S$ & $1897$ & $1650$ \\[2ex]
$K$ & $1S$ &  $481$ & $494$  & $K^*$ & $1S$ &  $907$ & $892$ \\
    & $2S$ & $1468$ & $1460$    &  & $2S$ & $1621$ & $1630$ \\
    & $3S$ & $1898$ & $1830$    &  & $3S$ & $1998$ & - \\[2ex]
$\eta'$ & $1S$ &  $828$ & $958$  & $\phi$ & $1S$ &  $1011$ & $1020$ \\
    & $2S$ & $1639$ & $1760$    &  & $2S$ & $1720$ & $1680$ \\
    & $3S$ & $2057$ & -    &  & $3S$ & $2097$ & $2170$ \\[2ex]
$B$ & $1S$ &  $5278$ & $5280$  & $B^*$ & $1S$ &  $5319$ & $5325$ \\
    & $2S$ & $5984$ & -    &  & $2S$ & $6005$ & - \\
    & $3S$ & $6373$ & -    &  & $3S$ & $6383$ & - \\[2ex]
$B_s$ & $1S$ &  $5355$ & $5367$  & $B^*_s$ & $1S$ &  $5400$ & $5415$ \\
    & $2S$ & $6017$ & -    &  & $2S$ & $6042$ & - \\
    & $3S$ & $6412$ & -    &  & $3S$ & $6432$ & - \\[2ex]
$\eta_b$ & $1S$ &  $9454$ & $9300$  & $\Upsilon$ & $1S$ &  $9505$ & $9460$ \\
    & $2S$ & $9985$ & -    &  & $2S$ & $10013$ & $10023$ \\
    & $3S$ & $10327$ & -    &  & $3S$ & $10345$ & $10355$
\end{tabular}
\end{ruledtabular}
\end{table*}

\begin{figure}[ht]
\epsfxsize=3.4in \epsfbox{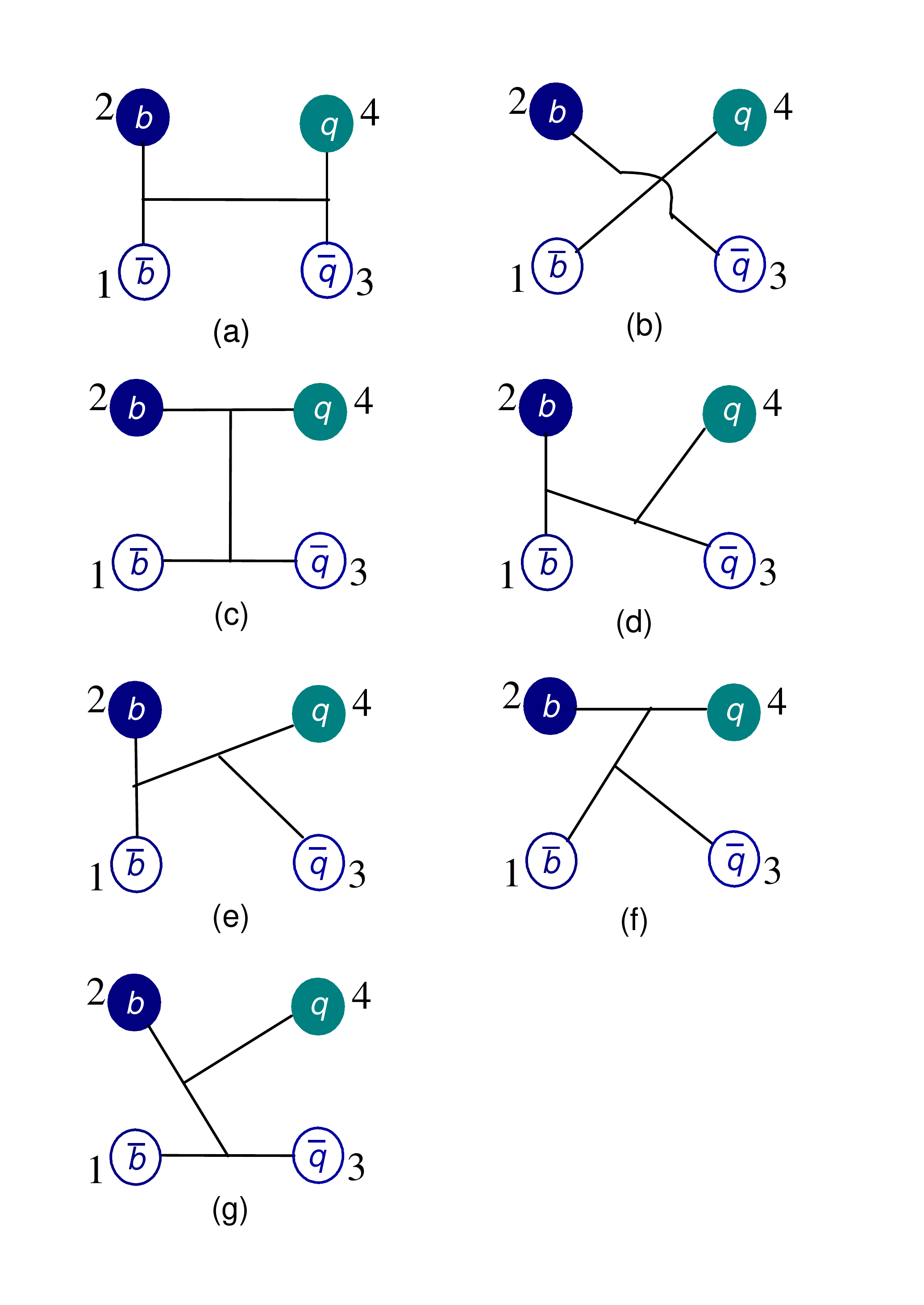}
\caption{\label{QQqq} Seven types of configurations in $b\bar{b}q\bar{q}$ $(q=u,\,d,\,s)$ tetraquarks. Panels $(a)$ and $(b)$ represent meson-meson structures, panel $(c)$ is diquark-antidiquark one and the other K-type structures are shown in panels $(d)$ to $(g)$.}
\end{figure}

Figure~\ref{QQqq} shows seven kinds of configurations for the $b\bar{b}q\bar{q}$ tetraquark system. In particular, panels~\ref{QQqq}(a) and (b) are the meson-meson structures, panel~\ref{QQqq}(c) is the diquark-antidiquark one, and the other K-type configurations are from panels (d) to (g). All of them, and their couplings, are considered in our investigation. However, for the purpose of solving a manageable 4-body problem, the K-type configurations are sometimes restricted, following a similar criteria as our investigation on the hidden-charm tetraquarks with strangeness~\cite{Yang:2021zhe}. Furthermore, it is important to note herein that just one configuration would be enough for the calculation, if all radial and orbital excited states were taken into account; however, this is obviously much less efficient and thus an economic way is to combine the different configurations in the ground state to perform the calculation.

The multiquark system's wave function at the quark level is an internal product of color, spin, flavor and space terms. Concerning the color degree-of-freedom, the colorless wave function of a 4-quark system in meson-meson configuration, as illustrated in Fig.~\ref{QQqq}(a) and (b), can be obtained by either two coupled color-singlet clusters, $1\otimes 1$:
\begin{align}
\label{Color1}
\chi^c_1 &= \frac{1}{3}(\bar{r}r+\bar{g}g+\bar{b}b)\times (\bar{r}r+\bar{g}g+\bar{b}b) \,,
\end{align}
or two coupled color-octet clusters, $8\otimes 8$:
\begin{align}
\label{Color2}
\chi^c_2 &= \frac{\sqrt{2}}{12}(3\bar{b}r\bar{r}b+3\bar{g}r\bar{r}g+3\bar{b}g\bar{g}b+3\bar{g}b\bar{b}g+3\bar{r}g\bar{g}r
\nonumber\\
&+3\bar{r}b\bar{b}r+2\bar{r}r\bar{r}r+2\bar{g}g\bar{g}g+2\bar{b}b\bar{b}b-\bar{r}r\bar{g}g
\nonumber\\
&-\bar{g}g\bar{r}r-\bar{b}b\bar{g}g-\bar{b}b\bar{r}r-\bar{g}g\bar{b}b-\bar{r}r\bar{b}b) \,.
\end{align}
The first color state is the so-called color-singlet channel and the second one is named as the hidden-color case.

The color wave functions associated to the diquark-antidiquark structure shown in Fig.~\ref{QQqq}(c) are the coupled color triplet-antitriplet clusters, $3\otimes \bar{3}$:
\begin{align}
\label{Color3}
\chi^c_3 &= \frac{\sqrt{3}}{6}(\bar{r}r\bar{g}g-\bar{g}r\bar{r}g+\bar{g}g\bar{r}r-\bar{r}g\bar{g}r+\bar{r}r\bar{b}b
\nonumber\\
&-\bar{b}r\bar{r}b+\bar{b}b\bar{r}r-\bar{r}b\bar{b}r+\bar{g}g\bar{b}b-\bar{b}g\bar{g}b
\nonumber\\
&+\bar{b}b\bar{g}g-\bar{g}b\bar{b}g) \,,
\end{align}
and the coupled color sextet-antisextet clusters, $6\otimes \bar{6}$:
\begin{align}
\label{Color4}
\chi^c_4 &= \frac{\sqrt{6}}{12}(2\bar{r}r\bar{r}r+2\bar{g}g\bar{g}g+2\bar{b}b\bar{b}b+\bar{r}r\bar{g}g+\bar{g}r\bar{r}g
\nonumber\\
&+\bar{g}g\bar{r}r+\bar{r}g\bar{g}r+\bar{r}r\bar{b}b+\bar{b}r\bar{r}b+\bar{b}b\bar{r}r
\nonumber\\
&+\bar{r}b\bar{b}r+\bar{g}g\bar{b}b+\bar{b}g\bar{g}b+\bar{b}b\bar{g}g+\bar{g}b\bar{b}g) \,.
\end{align}

Meanwhile, the colorless wave functions of the K-type structures shown in Fig.~\ref{QQqq}(d) to (g) are 
\begin{align}
\label{Color5}
\chi^c_5 &= \frac{1}{6\sqrt{2}}(\bar{r}r\bar{r}r+\bar{g}g\bar{g}g-2\bar{b}b\bar{b}b)+
\nonumber\\
&\frac{1}{2\sqrt{2}}(\bar{r}b\bar{b}r+\bar{r}g\bar{g}r+\bar{g}b\bar{b}g+\bar{g}r\bar{r}g+\bar{b}g\bar{g}b+\bar{b}r\bar{r}b)-
\nonumber\\
&\frac{1}{3\sqrt{2}}(\bar{g}g\bar{r}r+\bar{r}r\bar{g}g)+\frac{1}{6\sqrt{2}}(\bar{b}b\bar{r}r+\bar{b}b\bar{g}g+\bar{r}r\bar{b}b+\bar{g}g\bar{b}b) \,,
\end{align}
\begin{align}
\label{Color6}
\chi^c_6 &= \chi^c_1 \,,
\end{align}
\begin{align}
\label{Color7}
\chi^c_7 &= \chi^c_1 \,,
\end{align}
\begin{align}
\label{Color8}
\chi^c_8 &= \frac{1}{4}(1-\frac{1}{\sqrt{6}})\bar{r}r\bar{g}g-\frac{1}{4}(1+\frac{1}{\sqrt{6}})\bar{g}g\bar{g}g-\frac{1}{4\sqrt{3}}\bar{r}g\bar{g}r+
\nonumber\\
&\frac{1}{2\sqrt{2}}(\bar{r}b\bar{b}r+\bar{g}b\bar{b}g+\bar{b}g\bar{g}b+\bar{g}r\bar{r}g+\bar{b}r\bar{r}b)+
\nonumber\\
&\frac{1}{2\sqrt{6}}(\bar{r}r\bar{b}b-\bar{g}g\bar{b}b+\bar{b}b\bar{g}g+\bar{g}g\bar{r}r-\bar{b}b\bar{r}r) \,,
\end{align}
\begin{align}
\label{Color9}
\chi^c_9 &= \frac{1}{2\sqrt{6}}(\bar{r}b\bar{b}r+\bar{r}r\bar{b}b+\bar{g}b\bar{b}g+\bar{g}g\bar{b}b+\bar{r}g\bar{g}r+\bar{r}r\bar{g}g+
\nonumber\\
&\bar{b}b\bar{g}g+\bar{b}g\bar{g}b+\bar{g}g\bar{r}r+\bar{g}r\bar{r}g+\bar{b}b\bar{r}r+\bar{b}r\bar{r}b)+
\nonumber\\
&\frac{1}{\sqrt{6}}(\bar{r}r\bar{r}r+\bar{g}g\bar{g}g+\bar{b}b\bar{b}b) \,,
\end{align}
\begin{align}
\label{Color10}
\chi^c_{10} &= \frac{1}{2\sqrt{3}}(\bar{r}b\bar{b}r-\bar{r}r\bar{b}b+\bar{g}b\bar{b}g-\bar{g}g\bar{b}b+\bar{r}g\bar{g}r-\bar{r}r\bar{g}g-
\nonumber\\
&\bar{b}b\bar{g}g+\bar{b}g\bar{g}b-\bar{g}g\bar{r}r+\bar{g}r\bar{r}g-\bar{b}b\bar{r}r+\bar{b}r\bar{r}b) \,,
\end{align}
\begin{align}
\label{Color11}
\chi^c_{11} &= \chi^c_9 \,,
\end{align}
\begin{align}
\label{Color12}
\chi^c_{12} &= -\chi^c_{10} \,.
\end{align}

According to the nature of $SU(3)$ flavor, $b\bar{b}q\bar{q}$ $(q=u,\,d,\,s)$ tetraquark systems could be categorized into three parts, \emph{i.e.} $b\bar{b}q\bar{q}$, $b\bar{b}q\bar{s}$ and $b\bar{b}s\bar{s}$ $(q=u,\,d)$. The flavor wave function is then denoted as $\chi^{fi}_{I, M_I}$, where the superscript $f$ just indicates \emph{flavor} part of the wave function and $i=1,\,2$ and $3$ will refer to $b\bar{b}q\bar{q}$, $b\bar{b}q\bar{s}$ $(q=u, d)$ and $b\bar{b}s\bar{s}$ systems, respectively. We have isoscalar, $I=0$, and isovector, $I=1$, sectors in the $b\bar{b}q\bar{q}$ system, their flavor wave functions read as 
\begin{align}
&
\label{FWF0}
\chi_{0,0}^{fi} = -\frac{1}{\sqrt{2}}(\bar{b}b\bar{u}u+\bar{b}b\bar{d}d) \,, \\
&
\label{FWF1}
\chi_{1,0}^{fi} = -\frac{1}{\sqrt{2}}(\bar{b}b\bar{u}u-\bar{b}b\bar{d}d) \,.
\end{align}
Meanwhile, the flavor wave function of the other two 4-quark systems are simply $\bar{b}b\bar{s}q$ and $\bar{b}b\bar{s}s$, respectively. Note, too, that the third component of the isospin, $M_I$, is fixed to be zero for simplicity since the Hamiltonian does not have a flavor-dependent interaction which can distinguish the third component of the isospin quantum number.

We are going to consider $S$-wave ground states with spin ranging from $S=0$ to $2$. Therefore, the spin wave functions, $\chi^{\sigma_i}_{S, M_S}$, are given by ($M_S$ can be set to be equal to $S$ without loss of generality):
\begin{align}
\label{SWF1}
\chi_{0,0}^{\sigma_{u_1}}(4) &= \chi^\sigma_{00}\chi^\sigma_{00} \,, \\
\chi_{0,0}^{\sigma_{u_2}}(4) &= \frac{1}{\sqrt{3}}(\chi^\sigma_{11}\chi^\sigma_{1,-1}-\chi^\sigma_{10}\chi^\sigma_{10}+\chi^\sigma_{1,-1}\chi^\sigma_{11}) \,, \\
\chi_{0,0}^{\sigma_{u_3}}(4) &= \frac{1}{\sqrt{2}}\big((\sqrt{\frac{2}{3}}\chi^\sigma_{11}\chi^\sigma_{\frac{1}{2}, -\frac{1}{2}}-\sqrt{\frac{1}{3}}\chi^\sigma_{10}\chi^\sigma_{\frac{1}{2}, \frac{1}{2}})\chi^\sigma_{\frac{1}{2}, -\frac{1}{2}} \nonumber \\ 
&-(\sqrt{\frac{1}{3}}\chi^\sigma_{10}\chi^\sigma_{\frac{1}{2}, -\frac{1}{2}}-\sqrt{\frac{2}{3}}\chi^\sigma_{1, -1}\chi^\sigma_{\frac{1}{2}, \frac{1}{2}})\chi^\sigma_{\frac{1}{2}, \frac{1}{2}}\big) \,, \\
\chi_{0,0}^{\sigma_{u_4}}(4) &= \frac{1}{\sqrt{2}}(\chi^\sigma_{00}\chi^\sigma_{\frac{1}{2}, \frac{1}{2}}\chi^\sigma_{\frac{1}{2}, -\frac{1}{2}}-\chi^\sigma_{00}\chi^\sigma_{\frac{1}{2}, -\frac{1}{2}}\chi^\sigma_{\frac{1}{2}, \frac{1}{2}}) \,,
\end{align}
\begin{align}
\chi_{1,1}^{\sigma_{w_1}}(4) &= \chi^\sigma_{00}\chi^\sigma_{11} \,, \\ 
\chi_{1,1}^{\sigma_{w_2}}(4) &= \chi^\sigma_{11}\chi^\sigma_{00} \,, \\
\chi_{1,1}^{\sigma_{w_3}}(4) &= \frac{1}{\sqrt{2}} (\chi^\sigma_{11} \chi^\sigma_{10}-\chi^\sigma_{10} \chi^\sigma_{11}) \,, \\
\chi_{1,1}^{\sigma_{w_4}}(4) &= \sqrt{\frac{3}{4}}\chi^\sigma_{11}\chi^\sigma_{\frac{1}{2}, \frac{1}{2}}\chi^\sigma_{\frac{1}{2}, -\frac{1}{2}}-\sqrt{\frac{1}{12}}\chi^\sigma_{11}\chi^\sigma_{\frac{1}{2}, -\frac{1}{2}}\chi^\sigma_{\frac{1}{2}, \frac{1}{2}} \nonumber \\ 
&-\sqrt{\frac{1}{6}}\chi^\sigma_{10}\chi^\sigma_{\frac{1}{2}, \frac{1}{2}}\chi^\sigma_{\frac{1}{2}, \frac{1}{2}} \,, \\
\chi_{1,1}^{\sigma_{w_5}}(4) &= (\sqrt{\frac{2}{3}}\chi^\sigma_{11}\chi^\sigma_{\frac{1}{2}, -\frac{1}{2}}-\sqrt{\frac{1}{3}}\chi^\sigma_{10}\chi^\sigma_{\frac{1}{2}, \frac{1}{2}})\chi^\sigma_{\frac{1}{2}, \frac{1}{2}} \,, \\
\chi_{1,1}^{\sigma_{w_6}}(4) &= \chi^\sigma_{00}\chi^\sigma_{\frac{1}{2}, \frac{1}{2}}\chi^\sigma_{\frac{1}{2}, \frac{1}{2}} \,, \\
\label{SWF2}
\chi_{2,2}^{\sigma_{1}}(4) &= \chi^\sigma_{11}\chi^\sigma_{11} \,.
\end{align}
The superscripts $u_1,\ldots,u_4$ and $w_1,\ldots,w_6$ determine the spin wave function for each configuration of the $b\bar b q\bar q$ $(q=u,\,d,\,s)$ tetraquark system, their specific values are shown in Table~\ref{SpinIndex}. Furthermore, the expressions above are obtained by considering the coupling of two sub-clusters whose spin wave functions are given by trivial SU(2) algebra, and the necessary basis reads as
\begin{align}
\label{Spin}
\chi^\sigma_{11} &= \chi^\sigma_{\frac{1}{2}, \frac{1}{2}} \chi^\sigma_{\frac{1}{2}, \frac{1}{2}} \,, \\
\chi^\sigma_{1,-1} &= \chi^\sigma_{\frac{1}{2}, -\frac{1}{2}} \chi^\sigma_{\frac{1}{2}, -\frac{1}{2}} \,, \\
\chi^\sigma_{10} &= \frac{1}{\sqrt{2}}(\chi^\sigma_{\frac{1}{2}, \frac{1}{2}} \chi^\sigma_{\frac{1}{2}, -\frac{1}{2}}+\chi^\sigma_{\frac{1}{2}, -\frac{1}{2}} \chi^\sigma_{\frac{1}{2}, \frac{1}{2}}) \,, \\
\chi^\sigma_{00} &= \frac{1}{\sqrt{2}}(\chi^\sigma_{\frac{1}{2}, \frac{1}{2}} \chi^\sigma_{\frac{1}{2}, -\frac{1}{2}}-\chi^\sigma_{\frac{1}{2}, -\frac{1}{2}} \chi^\sigma_{\frac{1}{2}, \frac{1}{2}}) \,, 
\end{align}

\begin{table}[!t]
\caption{\label{SpinIndex} The values of the superscripts $u_1,\ldots,u_4$ and $w_1,\ldots,w_6$ that determine the spin wave function for each configuration of the $b\bar b q\bar q$ $(q=u,\,d,\,s)$ tetraquark system.}
\begin{ruledtabular}
\begin{tabular}{lcccccc}
& Di-meson & Diquark-antidiquark & $K_1$ & $K_2$ & $K_3$ & $K_4$ \\
\hline
$u_1$ & 1 & 3 & & & & \\
$u_2$ & 2 & 4 & & & & \\
$u_3$ &   &   & 5 & 7 &  9 & 11 \\
$u_4$ &   &   & 6 & 8 & 10 & 12 \\[2ex]
$w_1$ & 1 & 4 & & & & \\
$w_2$ & 2 & 5 & & & & \\
$w_3$ & 3 & 6 & & & & \\
$w_4$ &   &   & 7 & 10 & 13 & 16 \\
$w_5$ &   &   & 8 & 11 & 14 & 17 \\
$w_6$ &   &   & 9 & 12 & 15 & 18
\end{tabular}
\end{ruledtabular}
\end{table}

Among the different methods to solve the Schr\"odinger-like 4-body bound state equation, we use the Rayleigh-Ritz variational principle which is one of the most extended tools to solve eigenvalue problems because its simplicity and flexibility. Moreover, we use the complex-range method and thus the spatial wave function is written as follows:
\begin{equation}
\label{eq:WFexp}
\psi_{LM_L}(\theta)= \left[ \left[ \phi_{n_1l_1}(\vec{\rho}e^{i\theta}\,) \phi_{n_2l_2}(\vec{\lambda}e^{i\theta}\,)\right]_{l} \phi_{n_3l_3}(\vec{R}e^{i\theta}\,) \right]_{L M_L} \,,
\end{equation}
where the internal Jacobi coordinates are defined as
\begin{align}
\vec{\rho} &= \vec{x}_1-\vec{x}_2 \,, \\
\vec{\lambda} &= \vec{x}_3 - \vec{x}_4 \,, \\
\vec{R} &= \frac{m_1 \vec{x}_1 + m_2 \vec{x}_2}{m_1+m_2}- \frac{m_3 \vec{x}_3 + m_4 \vec{x}_4}{m_3+m_4} \,,
\end{align}
for the meson-meson configurations of panels~\ref{QQqq}(a) and (b); and as
\begin{align}
\vec{\rho} &= \vec{x}_1-\vec{x}_3 \,, \\
\vec{\lambda} &= \vec{x}_2 - \vec{x}_4 \,, \\
\vec{R} &= \frac{m_1 \vec{x}_1 + m_3 \vec{x}_3}{m_1+m_3}- \frac{m_2 \vec{x}_2 + m_4 \vec{x}_4}{m_2+m_4} \,,
\end{align}
for the diquark-antdiquark structure of panel~\ref{QQqq}(c). The remaining K-type configurations shown in panels~\ref{QQqq}(d) to \ref{QQqq}(g) are ($i, j, k, l$ take values according to the panels (d) to (g) of Fig.~\ref{QQqq}):
\begin{align}
\vec{\rho} &= \vec{x}_i-\vec{x}_j \,, \\
\vec{\lambda} &= \vec{x}_k- \frac{m_i \vec{x}_i + m_j \vec{x}_j}{m_i+m_j} \,, \\
\vec{R} &= \vec{x}_l- \frac{m_i \vec{x}_i + m_j \vec{x}_j+m_k \vec{x}_k}{m_i+m_j+m_k} \,.
\end{align}
It becomes obvious now that the center-of-mass kinetic term $T_{CM}$ can be completely eliminated for a non-relativistic system defined in any of the above sets of relative coordinates.

A crucial aspect of the Rayleigh-Ritz variational method is the basis expansion of the trial wave function. We are going to use the Gaussian expansion method (GEM)~\cite{Hiyama:2003cu} in which each relative coordinate is expanded in terms of Gaussian basis functions whose sizes are taken in geometric progression. This method has proven to be very efficient on solving the bound-state problem of multiquark systems~\cite{Yang:2015bmv} and the details on how the geometric progression is fixed can be found in, \emph{e.g.}, Ref.~\cite{Yang:2020atz}. Therefore, the form of the orbital wave functions, $\phi$'s, in Eq.~\eqref{eq:WFexp} is 
\begin{align}
&
\phi_{nlm}(\vec{r}e^{i\theta}\,) = N_{nl} (re^{i\theta})^{l} e^{-\nu_{n} (re^{i\theta})^2} Y_{lm}(\hat{r}) \,.
\end{align}
Since only $S$-wave states of $b\bar{b}q\bar{q}$ tetraquarks are investigated in this work, no laborious Racah algebra is needed while computing matrix elements; the value of the spherical harmonic function is just a constant, \emph{viz.} $Y_{00}=\sqrt{1/4\pi}$.

Finally, the complete wave-function that fulfills the Pauli principle is written as
\begin{equation}
\label{TPs}
\Psi_{JM_J,I,i,j,k}(\theta)=\left[ \left[ \psi_{L}(\theta) \chi^{\sigma_i}_{S}(4) \right]_{JM_J} \chi^{f_j}_I \chi^{c}_k \right] \,.
\end{equation}

%%%%%%%%%%%%%%%%%%%%%%%%%%%%%%%%%%%%%%%%%%%%%%%%%%%%%%%%%%%%%%%%%%%%%%%%%%%%%%%%

\section{Results}
\label{sec:results}

In the present calculation, we investigate all possible $S$-wave $b\bar{b}q\bar{q}$ $(q=u,\,d,\,s)$ tetraquarks by taking into account di-meson, diquark-antidiquark and K-type configurations. In our approach, a $b\bar{b}q\bar{q}$ tetraquark state has positive parity assuming that the angular momenta $l_1$, $l_2$ and $l_3$ in Eq.~\eqref{eq:WFexp} are all equal to zero. Accordingly, the total angular momentum, $J$, coincides with the total spin, $S$, and can take values of $0$, $1$ and $2$. Besides, the value of isospin, $I$, can be either $0$ or $1$ considering the quark content of $b\bar{b}q\bar{q}$ system in the $SU(2)$ flavor symmetry; however, it is $I=1/2$ for $b\bar{b}q\bar{s}$ and just $0$ for $b\bar{b}s\bar{s}$.

Tables~\ref{GresultCC1} to~\ref{GresultCC12} list our calculated results of the lowest-lying $b\bar{b}q\bar{q}$ tetraquark states. The allowed meson-meson, diquark-antidiquark and K-type configurations are listed in the first column; when possible, the experimental value of the non-interacting meson-meson threshold is labeled in parentheses. Each channel is assigned an index in the second column, it reflects a particular combination of spin ($\chi_J^{\sigma_i}$), flavor ($\chi_I^{f_j}$) and color ($\chi_k^c$) wave functions that are shown explicitly in the third column. The theoretical mass obtained in each channel is shown in the fourth column and the coupled result for each kind of configuration is presented in the last column. When a complete coupled-channels calculation is performed, last row of the table indicates the lowest-lying mass. When the CSM is used in the complete coupled-channels calculation, we show in Figs.~\ref{PP1} to~\ref{PP12}, the distribution of complex eigen-energies and, therein, the obtained resonance states are indicated inside circles.

Let us proceed now to describe in detail our theoretical findings for each sector of $b\bar{b}q\bar{q}$ tetraquarks.

\subsection{The $\mathbf{b\bar{b}q\bar{q}}\,(q=u,\,d)$ tetraquarks}

%Several resonances whose masses ranging from $10.58$ GeV to $11.42$ GeV are obtained in this tetraquark sector. 
Each iso-scalar and -vector sectors with total spin and parity $J^P=0^+$, $1^+$ and $2^+$ shall be discussed individually below. Particularly, due to symmetry properties of the Hamiltonian with respect to the K-type configurations of $b\bar{b}q\bar{q}$ tetraquarks, only $K_1$ and $K_3$ arrangements are sufficient to consider.

\begin{table}[!t]
\caption{\label{GresultCC1} Lowest-lying $b\bar{b}q\bar{q}$ tetraquark states with $I(J^P)=0(0^+)$ calculated within the real range formulation of the chiral quark model.
The allowed meson-meson, diquark-antidiquark and K-type configurations are listed in the first column; when possible, the experimental value of the non-interacting meson-meson threshold is labeled in parentheses. Each channel is assigned an index in the 2nd column, it reflects a particular combination of spin ($\chi_J^{\sigma_i}$), flavor ($\chi_I^{f_j}$) and color ($\chi_k^c$) wave functions that are shown explicitly in the 3rd column. The theoretical mass obatined in each channel is shown in the 4th column and the coupled result for each kind of configuration is presented in the 5th column.
When a complete coupled-channels calculation is performed, last row of the table indicates the lowest-lying mass.
(unit: MeV).}
\begin{ruledtabular}
\begin{tabular}{lcccc}
~~Channel   & Index & $\chi_J^{\sigma_i}$;~$\chi_I^{f_j}$;~$\chi_k^c$ & $M$ & Mixed~~ \\
        &   &$[i; ~j; ~k]$ &  \\[2ex]
$(\eta_b \eta)^1 (9848)$          & 1  & [1;~1;~1]  & $10143$ & \\
$(\Upsilon \omega)^1 (10242)$  & 2  & [2;~1;~1]   & $10201$ &  \\
$(B \bar{B})^1 (10560)$          & 3  & [1;~1;~1]  & $10556$ & \\
$(B^* \bar{B}^*)^1 (10650)$  & 4  & [2;~1;~1]   & $10638$ & $10143$ \\[2ex]
$(\eta_b \eta)^8$          & 5  & [1;~1;~2]  & $10959$ & \\
$(\Upsilon \omega)^8$  & 6  & [2;~1;~2]   & $10814$ &  \\
$(B \bar{B})^8$          & 7  & [1;~1;~2]  & $10680$ & \\
$(B^* \bar{B}^*)^8$  & 8  & [2;~1;~2]   & $10666$ & $10589$ \\[2ex]
$(bq)(\bar{q}\bar{b})$      & 9   & [3;~1;~3]  & $10774$ & \\
$(bq)(\bar{q}\bar{b})$      & 10   & [3;~1;~4]  & $10729$ & \\
$(bq)^*(\bar{q}\bar{b})^*$  & 11  & [4;~1;~3]   & $10829$ & \\
$(bq)^*(\bar{q}\bar{b})^*$  & 12  & [4;~1;~4]   & $10708$ & $10514$ \\[2ex]
$K_1$  & 13  & [5;~1;~5]   & $10817$ & \\
  & 14  & [6;~1;~5]   & $10960$ & \\
  & 15  & [5;~1;~6]   & $10538$ & \\
  & 16  & [6;~1;~6]   & $10484$ & $10483$ \\[2ex]
$K_3$  & 17  & [9;~1;~9]   & $10700$ & \\
  & 18  & [10;~1;~9]   & $10722$ & \\
  & 19  & [9;~1;~10]   & $10825$ & \\
  & 20  & [10;~1;~10]   & $10767$ & $10507$ \\[2ex]
\multicolumn{4}{c}{Complete coupled-channels:} & $10143$
\end{tabular}
\end{ruledtabular}
\end{table}

\begin{figure}[!t]
\includegraphics[clip, trim={3.0cm 1.5cm 3.0cm 1.0cm}, width=0.45\textwidth]{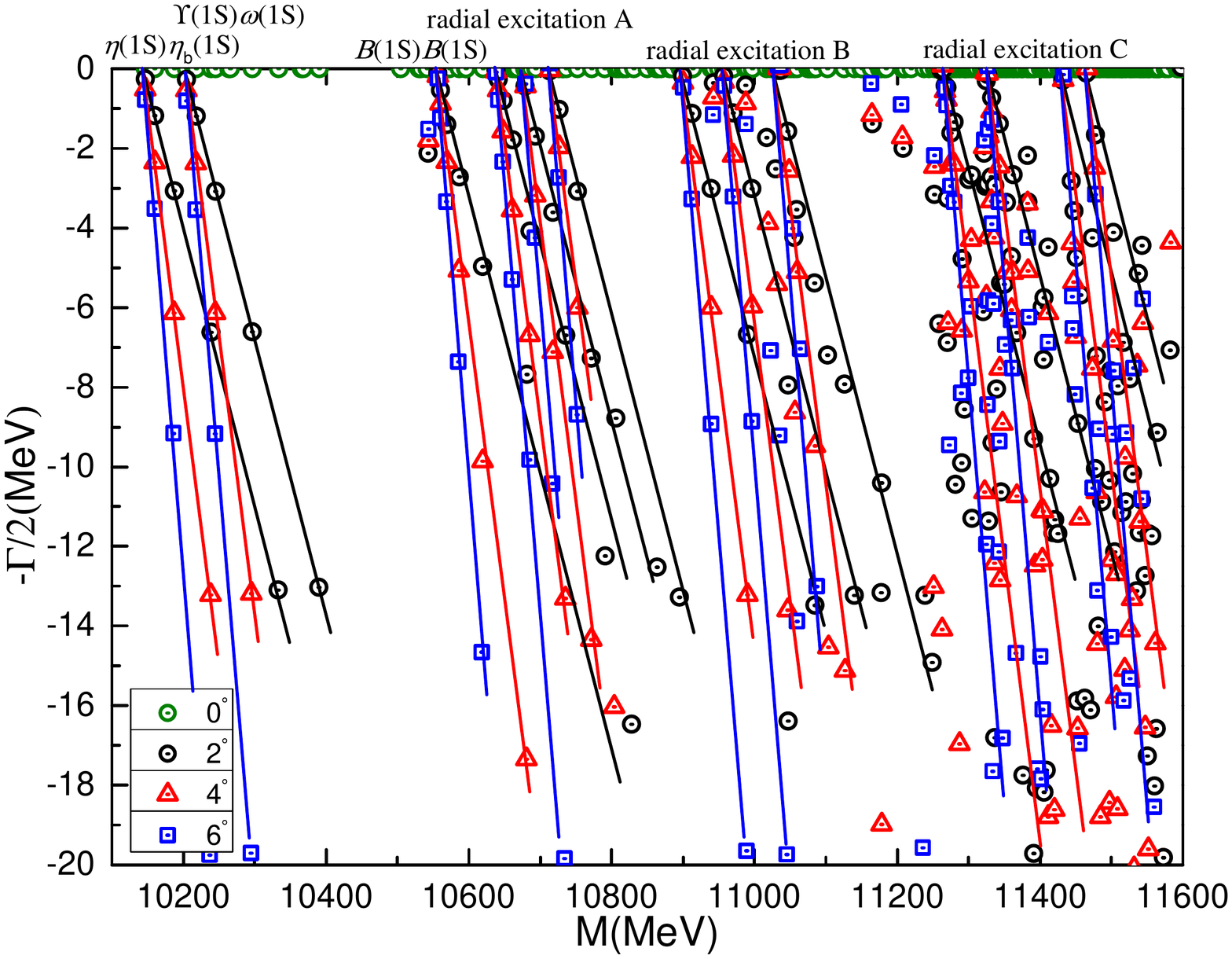} \\
\includegraphics[clip, trim={3.0cm 1.5cm 3.0cm 1.0cm}, width=0.45\textwidth]{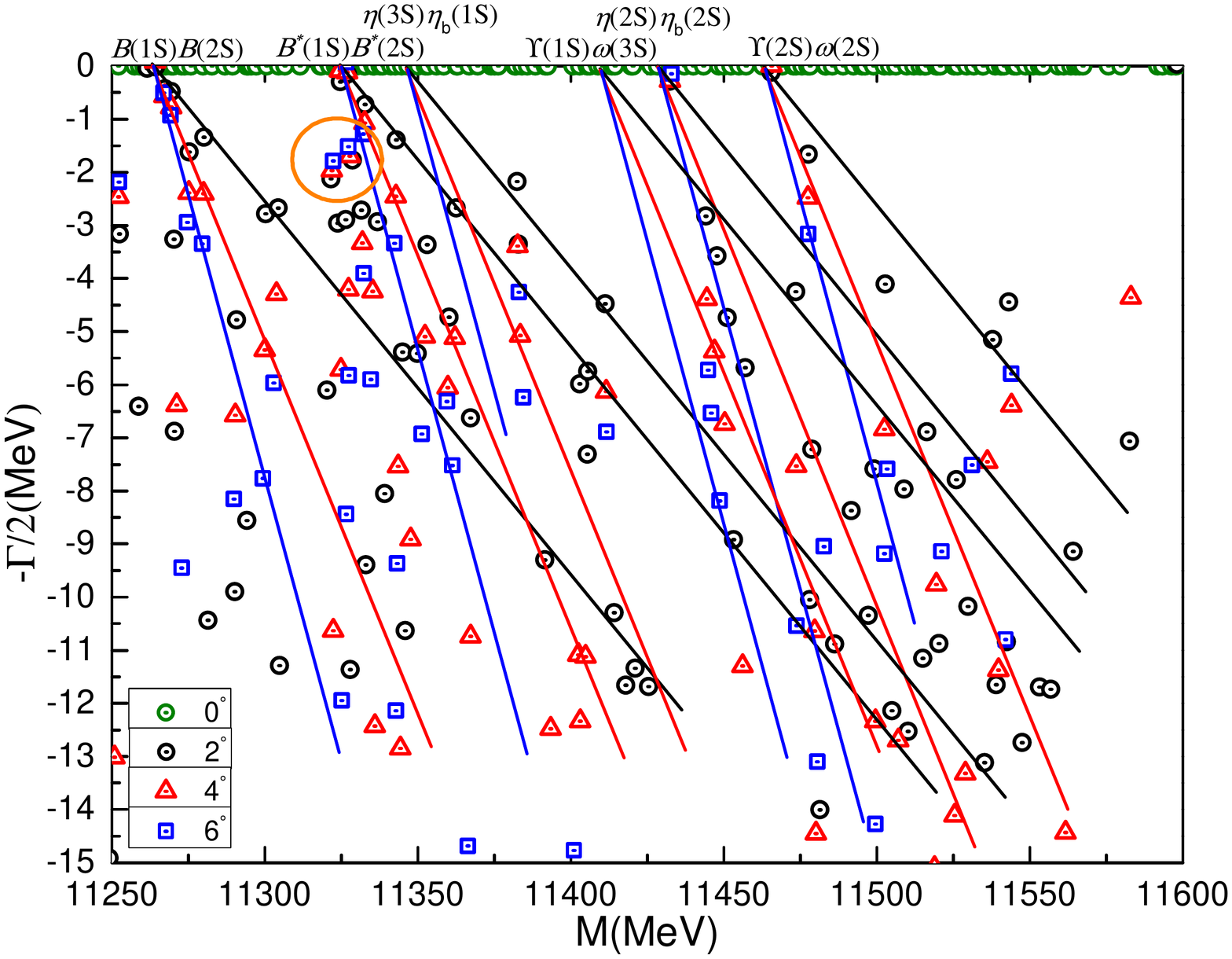}
\caption{\label{PP1} {\it Top panel:} The complete coupled-channels calculation of $b\bar{b}q\bar{q}$ tetraquark system with $I(J^P)=0(0^+)$ quantum numbers. {\it Bottom panel:} Enlarged top panel, with real values of energy ranging from $11.25\,\text{GeV}$ to $11.60\,\text{GeV}$.
 We use the complex-scaling method of the chiral quark model varying $\theta$ from $0^\circ$ to $6^\circ$.}
\end{figure}

{\bf The $\bm{I(J^P)=0(0^+)}$ sector:} 
Four meson-meson channels, $\eta_b \eta$, $\Upsilon \omega$, $B \bar{B}$ and $B^* \bar{B}^*$ in both color-singlet and hidden-color configurations, four diquark-antidiquark channels, along with K-type configurations are individually studied in Table~\ref{GresultCC1}. One can see that the formation of a bound state is not possible in each single channel calculation. The four singlet-color meson-meson states are located above their thresholds with masses at 10.14 GeV, 10.20 GeV, 10.55 GeV and 10.64 GeV, respectively. The other tetraquark configurations: hidden-color meson-meson, diquark-antidiquark, and K-type channels, present states in an energy region which ranges from 10.70 GeV to 10.96 GeV. After a coupled-channels calculation among each type of configuration is performed, the lowest mass is around 10.5 GeV for the three exotic structures hidden-color, diquark-antidiquark and K-type. However, the lowest mass in singlet-color coupled-channels calculation remains at the $\eta_b \eta$ theoretical threshold value of 10.14 GeV. This result also holds for the fully coupled-channels computation in a real-range investigation.

In a further step, a complete coupled-channels calculation is performed using the complex-range method and its output is shown in Fig.~\ref{PP1}. Therein, with a rotated angle ranging from $0^\circ$ to $6^\circ$, the scattering states of two mesons are well presented within an interval of $10.10-11.60$ GeV. In particular, apart from the three lower scattering states, $\eta_b(1S) \eta(1S)$, $\Upsilon(1S)\omega(1S)$ and $B(1S)\bar{B}(1S)$, presented in the top panel, the other dimeson channels, listed in Table~\ref{GresultCC1}, along with their radial excitations are generally located in the regions of $A$, $B$ and $C$ of Fig.~\ref{PP1}, respectively. In particular, there are three channels, $B^*(1S)\bar{B}^*(1S)$, $\eta_b(2S)\eta(1S)$ and $\Upsilon(2S)\omega(1S)$ in the region $A$, another three in $B$ which are $\eta_b(1S)\eta(2S)$, $\Upsilon(1S)\omega(2S)$ and $\Upsilon(3S)\omega(1S)$; and six radial excitation channels of region $C$ are clearly shown in the bottom panel of Fig.~\ref{PP1}. Therein, two narrow resonance poles are obtained above the $B(1S)\bar{B}(2S)$ threshold lines. Their calculated complex energies, which are denoted as $M+i\Gamma$, are $11322+i3.9$ MeV and $11328+i3.4$ MeV, respectively. The dominant decay channel of these almost degenerate resonances could be the $B(1S)\bar{B}(2S)$ state.

%%%%%%%%%%%%%%%%%%%%%%%%%%%%%%%%%%%%%%%%

\begin{table}[!t]
\caption{\label{GresultCC2} Lowest-lying $b\bar{b}q\bar{q}$ tetraquark states with $I(J^P)=0(1^+)$ calculated within the real range formulation of the chiral quark model. The results are similarly organized as those in Table~\ref{GresultCC1}.
(unit: MeV).}
\begin{ruledtabular}
\begin{tabular}{lcccc}
~~Channel   & Index & $\chi_J^{\sigma_i}$;~$\chi_I^{f_j}$;~$\chi_k^c$ & $M$ & Mixed~~ \\
        &   &$[i; ~j; ~k]$ &  \\[2ex]
$(\eta_b \omega)^1 (10082)$          & 1  & [1;~1;~1]  & $10150$ & \\
$(\Upsilon \eta)^1 (10008)$  & 2  & [2;~1;~1]   & $10194$ &  \\
$(\Upsilon \omega)^1 (10242)$  & 3  & [3;~1;~1]   & $10201$ &  \\
$(B \bar{B}^*)^1 (10605)$          & 4  & [1;~1;~1]  & $10597$ & \\
$(B^* \bar{B}^*)^1 (10650)$  & 5  & [3;~1;~1]   & $10638$ & $10150$ \\[2ex]
$(\eta_b \omega)^8$  & 6  & [1;~1;~2]  & $10831$ & \\
$(\Upsilon \eta)^8$    & 7  & [2;~1;~2]   & $10960$ &  \\
$(\Upsilon \omega)^8$  & 8  & [3;~1;~2]   & $10823$ &  \\
$(B \bar{B}^*)^8$          & 9 & [1;~1;~2]  & $10681$ & \\
$(B^* \bar{B}^*)^8$      & 10  & [3;~1;~2]   & $10667$ & $10587$ \\[2ex]
$(bq)(\bar{q}\bar{b})^*$      & 11   & [4;~1;~3]  & $10788$ & \\
$(bq)(\bar{q}\bar{b})^*$      & 12   & [4;~1;~4]  & $10807$ & \\
$(bq)^*(\bar{q}\bar{b})^*$  & 13  & [6;~1;~3]   & $10725$ & \\
$(bq)^*(\bar{q}\bar{b})^*$  & 14  & [6;~1;~4]   & $10705$ & $10524$ \\[2ex]
$K_1$  & 15  & [7;~1;~5]   & $10922$ & \\
  & 16  & [8;~1;~5]   & $10878$ & \\
  & 17  & [9;~1;~5]   & $10835$ & \\
  & 18  & [7;~1;~6]   & $10536$ & \\
  & 19  & [8;~1;~6]   & $10537$ & \\
  & 20  & [9;~1;~6]   & $10487$ & $10487$ \\[2ex]
$K_3$  & 21  & [13;~1;~9]   & $10718$ & \\
  & 22  & [14;~1;~9]   & $10677$ & \\
  & 23  & [15;~1;~9]   & $10717$ & \\
  & 24  & [13;~1;~10]   & $10813$ & \\
  & 25  & [14;~1;~10]   & $10744$ & \\
  & 26  & [15;~1;~10]   & $10780$ & $10511$ \\[2ex]
\multicolumn{4}{c}{Complete coupled-channels:} & $10150$
\end{tabular}
\end{ruledtabular}
\end{table}

\begin{figure}[!t]
\includegraphics[clip, trim={3.0cm 1.9cm 3.0cm 1.0cm}, width=0.45\textwidth]{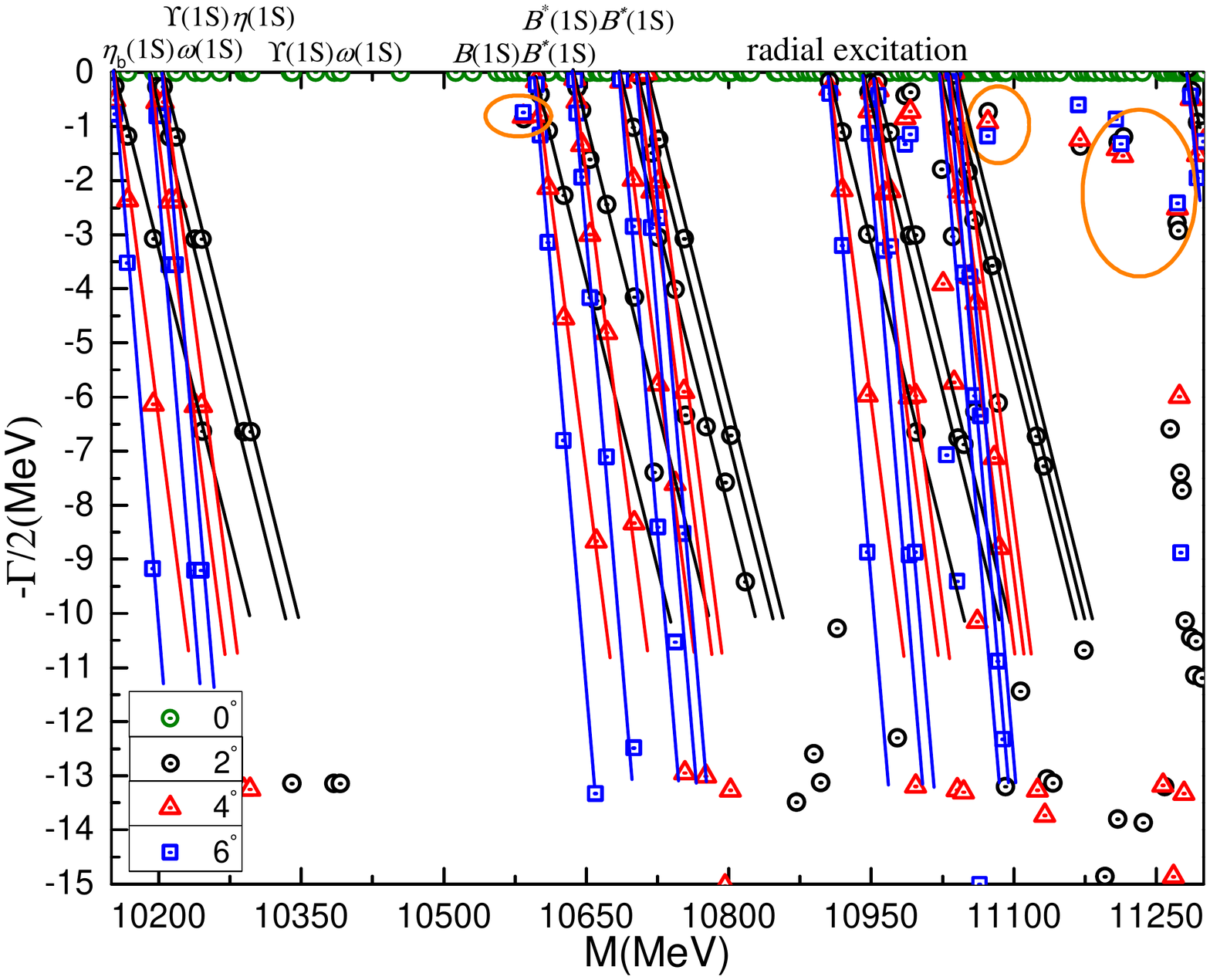} \\
\includegraphics[clip, trim={3.0cm 1.9cm 3.0cm 1.0cm}, width=0.45\textwidth]{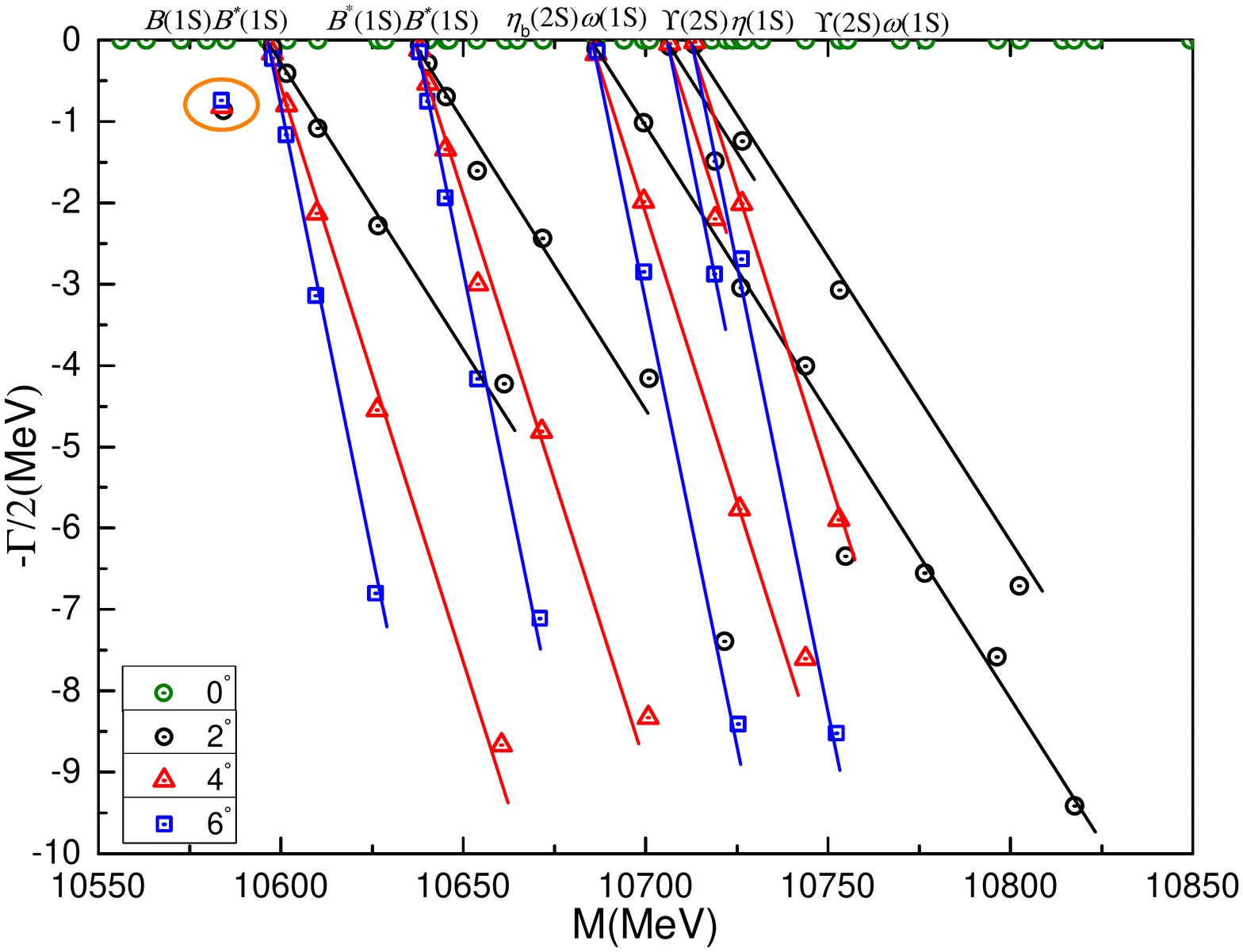} \\
\includegraphics[clip, trim={3.0cm 1.9cm 3.0cm 1.0cm}, width=0.45\textwidth]{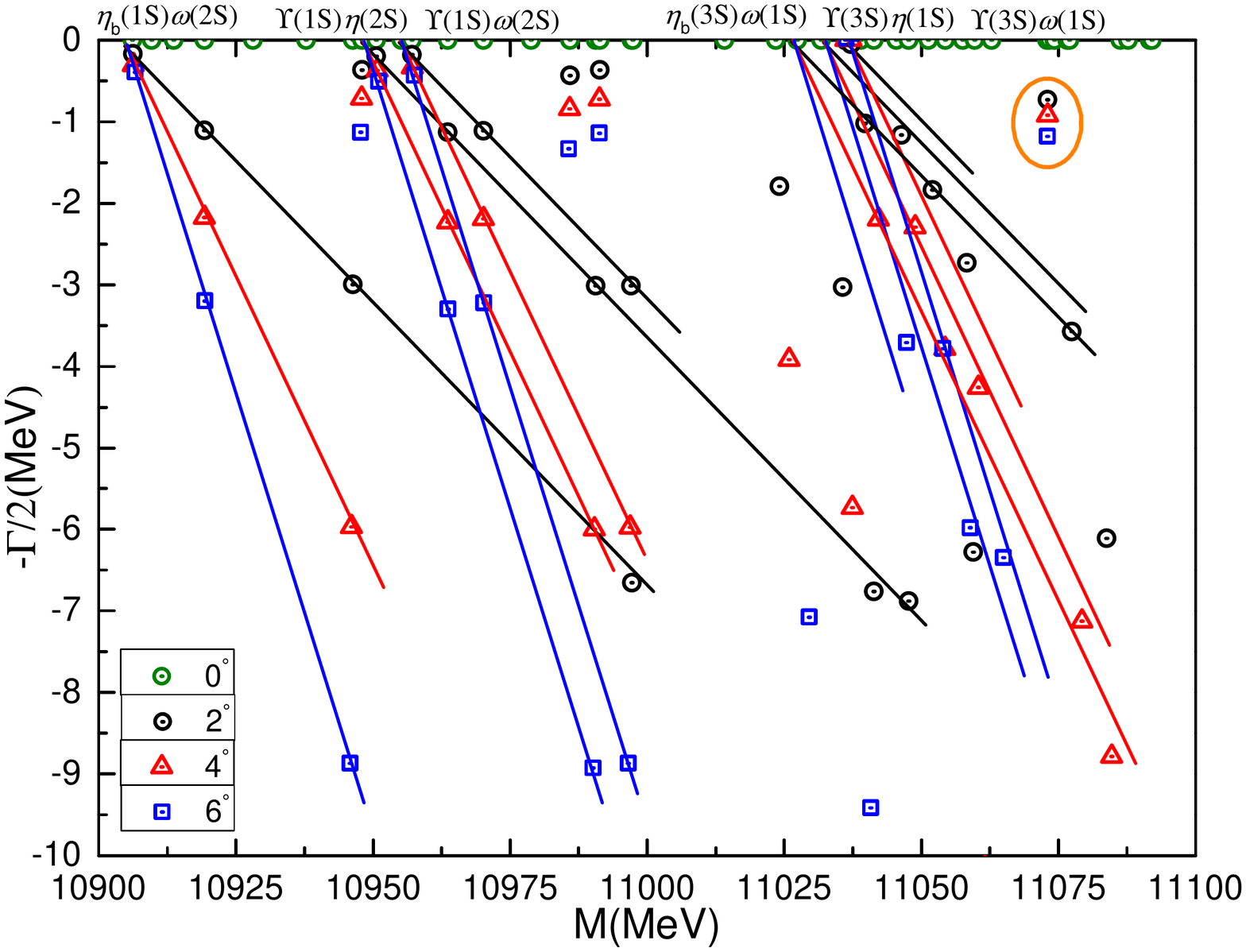}
\caption{\label{PP2} The complete coupled-channels calculation of $b\bar{b}q\bar{q}$ tetraquark system with $I(J^P)=0(1^+)$. Particularly, middle and bottom panels are the enlarged parts of dense energy region from $10.55\,\text{GeV}$ to $11.10\,\text{GeV}$.}
\end{figure}

{\bf The $\bm{I(J^P)=0(1^+)}$ sector:} There are 26 channels in this sector which includes five color-singlet meson-meson configurations, another five in the hidden-color meson-meson ones, four more in diquark-antidiquark arrangement, and 12 K-type configurations. Table~\ref{GresultCC2} shows the obtained masses in each channels which are always above the lowest meson-meson threshold. The five di-meson structures in color-singlet configuration have masses within an energy region of $10.15-10.60$ GeV. 
Furthermore, masses of the hidden-color $(b\bar{b})$$(q\bar{q})$ configurations are $\sim$10.90 GeV and $\sim$10.67 GeV for the $(q\bar{b})$$(b\bar{q})$ ones, respectively; diquark-antidiquark channels are characterized with masses ranging from $10.70$ GeV to $10.81$ GeV; and, for the K-type configurations, each calculated channel presents a mass lying in an interval of $10.48-10.92$ GeV.

If one considers the coupling between channels of the same kind of tetraquark configuration, the lowest masses are located at $\sim$10.5 GeV for the exotic structures hidden-color, diquark-antidiquark and K-type. However, the induced binding due to coupling is extremely weak in color-singlet di-meson case and the lowest mass remains at 10.15 GeV of the $\eta_b\omega$ theoretical threshold value. Furthermore, the same result is also obtained in a real-range calculation when all of the channels listed in Table~\ref{GresultCC2} are considered. Hence, bound state is excluded.

Figure~\ref{PP2} presents the calculated results of fully coupled-channels case with the complex-range method employed. The general distributions of complex energies of $b\bar{b}q\bar{q}$ system are presented in the top panel of Fig.~\ref{PP2}, whose energy range is from 10.10 to 11.30 GeV. Within a rotated angle ranging from $0^\circ$ to $6^\circ$, the scattering nature of $\eta_b \omega$, $\Upsilon \eta$, $\Upsilon \omega$ and $B^{(*)}\bar{B}^{(*)}$ are clearly shown. However, four stable poles are obtained against the variation of rotated angle $\theta$. Their computed masses$(M)$ and widths$(\Gamma)$ read $10584+i1.6$ MeV, $11073+i1.8$ MeV, $11209+i2.8$ MeV and $11272+i5.0$ MeV, respectively. 

In order to have a better analysis of these resonances, two regions of dense radial excitations whose energy intervals are $10.55-10.85$ GeV and $10.90-11.10$ GeV, respectively, are plotted in the middle and bottom panels of Fig.~\ref{PP2}. From the middle panel, we could find that one resonance pole is quite close to the $B(1S)\bar{B}^*(1S)$ threshold line(s). Apparently, this resonance with $I(J^P)=0(1^+)$ has a mass similar to the the exotic state $Z_b(10610)$ but cannot be assigned because the experimental candidate has isospin one.

Additionally, in the high energy region of the bottom panel, the radial excited states of $\eta_b \omega$, $\Upsilon \eta$ and $\Upsilon \omega$ are generally presented. Moreover, there is also an indication of a resonance above the $\Upsilon(3S)\omega(1S)$ threshold, and this pole is located at $11073+i1.8$ MeV. A similar conclusion is obtained at around 11.2 GeV, therein, two resonances with $11209+i2.8$ MeV and $11272+i5.0$ MeV are clearly shown in the big circle of top panel. The $\Upsilon(3S)\omega(1S)$ should be the dominant decay channel of these three resonances and they are expected to be identified in the future high energy experiments with mass $\sim$11.2 GeV and quantum numbers $I(J^P)=0(1^+)$.

%%%%%%%%%%%%%%%%%%%%%%%%%%%%%%%%%%%%%%%%

\begin{table}[!t]
\caption{\label{GresultCC3} Lowest-lying $b\bar{b}q\bar{q}$ tetraquark states with $I(J^P)=0(2^+)$ calculated within the real range formulation of the chiral quark model. The results are similarly organized as those in Table~\ref{GresultCC1}.
(unit: MeV).}
\begin{ruledtabular}
\begin{tabular}{lcccc}
~~Channel   & Index & $\chi_J^{\sigma_i}$;~$\chi_I^{f_j}$;~$\chi_k^c$ & $M$ & Mixed~~ \\
        &   &$[i; ~j; ~k]$ &  \\[2ex]
$(\Upsilon \omega)^1 (10242)$  & 1  & [1;~1;~1]   & $10201$ &  \\
$(B^* \bar{B}^*)^1 (10650)$  & 2  & [1;~1;~1]   & $10626$ & $10201$ \\[2ex]
$(\Upsilon \omega)^8$  & 3  & [1;~1;~2]   & $10840$ &  \\
$(B^* \bar{B}^*)^8$  & 4  & [1;~1;~2]   & $10670$ & $10604$ \\[2ex]
$(bq)^*(\bar{q}\bar{b})^*$  & 5  & [1;~1;~3]   & $10756$ & \\
$(bq)^*(\bar{q}\bar{b})^*$  & 6  & [1;~1;~4]   & $10699$ & $10544$ \\[2ex]
$K_1$  & 7  & [1;~1;~5]   & $10845$ & \\
  & 8  & [1;~1;~6]   & $10538$ & $10538$ \\[2ex]
$K_3$  & 9  & [1;~1;~9]   & $10690$ & \\
  & 10  & [1;~1;~10]   & $10743$ & $10543$ \\[2ex]
\multicolumn{4}{c}{Complete coupled-channels:} & $10201$
\end{tabular}
\end{ruledtabular}
\end{table}

\begin{figure}[!t]
\includegraphics[width=0.49\textwidth, trim={2.3cm 2.0cm 2.0cm 1.0cm}]{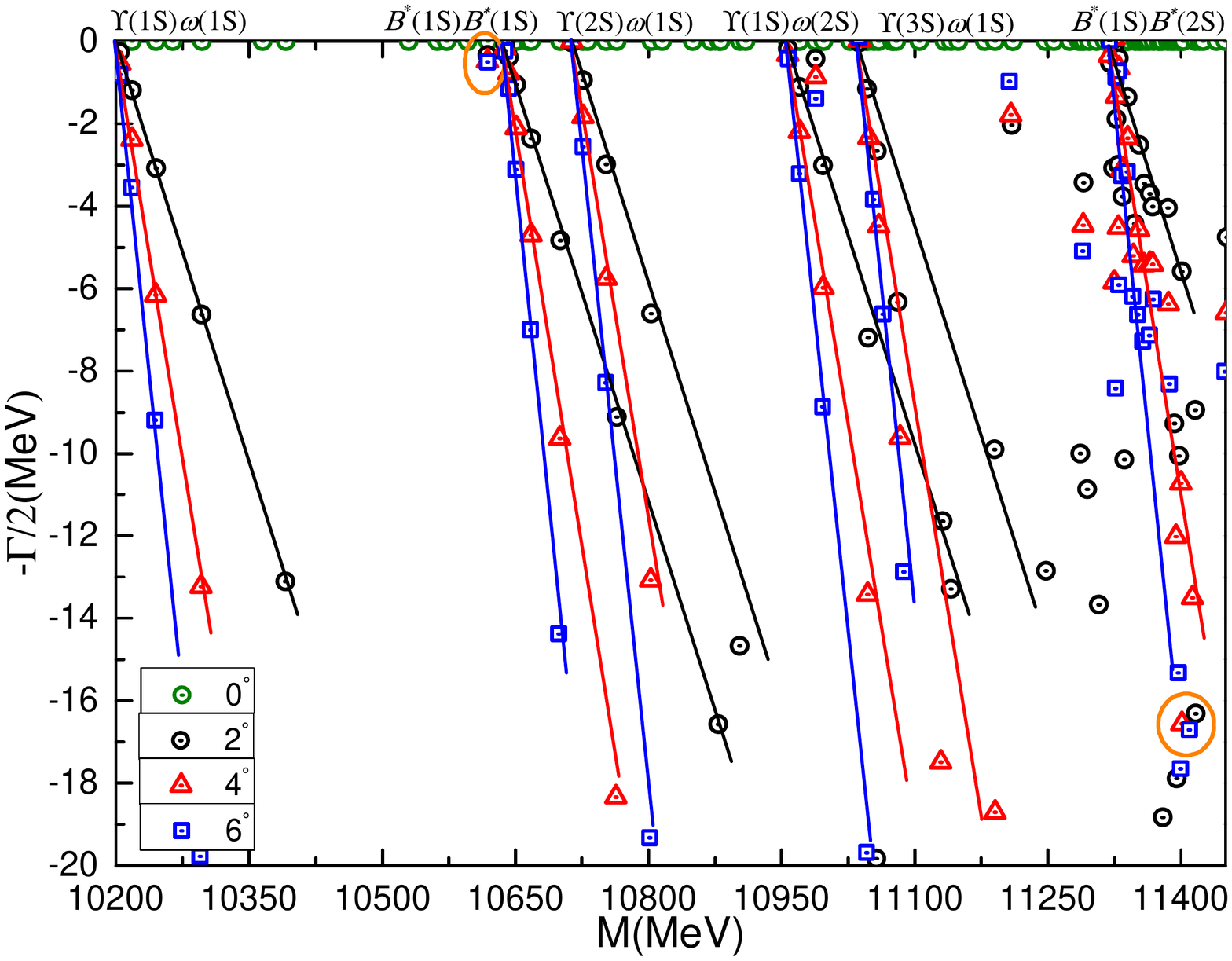}
\caption{\label{PP3} The complete coupled-channels calculation of the $b\bar{b}q\bar{q}$ tetraquark system with $I(J^P)=0(2^+)$ quantum numbers. We use the complex-scaling method of the chiral quark model varying $\theta$ from $0^\circ$ to $6^\circ$.}
\end{figure}

{\bf The $\bm{I(J^P)=0(2^+)}$ state:} Table~\ref{GresultCC3} shows that two meson-meson structures, $\Upsilon\omega$ and $B^* \bar{B}^*$ (in both color-singlet and hidden-color configurations), one $(bq)^*(\bar{q}\bar{b})^*$ diquark-antidiquark arrangement and four K-type configurations contribute to the $I(J^P)=0(2^+)$ state. 

Firstly, in each single channel calculation, a $B^* \bar{B}^*$ bound state, which binding energy is $-12$ MeV, is found in its color-singlet channel, and the modified mass is $10638$ MeV. Although this value is quite compatible with $Z_b(10650)$ mass, a conclusion can not be drawn herein for coupling effect is not considered temporarily and the spin-parity conflicts with $1^+$ experimentally. Meanwhile, the lowest channel, $\Upsilon \omega$, is of scattering nature with theoretical mass at $10.20$ GeV. The other exotic configurations: hidden-color dimeson, diquark-antidiquark and K-type, are generally located above 10.7 GeV, except for one $K_1$ channel with a calculated mass at $10538$ MeV.

When real-range coupled-channel computations are performed in each kind of structure, firstly, in the color-singlet channels, the lowest energy remains at $\Upsilon \omega$ threshold value of $10.20$ GeV (this result is unchanged even in a fully coupled case), and masses of the other four multiquark arrangements are around $10.54$ GeV.
 
In a further step, Fig.~\ref{PP3} shows the calculated results obtained in a complex-range investigation. The $\Upsilon \omega$ and $B^* \bar{B}^*$ thresholds, considering each meson in both ground and radial excitation, are well presented within an energy interval of $10.20-11.50$ GeV. Moreover, two resonance states are obtained (circled in the figure) at $10618+i1.0$ MeV and $11416+i32.6$ MeV, respectively. The narrow state is quite close to $B^*(1S)\bar{B}^*(1S)$ threshold line, whereas the wide resonance is located at the $B^*(1S)\bar{B}^*(2S)$ threshold line. 

%%%%%%%%%%%%%%%%%%%%%%%%%%%%%%%%%%%%%%%%

\begin{table}[!t]
\caption{\label{GresultCC4} Lowest-lying $b\bar{b}q\bar{q}$ tetraquark states with $I(J^P)=1(0^+)$ calculated within the real range formulation of the chiral quark model. The results are similarly organized as those in Table~\ref{GresultCC1}.
(unit: MeV).}
\begin{ruledtabular}
\begin{tabular}{lcccc}
~~Channel   & Index & $\chi_J^{\sigma_i}$;~$\chi_I^{f_j}$;~$\chi_k^c$ & $M$ & Mixed~~ \\
        &   &$[i; ~j; ~k]$ &  \\[2ex]
$(\eta_b \pi)^1 (9440)$          & 1  & [1;~1;~1]  & $9603$ & \\
$(\Upsilon \rho)^1 (10230)$  & 2  & [2;~1;~1]   & $10277$ &  \\
$(B \bar{B})^1 (10560)$          & 3  & [1;~1;~1]  & $10556$ & \\
$(B^* \bar{B}^*)^1 (10650)$  & 4  & [2;~1;~1]   & $10634$ & $9603$ \\[2ex]
$(\eta_b \pi)^8$          & 5  & [1;~1;~2]  & $10830$ & \\
$(\Upsilon \rho)^8$  & 6  & [2;~1;~2]   & $10875$ &  \\
$(B \bar{B})^8$          & 7  & [1;~1;~2]  & $10680$ & \\
$(B^* \bar{B}^*)^8$  & 8  & [2;~1;~2]   & $10601$ & $10455$ \\[2ex]
$(bq)(\bar{q}\bar{b})$      & 9   & [3;~1;~3]  & $10774$ & \\
$(bq)(\bar{q}\bar{b})$      & 10   & [3;~1;~4]  & $10740$ & \\
$(bq)^*(\bar{q}\bar{b})^*$  & 11  & [4;~1;~3]   & $10729$ & \\
$(bq)^*(\bar{q}\bar{b})^*$  & 12  & [4;~1;~4]   & $10622$ & $10383$ \\[2ex]
$K_1$  & 13  & [5;~1;~5]   & $10868$ & \\
  & 14  & [6;~1;~5]   & $10832$ & \\
  & 15  & [5;~1;~6]   & $10570$ & \\
  & 16  & [6;~1;~6]   & $10349$ & $10348$ \\[2ex]
$K_3$  & 17  & [9;~1;~9]   & $10607$ & \\
  & 18  & [10;~1;~9]   & $10722$ & \\
  & 19  & [9;~1;~10]   & $10726$ & \\
  & 20  & [10;~1;~10]   & $10767$ & $10371$ \\[2ex]
\multicolumn{4}{c}{Complete coupled-channels:} & $9603$
\end{tabular}
\end{ruledtabular}
\end{table}

\begin{figure}[!t]
\includegraphics[width=0.49\textwidth, trim={2.3cm 2.0cm 2.0cm 1.0cm}]{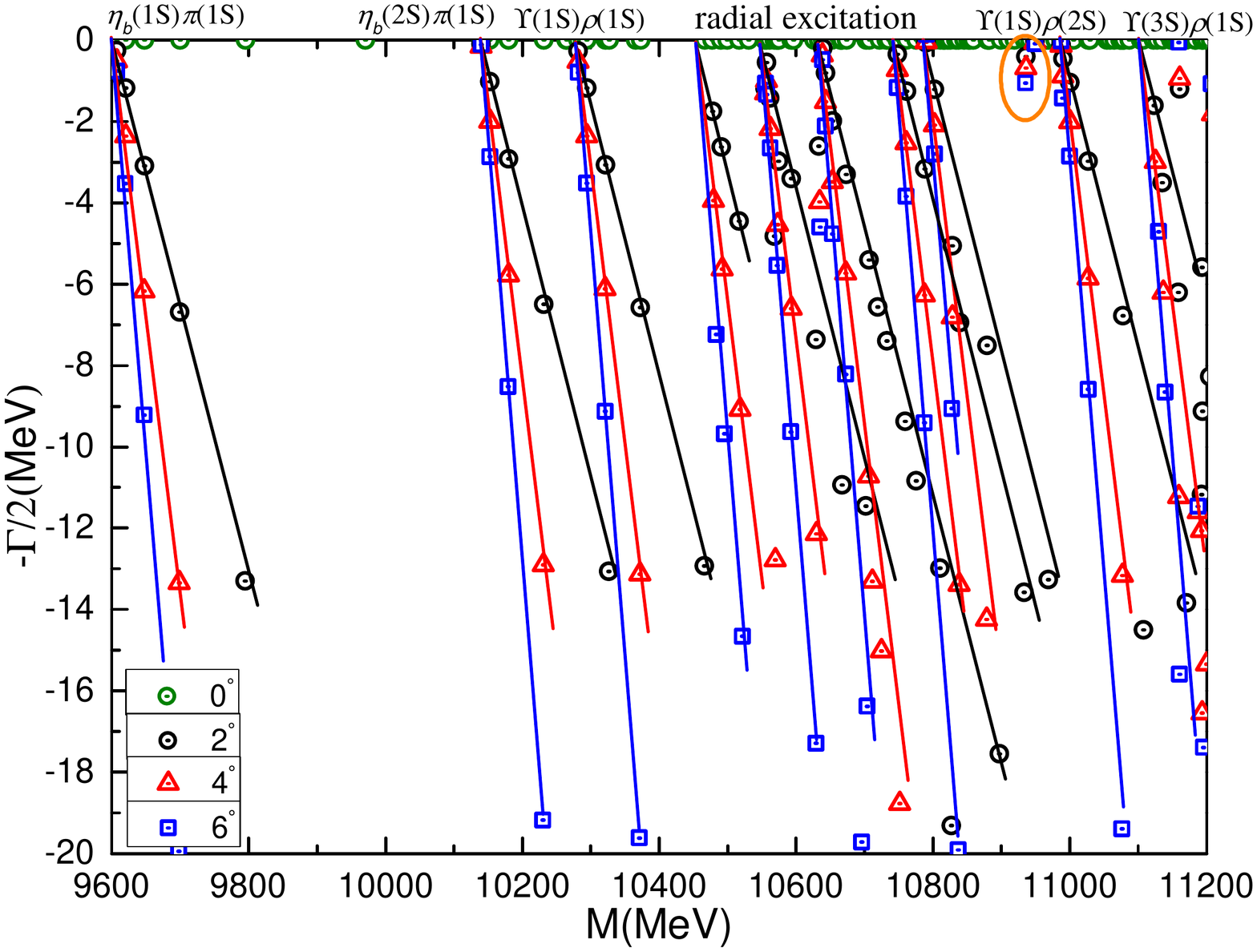}
\caption{\label{PP4} The complete coupled-channels calculation of the $b\bar{b}q\bar{q}$ tetraquark system with $I(J^P)=1(0^+)$ quantum numbers. We use the complex-scaling method of the chiral quark model varying $\theta$ from $0^\circ$ to $6^\circ$.}
\end{figure}

{\bf The $\bm{I(J^P)=1(0^+)}$ sector:} Table~\ref{GresultCC4} lists our results for the isovector $b\bar{b}q\bar{q}$ tetraquark with spin-parity $J^P=0^+$. As in the case of the $I(J^P)=0(0^+)$ sector, 20 channels are under investigation and no one shows a bound state, except the color-singlet channel $B^* \bar{B}^*$ with a very small binding energy $E_b=-4$ MeV. Apart from the lowest channel $\eta_b \pi$ $(M=9603)$, masses of color-singlet meson-meson configurations are located at $10.28-10.63$ GeV. The other color configurations are in a mass region between $10.57$ GeV and $10.87$ GeV, note however that $K_1$ channel has a mass of $10.35$ GeV. Furthermore, in each exotic configuration's coupled-channel computation, the lowest mass tends to $10.35$ GeV, and the scattering state of $\eta_b \pi$ at $9603$ MeV remains unchanged in various analysis of this kind.

The complete coupled-channels calculation has been also extended to the complex-range. Figure~\ref{PP4} shows the distribution of complex energies within an energy range $9.6-11.2$ GeV. The scattering nature of $\eta_b\pi$, $\Upsilon \rho$ and $B^{(*)} \bar{B}^{(*)}$, both in ground and radial excitation states, is clearly presented. In particular, there are five scattering states, $\eta_b(3S)\pi(1S)$, $B(1S)\bar{B}(1S)$, $B^*(1S) \bar{B}^*(1S)$ $\eta_b(1S)\pi(2S)$ and $\Upsilon(2S)\rho(1S)$, located in an energy region from $10.4$ GeV to $10.9$ GeV. With respect to the shallow $B^* \bar{B}^*$ bound state, whose mass is $10634$ MeV, found in a color-singlet channel calculation, it disappears in the complete coupled-channel situation. Besides, one narrow resonance circled in Fig.~\ref{PP4} is obtained with a complex energy of $10935+i1.4$ MeV. This pole locates between the threshold lines of $\Upsilon(2S)\rho(1S)$ and $\Upsilon(1S)\rho(2S)$, hence the dominate decay channel should be $\Upsilon(2S)\rho(1S)$.

\begin{table}[!t]
\caption{\label{GresultCC5} Lowest-lying $b\bar{b}q\bar{q}$ tetraquark states with $I(J^P)=1(1^+)$ calculated within the real range formulation of the chiral quark model. The results are similarly organized as those in Table~\ref{GresultCC1}.
(unit: MeV).}
\begin{ruledtabular}
\begin{tabular}{lcccc}
~~Channel   & Index & $\chi_J^{\sigma_i}$;~$\chi_I^{f_j}$;~$\chi_k^c$ & $M$ & Mixed~~ \\
        &   &$[i; ~j; ~k]$ &  \\[2ex]
$(\eta_b \rho)^1 (10070)$          & 1  & [1;~1;~1]  & $10226$ & \\
$(\Upsilon \pi)^1 (9600)$  & 2  & [2;~1;~1]   & $9654$ &  \\
$(\Upsilon \rho)^1 (10230)$  & 3  & [3;~1;~1]   & $10277$ &  \\
$(B \bar{B}^*)^1 (10605)$          & 4  & [1;~1;~1]  & $10597$ & \\
$(B^* \bar{B}^*)^1 (10650)$  & 5  & [3;~1;~1]   & $10636$ & $9654$ \\[2ex]
$(\eta_b \rho)^8$  & 6  & [1;~1;~2]  & $10890$ & \\
$(\Upsilon \pi)^8$    & 7  & [2;~1;~2]   & $10830$ &  \\
$(\Upsilon \rho)^8$  & 8  & [3;~1;~2]   & $10883$ &  \\
$(B \bar{B}^*)^8$          & 9 & [1;~1;~2]  & $10681$ & \\
$(B^* \bar{B}^*)^8$      & 10  & [3;~1;~2]   & $10637$ & $10518$ \\[2ex]
$(bq)(\bar{q}\bar{b})^*$      & 11   & [4;~1;~3]  & $10788$ & \\
$(bq)^*(\bar{q}\bar{b})$      & 12   & [4;~1;~4]  & $10762$ & \\
$(bq)^*(\bar{q}\bar{b})^*$  & 13  & [6;~1;~3]   & $10725$ & \\
$(bq)^*(\bar{q}\bar{b})^*$  & 14  & [6;~1;~4]   & $10664$ & $10461$ \\[2ex]
$K_1$  & 15  & [7;~1;~5]   & $10848$ & \\
  & 16  & [8;~1;~5]   & $10862$ & \\
  & 17  & [9;~1;~5]   & $10884$ & \\
  & 18  & [7;~1;~6]   & $10471$ & \\
  & 19  & [8;~1;~6]   & $10525$ & \\
  & 20  & [9;~1;~6]   & $10519$ & $10400$ \\[2ex]
$K_3$  & 21  & [13;~1;~9]   & $10644$ & \\
  & 22  & [14;~1;~9]   & $10705$ & \\
  & 23  & [15;~1;~9]   & $10717$ & \\
  & 24  & [13;~1;~10]   & $10730$ & \\
  & 25  & [14;~1;~10]   & $10778$ & \\
  & 26  & [15;~1;~10]   & $10780$ & $10406$ \\[2ex]
\multicolumn{4}{c}{Complete coupled-channels:} & $9654$
\end{tabular}
\end{ruledtabular}
\end{table}

\begin{figure}[!t]
\includegraphics[clip, trim={3.0cm 2.0cm 3.0cm 1.0cm}, width=0.45\textwidth]{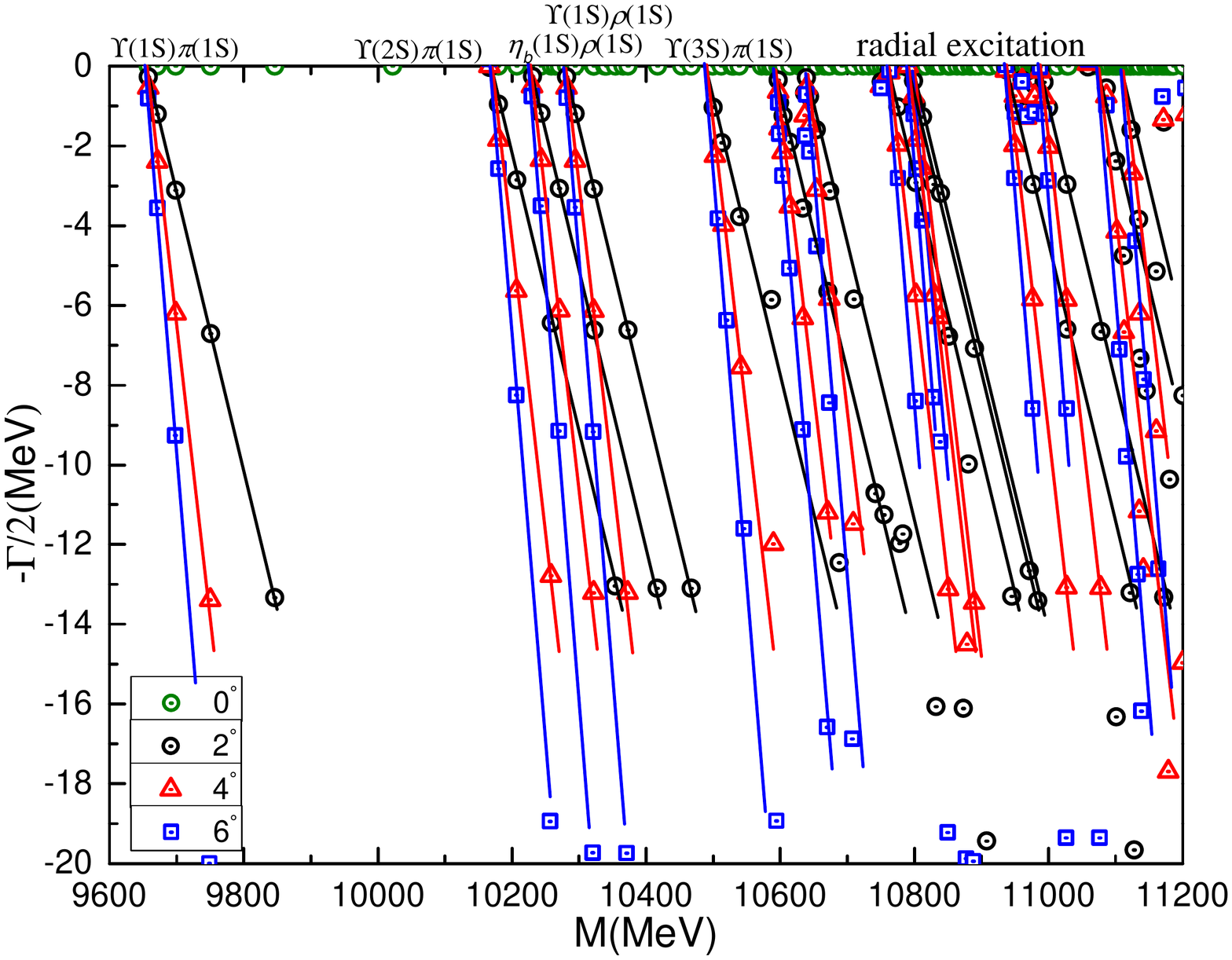} \\
\includegraphics[clip, trim={3.0cm 2.0cm 3.0cm 1.0cm}, width=0.45\textwidth]{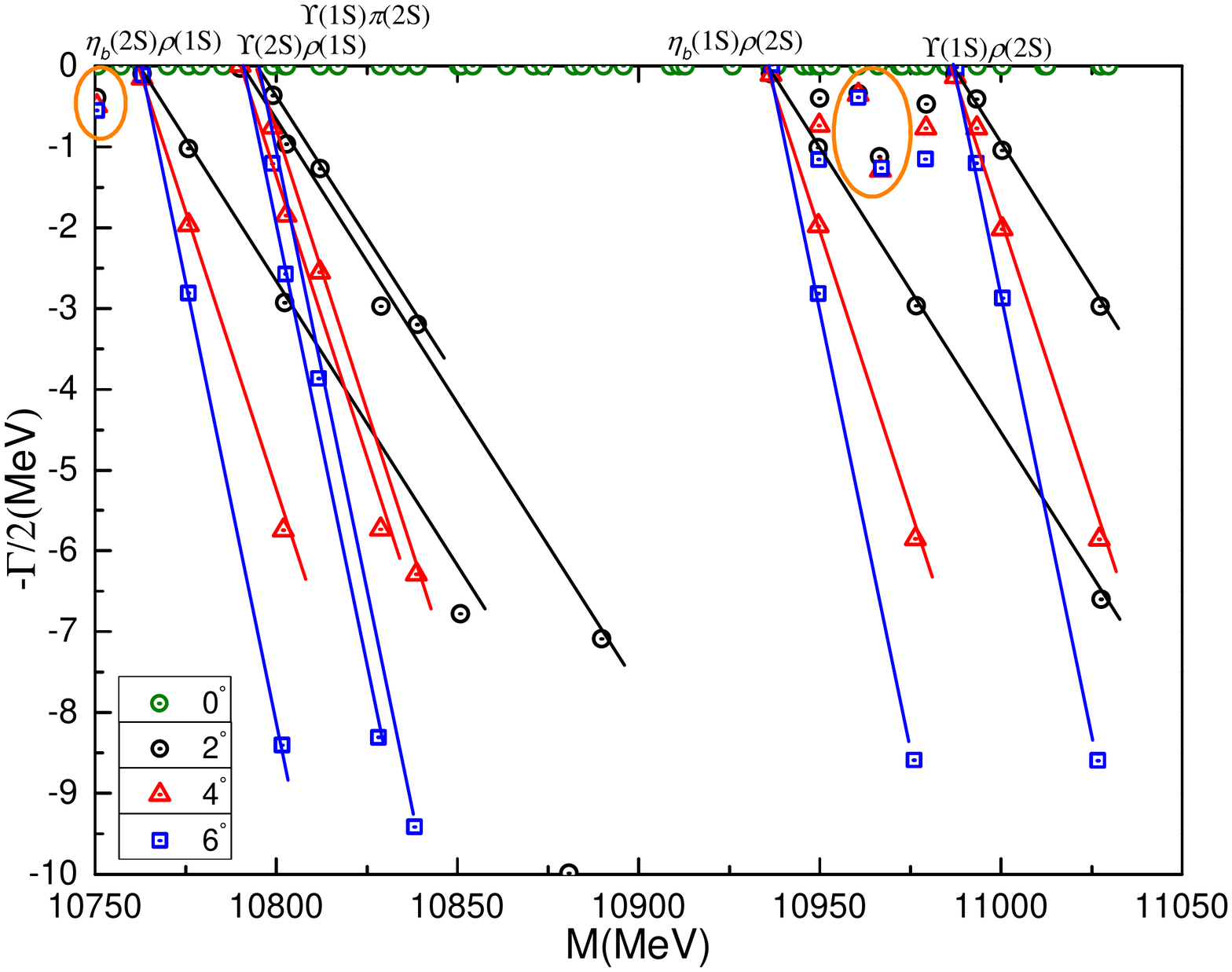}
\caption{\label{PP5} {\it Top panel:} The complete coupled-channels calculation of the $b\bar{b}q\bar{q}$ tetraquark system with $I(J^P)=1(1^+)$ quantum numbers. {\it Bottom panel:} Enlarged top panel, with real values of energy ranging from $10.75\,\text{GeV}$ to $11.05\,\text{GeV}$. We use the complex-scaling method of the chiral quark model varying $\theta$ from $0^\circ$ to $6^\circ$.}
\end{figure}

{\bf The $\bm{I(J^P)=1(1^+)}$ sector:} The numerical analysis of this case is quite similar to the $I(J^P)=0(1^+)$ because 26 channels must be also explored. From Table~\ref{GresultCC5}, we could find that the $B^* \bar{B}^*$ color-singlet channel is still weakly bound with $E_b=-2$ MeV, then the modified mass is $10648$ MeV, and the other meson-meson configurations in this channel are all of scattering nature. Accordingly, the $Z_b(10650)$ could be identified as a $B^* \bar{B}^*$ shallow bound state. However, this state is quite unstable, and scattered in a complete coupled-channels study that is discussed in the next paragraph. Meanwhile, the single channel masses of exotic structures are in a region of $10.47-10.89$ GeV, and coupled-channel mechanisms help little in pushing down the lowest masses, being all close to $10.4$ GeV.  

In a complex-range investigation of the complete coupled-channel calculation, where the rotated angle $\theta$ is still varied from $0^\circ$ to $6^\circ$, the complex energies are presented in Fig.~\ref{PP5}. In particular, within $9.6-11.2$ GeV, the scattering states of $\Upsilon \pi$, $\Upsilon \rho$, $\eta_b \rho$ and $B^{(*)} \bar{B}^{(*)}$ are generally shown in the top panel. As mentioned above, it seems that no stable pole is obtained; in particular, the previously obtained $B^* \bar{B}^*$ shallow bound-state is unavailable. However, due to a dense distribution of radial excitation states in the energy region $10.75-11.05$ GeV, an enlarged part is plotted in the bottom panel of Fig.~\ref{PP5}. Therein, one could find three stable resonance poles. The calculated masses and widths read $10750+i1.0$ MeV, $10960+i0.7$ MeV and $10967+i2.6$ MeV, respectively. Besides, according to their distributions, the dominant channel of the lowest resonance should be $B^*(1S) \bar{B}^*(1S)$ and the other two should be $\eta_b(1S)\rho(2S)$.

\begin{table}[!t]
\caption{\label{GresultCC6} Lowest-lying $b\bar{b}q\bar{q}$ tetraquark states with $I(J^P)=1(2^+)$ calculated within the real range formulation of the chiral quark model. The results are similarly organized as those in Table~\ref{GresultCC1}.
(unit: MeV).}
\begin{ruledtabular}
\begin{tabular}{lcccc}
~~Channel   & Index & $\chi_J^{\sigma_i}$;~$\chi_I^{f_j}$;~$\chi_k^c$ & $M$ & Mixed~~ \\
        &   &$[i; ~j; ~k]$ &  \\[2ex]
$(\Upsilon \rho)^1 (10230)$  & 1  & [1;~1;~1]   & $10277$ &  \\
$(B^* \bar{B}^*)^1 (10650)$  & 2  & [1;~1;~1]   & $10638$ & $10277$ \\[2ex]
$(\Upsilon \rho)^8$  & 3  & [1;~1;~2]   & $10898$ &  \\
$(B^* \bar{B}^*)^8$  & 4  & [1;~1;~2]   & $10698$ & $10636$ \\[2ex]
$(bq)^*(\bar{q}\bar{b})^*$  & 5  & [1;~1;~3]   & $10802$ & \\
$(bq)^*(\bar{q}\bar{b})^*$  & 6  & [1;~1;~4]   & $10737$ & $10576$ \\[2ex]
$K_1$  & 7  & [1;~1;~5]   & $10893$ & \\
  & 8  & [1;~1;~6]   & $10570$ & $10570$ \\[2ex]
$K_3$  & 9  & [1;~1;~9]   & $10731$ & \\
  & 10  & [1;~1;~10]   & $10794$ & $10575$ \\[2ex]
\multicolumn{4}{c}{Complete coupled-channels:} & $10277$
\end{tabular}
\end{ruledtabular}
\end{table}

\begin{figure}[!t]
\includegraphics[width=0.49\textwidth, trim={2.3cm 2.0cm 2.0cm 1.0cm}]{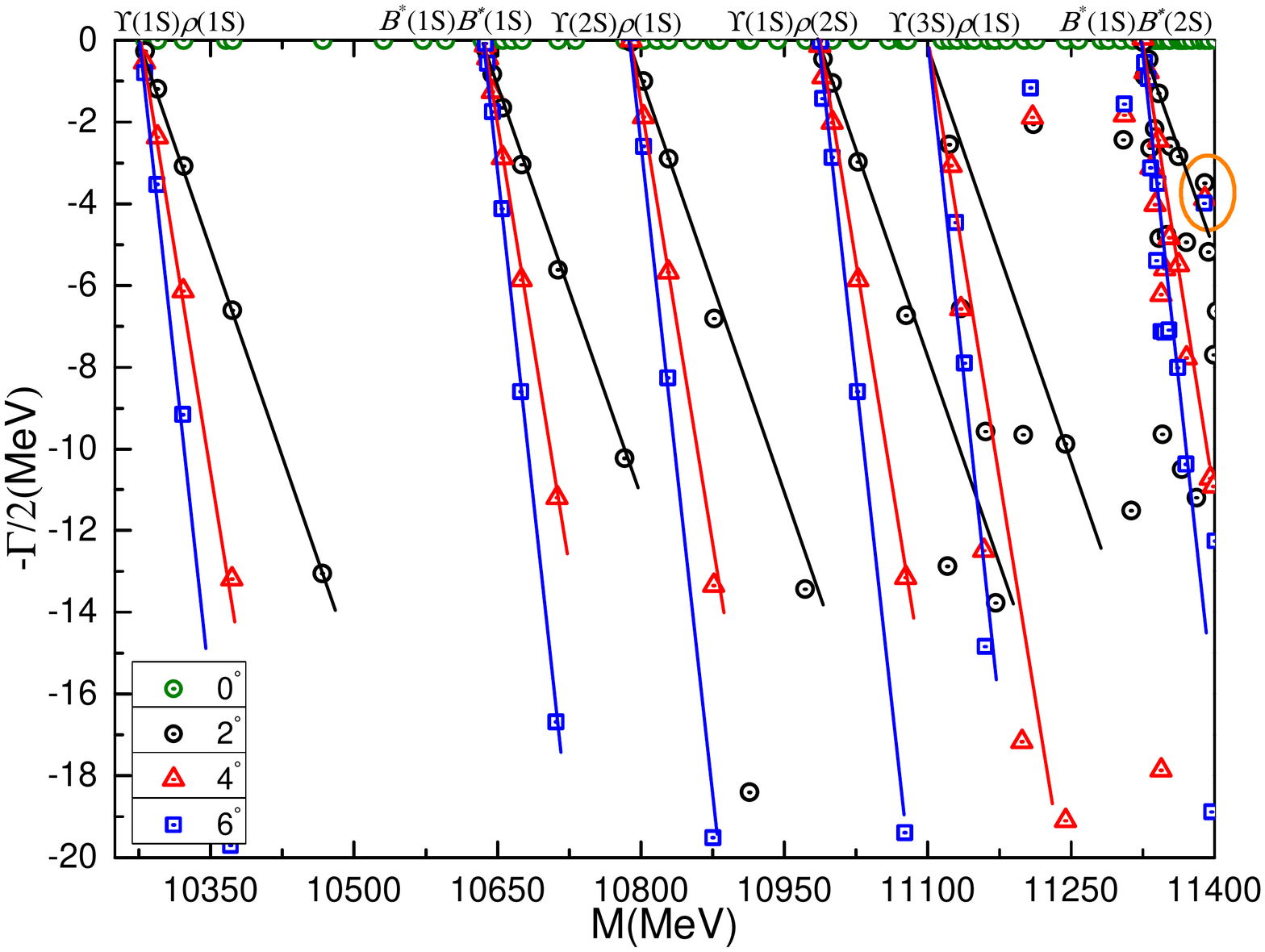}
\caption{\label{PP6} The complete coupled-channels calculation of the $b\bar{b}q\bar{q}$ tetraquark system with $I(J^P)=1(2^+)$ quantum numbers. We use the complex-scaling method of the chiral quark model varying $\theta$ from $0^\circ$ to $6^\circ$.}
\end{figure}

{\bf The $\bm{I(J^P)=1(2^+)}$ sector:} The real- and complex-range results for the highest spin and isospin $b\bar{b}q\bar{q}$ channel are shown in Table~\ref{GresultCC6} and Fig.~\ref{PP6}, respectively. The $\Upsilon \rho$ and $B^* \bar{B}^*$ cases are both scattering states. Furthermore, those belonging to the hidden-color meson-meson, diquark-antidiquark and K-type configurations are generally located in the energy range $10.6-10.9$ GeV. In the coupled-channel calculation of each quark-arrangement configuration, the masses are $\sim10.57$ GeV, and the scattering nature of the lowest channel $\Upsilon \rho$ remains at $10277$ MeV.

In Fig.~\ref{PP6}, where a fully coupled-channel calculation in complex-range is performed, the distributions of complex energies of $\Upsilon \rho$ and $B^* \bar{B}^*$ are well presented. Moreover, a resonance state is found with mass and decay width equal to $11390+i7.8$ MeV; obviously, the $B^*(1S) \bar{B}^*(2S)$ should be its dominant channel.

%%%%%%%%%%%%%%%%%%%%%%%%%%%%%%%%%%%%%%%%%%%%%%%%%%%%%%%%%%%%%%%%%%%%%%%%%%%%%%%%

\subsection{The $\mathbf{b\bar{b}u\bar{s}}$ tetraquarks}

A natural continuation of the investigation performed above is the analysis of the $b\bar{b}u\bar{s}$ tetraquark system with spin-parity $J^P=0^+$, $1^+$ and $2^+$, and isospin $I=\frac{1}{2}$. A similar situation that the one discussed in the hidden-charm tetraquarks with strangeness~\cite{Yang:2021zhe} is found for the $b\bar{b}u\bar{s}$ tetraquarks. In summary, several narrow resonances whose masses are around $11.1$ GeV are obtained, and further details can be found below.

\begin{table}[!t]
\caption{\label{GresultCC7} Lowest-lying $b\bar{b}u\bar{s}$ tetraquark states with $I(J^P)=\frac{1}{2}(0^+)$ calculated within the real range formulation of the chiral quark model. The results are similarly organized as those in Table~\ref{GresultCC1}.
(unit: MeV).}
\begin{ruledtabular}
\begin{tabular}{lcccc}
~~Channel   & Index & $\chi_J^{\sigma_i}$;~$\chi_I^{f_j}$;~$\chi_k^c$ & $M$ & Mixed~~ \\
        &   &$[i; ~j; ~k]$ &  \\[2ex]
$(\eta_b K)^1 (9794)$          & 1  & [1;~2;~1]  & $9935$ & \\
$(\Upsilon K^*)^1 (10352)$  & 2  & [2;~2;~1]   & $10412$ &  \\
$(B B_s)^1 (10647)$          & 3  & [1;~2;~1]  & $10633$ & \\
$(B^* B^*_s)^1 (10740)$  & 4  & [2;~2;~1]   & $10719$ & $9935$ \\[2ex]
$(\eta_b K)^8$          & 5  & [1;~2;~2]  & $10963$ & \\
$(\Upsilon K^*)^8$  & 6  & [2;~2;~2]   & $10978$ &  \\
$(B B_s)^8$          & 7  & [1;~2;~2]  & $10838$ & \\
$(B^* B^*_s)^8$  & 8  & [2;~2;~2]   & $10776$ & $10652$ \\[2ex]
$(bu)(\bar{s}\bar{b})$      & 9   & [3;~2;~3]  & $10875$ & \\
$(bu)(\bar{s}\bar{b})$      & 10   & [4;~2;~3]  & $10853$ & \\
$(bu)^*(\bar{s}\bar{b})^*$  & 11  & [3;~2;~4]   & $10844$ & \\
$(bu)^*(\bar{s}\bar{b})^*$  & 12  & [4;~2;~4]   & $10757$ & $10572$ \\[2ex]
$K_1$  & 13  & [5;~2;~5]   & $10970$ & \\
  & 14  & [6;~2;~5]   & $10962$ & \\
  & 15  & [5;~2;~6]   & $10701$ & \\
  & 16  & [6;~2;~6]   & $10540$ & $10539$ \\[2ex]
$K_2$  & 17  & [7;~2;~7]   & $10696$ & \\
  & 18  & [8;~2;~7]   & $10533$ & \\
  & 19  & [7;~2;~8]   & $10970$ & \\
  & 20  & [8;~2;~8]   & $10961$ & $10533$ \\[2ex]
$K_3$  & 21  & [9;~2;~9]   & $10742$ & \\
  & 22  & [10;~2;~9]   & $10836$ & \\
  & 23  & [9;~2;~10]   & $10839$ & \\
  & 24  & [10;~2;~10]   & $10867$ & $10562$ \\[2ex]
$K_4$  & 25  & [11;~2;~11]   & $10751$ & \\
  & 26  & [12;~2;~11]   & $10844$ & \\
  & 27  & [11;~2;~12]   & $10842$ & \\
  & 28  & [12;~2;~12]   & $10870$ & $10561$ \\[2ex]
\multicolumn{4}{c}{Complete coupled-channels:} & $9935$
\end{tabular}
\end{ruledtabular}
\end{table}

\begin{figure}[!t]
\includegraphics[width=0.49\textwidth, trim={2.3cm 2.0cm 2.0cm 1.0cm}]{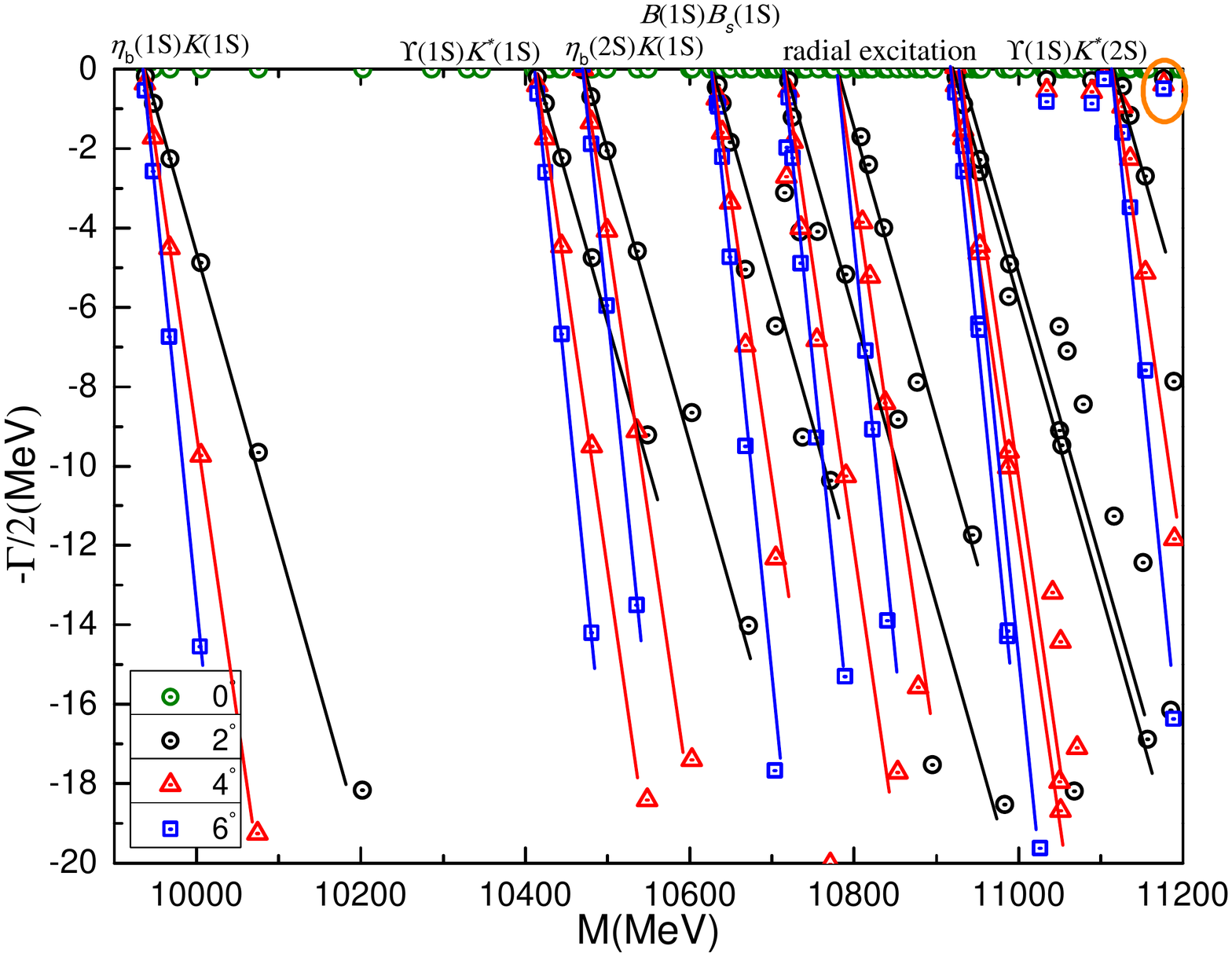}
\caption{\label{PP7} The complete coupled-channels calculation of the $b\bar{b}u\bar{s}$ tetraquark system with $I(J^P)=\frac{1}{2}(0^+)$ quantum numbers. We use the complex-scaling method of the chiral quark model varying $\theta$ from $0^\circ$ to $6^\circ$.}
\end{figure}

{\bf The $\bm{I(J^P)=\frac{1}{2}(0^+)}$ sector:} Table~\ref{GresultCC7} summarizes all contributing channels, they are 28 and include meson-meson, diquark-antidiquark and K-type structures. The color-singlet channels, which include $\eta_b K$, $\Upsilon K^*$ and $B^{(*)}B^{(*)}_s$,  are all of scattering nature. Namely, their masses are just located at the values of corresponding theoretical thresholds. Besides, the remaining channels of hidden-color, diquark-antidiquark and K-type configurations are generally located in the interval $10.6-10.9$ GeV.

If a coupled-channel calculation is performed in each individual configuration, the lowest masses of exotic configurations are close to $10.56$ GeV, except for hidden-color channels which is located at $10.65$ GeV. Meanwhile, the coupled mass of color-singlet channels is still at the $\eta_b K$ theoretical threshold value $9935$ MeV, and this fact is not changed even in a complete coupled-channel calculation in real-range formulation.

When a fully coupled-channel calculation is performed in complex-range, Fig.~\ref{PP7} shows the obtained results. Therein, with an energy interval $9.0-11.2$ GeV, the scattering nature of $\eta_b K$, $\Upsilon K^*$ and $B^{(*)}B^{(*)}_s$ both in ground and radial excitation states is clearly presented. Besides, one quite narrow resonance is found and we circle it above the $\Upsilon(1S)K^*(2S)$ threshold line(s); the calculated complex energy is $11176+i0.8$ MeV and it should be stable against meson-meson strong decay processes.

\begin{table*}[!t]
\caption{\label{GresultCC8} Lowest-lying $b\bar{b}u\bar{s}$ tetraquark states with $I(J^P)=\frac{1}{2}(1^+)$ calculated within the real range formulation of the chiral quark model. The results are similarly organized as those in Table~\ref{GresultCC1}.
(unit: MeV).}
\begin{ruledtabular}
\begin{tabular}{lccccccccc}
~~Channel   & Index & $\chi_J^{\sigma_i}$;~$\chi_I^{f_j}$;~$\chi_k^c$ & $M$ & Mixed~~ & ~~Channel   & Index & $\chi_J^{\sigma_i}$;~$\chi_I^{f_j}$;~$\chi_k^c$ & $M$ & Mixed~~\\
        &   &$[i; ~j; ~k]$ &  &  &  &   &$[i; ~j; ~k]$ &  \\[2ex]
$(\eta_b K^*)^1 (10192)$          & 1  & [1;~2;~1]  & $10361$ &  & $K_1$  & 19  & [7;~2;~5]   & $10969$ & \\
$(\Upsilon K)^1 (9954)$  & 2  & [2;~2;~1]   & $9986$ &  & & 20  & [8;~2;~5]   & $10972$ & \\
$(\Upsilon K^*)^1 (10352)$  & 3  & [3;~2;~1]   & $10412$ &  & & 21  & [9;~2;~5]   & $10986$ & \\
$(B^* B_s)^1 (10692)$          & 4  & [1;~2;~1]  & $10674$ &  & & 22  & [7;~2;~6]   & $10635$ & \\
$(B B^*_s)^1 (10695)$  & 5  & [2;~2;~1]   & $10678$ &   & & 23  & [8;~2;~6]   & $10670$ & \\
$(B^* B^*_s)^1 (10740)$  & 6  & [3;~2;~1]   & $10719$ & $9986$  & & 24  & [9;~2;~6]   & $10650$ & $10591$ \\[2ex]
$(\eta_b K^*)^8$  & 7  & [1;~2;~2]  & $10992$ &  & $K_2$  & 25  & [10;~2;~7]   & $10629$ & \\
$(\Upsilon K)^8$    & 8  & [2;~2;~2]   & $10964$ &   & & 26  & [11;~2;~7]   & $10665$ & \\
$(\Upsilon K^*)^8$  & 9  & [3;~2;~2]   & $10985$ &  & & 27  & [12;~2;~7]   & $10645$ & \\
$(B^* B_s)^8$          & 10 & [1;~2;~2]  & $10840$ &  & & 28  & [10;~2;~8]   & $10966$ & \\
$(B B^*_s)^8$      & 11  & [2;~2;~2]   & $10839$ &   & & 29  & [11;~2;~8]   & $10974$ & \\
$(B^* B^*_s)^8$      & 12  & [3;~2;~2]   & $10801$ & $10675$  &  & 30  & [12;~2;~8]   & $10986$ & $10584$ \\[2ex]
$(bu)(\bar{s}\bar{b})^*$      & 13   & [4;~2;~3]  & $10888$ &  & $K_3$  & 31  & [13;~2;~9]   & $10773$ & \\
$(bu)(\bar{s}\bar{b})^*$      & 14   & [5;~2;~3]  & $10889$ &  & & 32  & [14;~2;~9]   & $10809$ & \\
$(bu)^*(\bar{s}\bar{b})$      & 15   & [6;~2;~3]  & $10866$ &  & & 33  & [15;~2;~9]   & $10831$ & \\
$(bu)^*(\bar{s}\bar{b})$      & 16   & [4;~2;~4]  & $10841$ &  & & 34  & [13;~2;~10]   & $10839$ & \\
$(bu)^*(\bar{s}\bar{b})^*$  & 17  & [5;~2;~4]   & $10840$ &  & & 35  & [14;~2;~10]   & $10868$ & \\
$(bu)^*(\bar{s}\bar{b})^*$  & 18  & [6;~2;~4]   & $10787$ & $10600$  & & 36  & [15;~2;~10]   & $10881$ & $10592$ \\[2ex]
&  &   & &  &  $K_4$  & 37  & [16;~2;~11]   & $10783$ & \\
&  &  &   & &   & 38  & [17;~2;~11]   & $10815$ & \\
&  &  &   & &     & 39  & [18;~2;~11]   & $10840$ & \\
&  &  &   & &     & 40  & [16;~2;~12]   & $10844$ & \\
&  &  &   & &     & 41  & [17;~2;~12]   & $10870$ & \\
&  &  &   & &     & 42  & [18;~2;~12]   & $10882$ & $10596$ \\[2ex]
\multicolumn{9}{c}{Complete coupled-channels:} & $9986$
\end{tabular}
\end{ruledtabular}
\end{table*}

\begin{figure}[!t]
\includegraphics[clip, trim={3.0cm 1.9cm 3.0cm 1.0cm}, width=0.45\textwidth]{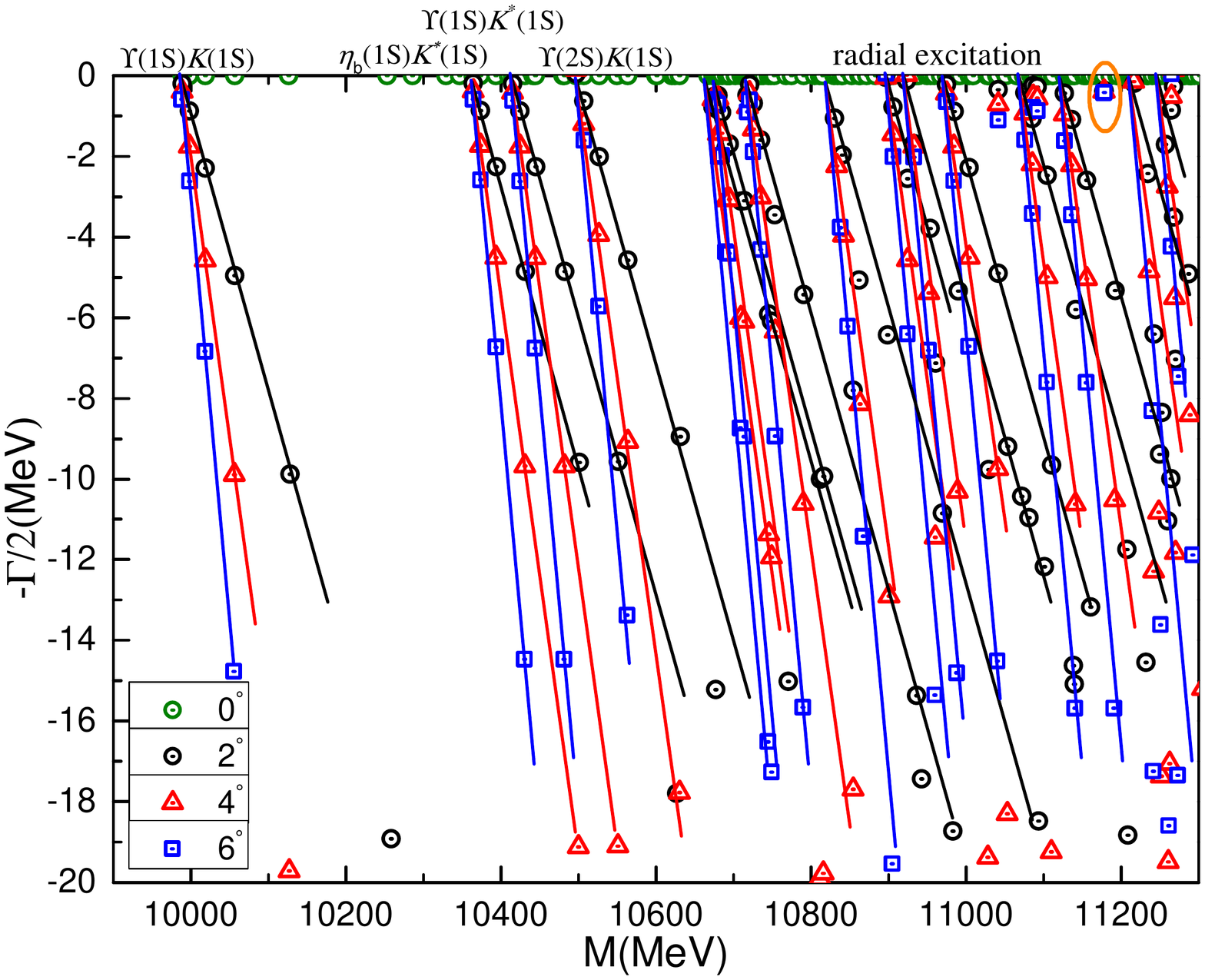} \\
\includegraphics[clip, trim={3.0cm 1.9cm 3.0cm 1.0cm}, width=0.45\textwidth]{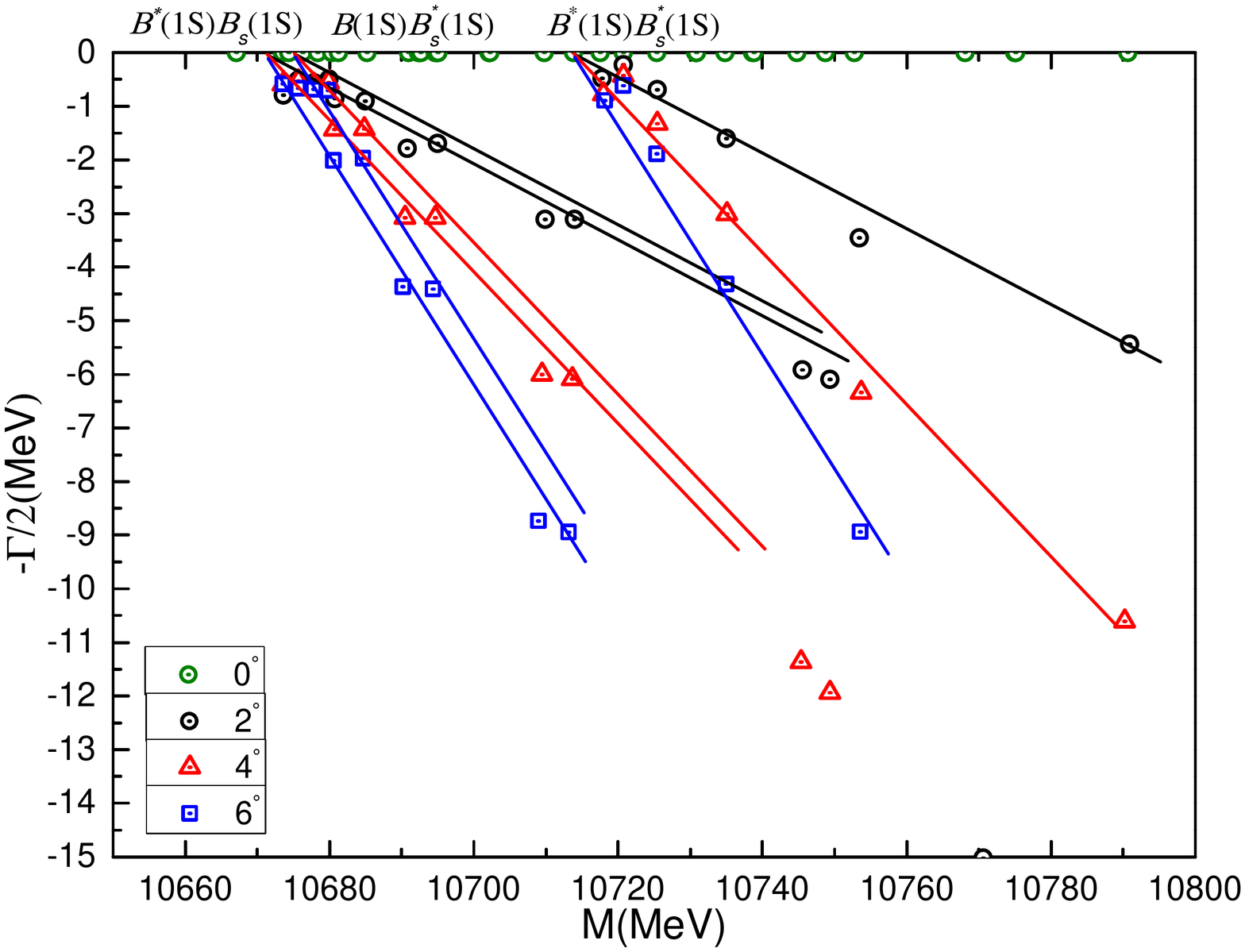} \\
\includegraphics[clip, trim={3.0cm 1.9cm 3.0cm 1.0cm}, width=0.45\textwidth]{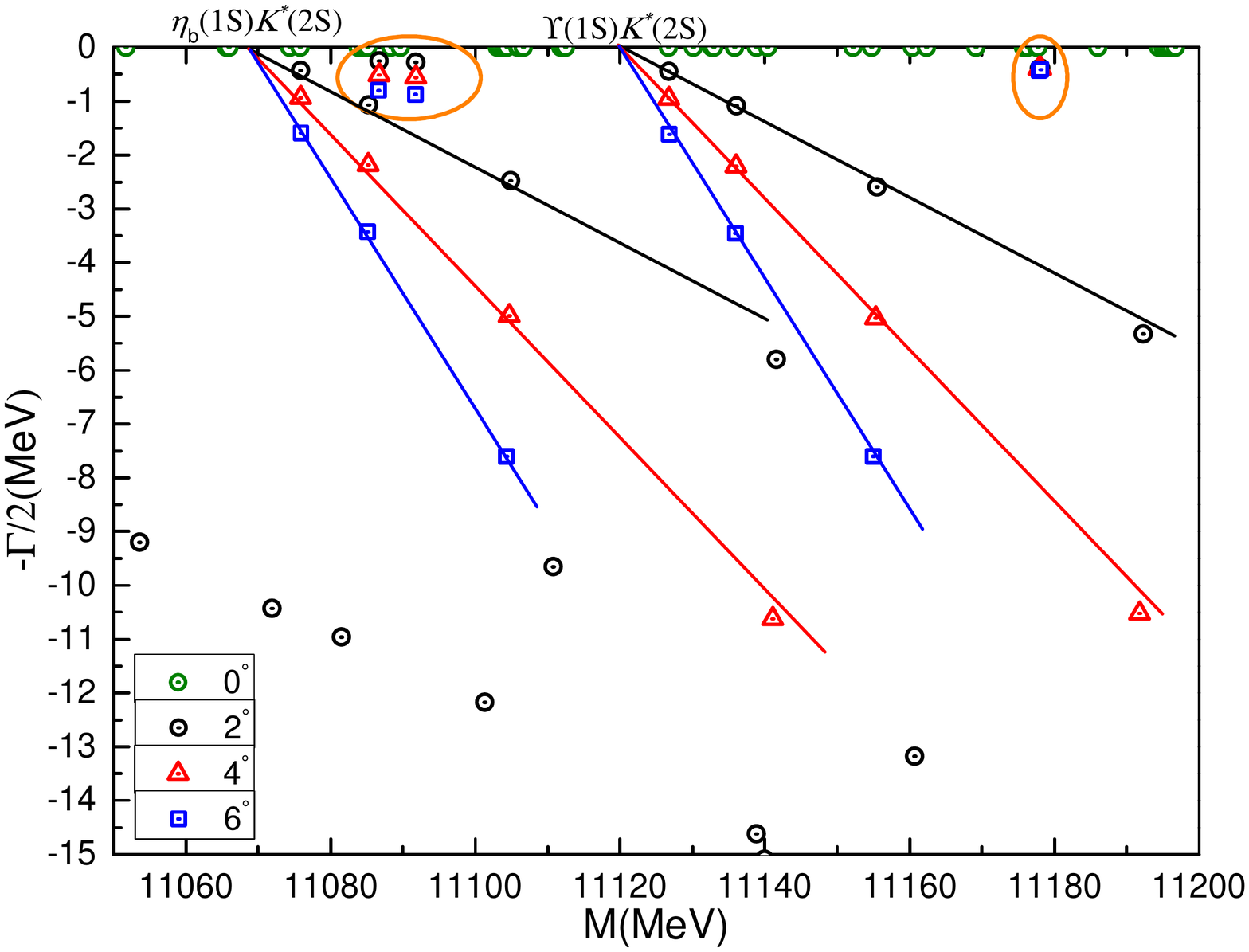}
\caption{\label{PP8} The complete coupled-channels calculation of the $b\bar{b}u\bar{s}$ tetraquark system with $I(J^P)=\frac{1}{2}(1^+)$ quantum numbers. Particularly, middle and bottom panels are the enlarged parts of dense energy region from $10.65\,\text{GeV}$ to $11.20\,\text{GeV}$.}
\end{figure}

{\bf The $\bm{I(J^P)=\frac{1}{2}(1^+)}$ sector:} 42 channels shown in Table~\ref{GresultCC8} are under investigation; namely, six meson-meson channels in both color-singlet and hidden-color configurations, six diquark-antidiquark structures, and 24 K-type arrangements. Firstly, in the single-channel calculation, no bound state is obtained. The lowest one is $\Upsilon K$ scattering state with mass at $9986$ MeV, and the other color-singlet channels are generally located in an interval $10.3-10.7$ GeV. Besides, masses of exotic configurations distribute from $10.6$ GeV to $10.9$ GeV. Although the coupled mass in hidden-color, diquark-antidiquark, and K-type configurations tends to $10.59$ GeV, this strong coupling effect does not hold for color-singlet channels study, and the lowest mass remains at $\Upsilon K$ threshold value.

Fig.~\ref{PP8} shows our findings using the complex-range method to the complete coupled-channel calculation. Therein, in the top panel, the scattering states of $\eta_b K^*$, $\Upsilon K^{(*)}$ and $B^{(*)}B^{(*)}_s$ are clearly identified in the energy region $9.0-11.3$ GeV. Since there are dense distributions of radial excited states, two energy intervals, $10.65-10.80$ GeV and $11.05-11.20$ GeV, are plotted in the middle and bottom panels, respectively. Apart from three scattering states of $B^*(1S)B_s(1S)$, $B(1S)B^*_s(1S)$ and $B^*(1S)B^*_s(1S)$ obtained in the middle panel, no resonance is available. Three narrow resonance poles are found in the bottom panel. Their complex energies are $11178+i0.8$ GeV, $11086+i1.0$ GeV and $11092+i1.1$ GeV. By considering the location of the resonance poles, the $\eta_b(1S)K^*(2S)$ should be the dominant decay channel of the two lower-energy resonances, whereas the $\Upsilon(1S)K^*(2S)$ channel must play a dominant role for the remaining resonance.

\begin{table}[!t]
\caption{\label{GresultCC9} Lowest-lying $b\bar{b}u\bar{s}$ tetraquark states with $I(J^P)=\frac{1}{2}(2^+)$ calculated within the real range formulation of the chiral quark model. The results are similarly organized as those in Table~\ref{GresultCC1}.
(unit: MeV).}
\begin{ruledtabular}
\begin{tabular}{lcccc}
~~Channel   & Index & $\chi_J^{\sigma_i}$;~$\chi_I^{f_j}$;~$\chi_k^c$ & $M$ & Mixed~~ \\
        &   &$[i; ~j; ~k]$ &  \\[2ex]
$(\Upsilon K^*)^1 (10352)$  & 1  & [1;~2;~1]   & $10412$ &  \\
$(B^* B^*_s)^1 (10740)$  & 2  & [1;~2;~1]   & $10719$ & $10412$ \\[2ex]
$(\Upsilon K^*)^8$  & 3  & [1;~2;~2]   & $11000$ &  \\
$(B^* B^*_s)^8$  & 4  & [1;~2;~2]   & $10843$ & $10785$ \\[2ex]
$(bu)^*(\bar{s}\bar{b})^*$  & 5  & [1;~2;~3]   & $10890$ & \\
$(bu)^*(\bar{s}\bar{b})^*$  & 6  & [1;~2;~4]   & $10839$ & $10707$ \\[2ex]
$K_1$  & 7  & [1;~2;~5]   & $10994$ & \\
  & 8  & [1;~2;~6]   & $10701$ & $10701$ \\[2ex]
$K_2$  & 9  & [1;~2;~7]   & $10696$ & \\
  & 10  & [1;~2;~6]   & $10994$ & $10696$ \\[2ex]
$K_3$  & 11  & [1;~2;~9]   & $10830$ & \\
  & 12  & [1;~2;~10]   & $10881$ & $10707$ \\[2ex]
$K_4$  & 13  & [1;~2;~11]   & $10838$ & \\
  & 14  & [1;~2;~12]   & $10883$ & $10705$ \\[2ex]
\multicolumn{4}{c}{Complete coupled-channels:} & $10412$
\end{tabular}
\end{ruledtabular}
\end{table}

\begin{figure}[!t]
\includegraphics[width=0.49\textwidth, trim={2.3cm 2.0cm 2.0cm 1.0cm}]{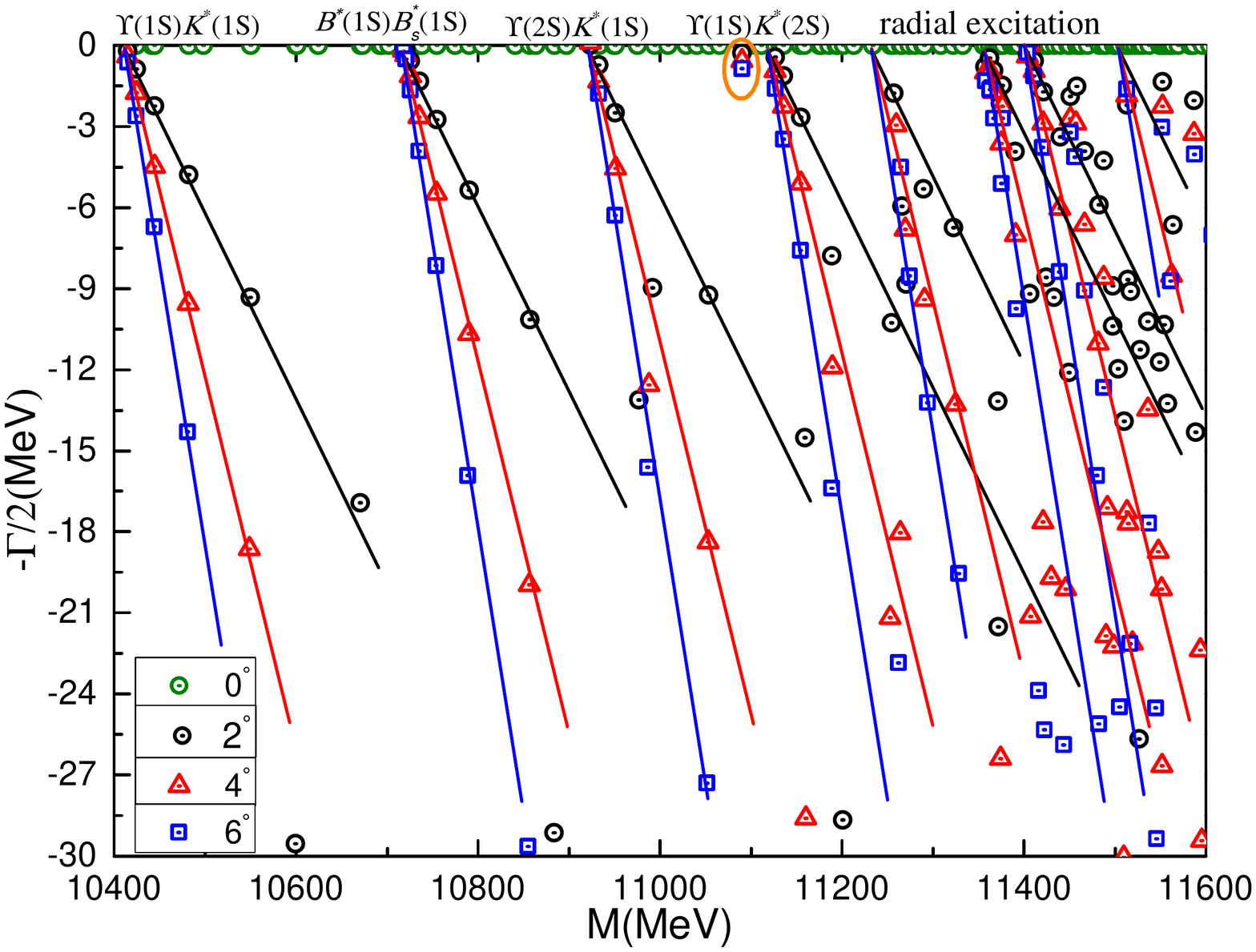}
\caption{\label{PP9} The complete coupled-channels calculation of the $b\bar{b}u\bar{s}$ tetraquark system with $I(J^P)=\frac{1}{2}(2^+)$ quantum numbers. We use the complex-scaling method of the chiral quark model varying $\theta$ from $0^\circ$ to $6^\circ$.}
\end{figure}

{\bf The $\bm{I(J^P)=\frac{1}{2}(2^+)}$ sector:} From Table~\ref{GresultCC9}, we can see that the 14 contributing channels are all above $10.41$ GeV and the lowest mass is $10412$ MeV, which is just the theoretical value of $\Upsilon K^*$ threshold. Moreover, this fact remains unchanged when a complete coupled-channel calculation is performed in real-range formulation.

In a complex-range investigation in which the rotated angle is varied from $0^\circ$ to $6^\circ$, the scattering states of $\Upsilon K^*$ and $B^*B^*_s$ are well identified in Fig.~\ref{PP9}. Particularly, they are located within an energy interval $10.4-11.6$ GeV. It is worth noting that a stable resonance is obtained above the threshold lines of  $\Upsilon(2S)K^*(1S)$, and the calculated pole is $11090+i1.1$ MeV.

\subsection{The $\mathbf{b\bar{b}s\bar{s}}$ tetraquarks}

The last sector of hidden-bottom tetraquarks in our investigation is the $b\bar{b}s\bar{s}$ system. Three spin-parity states, $J^P=0^+$, $1^+$ and $2^+$, with isospin $I=0$, are studied. As in the $b\bar{b}q\bar{q}$ $(q=u,\,d)$ system, due to symmetry properties of the Hamiltonian with respect to the K-type configurations, only $K_1$ and $K_3$ arrangements are sufficient to consider. In summary, only one extremely narrow resonance is obtained in the $I(J^P)=0(1^+)$ channel.

\begin{table}[!t]
\caption{\label{GresultCC10} Lowest-lying $b\bar{b}s\bar{s}$ tetraquark states with $I(J^P)=0(0^+)$ calculated within the real range formulation of the chiral quark model. The results are similarly organized as those in Table~\ref{GresultCC1}.
(unit: MeV).}
\begin{ruledtabular}
\begin{tabular}{lcccc}
~~Channel   & Index & $\chi_J^{\sigma_i}$;~$\chi_I^{f_j}$;~$\chi_k^c$ & $M$ & Mixed~~ \\
        &   &$[i; ~j; ~k]$ &  \\[2ex]
$(\eta_b \eta')^1 (9848)$          & 1  & [1;~3;~1]  & $10282$ & \\
$(\Upsilon \phi)^1 (10242)$  & 2  & [2;~3;~1]   & $10516$ &  \\
$(B_s \bar{B}_s)^1 (10560)$          & 3  & [1;~3;~1]  & $10710$ & \\
$(B^*_s \bar{B}^*_s)^1 (10650)$  & 4  & [2;~3;~1]   & $10800$ & $10282$ \\[2ex]
$(\eta_b \eta')^8$          & 5  & [1;~3;~2]  & $11102$ & \\
$(\Upsilon \phi)^8$  & 6  & [2;~3;~2]   & $11068$ &  \\
$(B_s \bar{B}_s)^8$          & 7  & [1;~3;~2]  & $10983$ & \\
$(B^*_s \bar{B}^*_s)^8$  & 8  & [2;~3;~2]   & $10938$ & $10822$ \\[2ex]
$(bs)(\bar{s}\bar{b})$      & 9   & [3;~3;~3]  & $10972$ & \\
$(bs)(\bar{s}\bar{b})$      & 10   & [3;~3;~4]  & $10967$ & \\
$(bs)^*(\bar{s}\bar{b})^*$  & 11  & [4;~3;~3]   & $10952$ & \\
$(bs)^*(\bar{s}\bar{b})^*$  & 12  & [4;~3;~4]   & $10886$ & $10745$ \\[2ex]
$K_1$  & 13  & [5;~3;~5]   & $11064$ & \\
  & 14  & [6;~3;~5]   & $11097$ & \\
  & 15  & [5;~3;~6]   & $10815$ & \\
  & 16  & [6;~3;~6]   & $10712$ & $10711$ \\[2ex]
$K_3$  & 17  & [9;~3;~9]   & $10880$ & \\
  & 18  & [10;~3;~9]   & $10951$ & \\
  & 19  & [9;~3;~10]   & $10958$ & \\
  & 20  & [10;~3;~10]   & $10967$ & $10740$ \\[2ex]
\multicolumn{4}{c}{Complete coupled-channels:} & $10282$
\end{tabular}
\end{ruledtabular}
\end{table}

\begin{figure}[!t]
\includegraphics[clip, trim={3.0cm 2.0cm 3.0cm 1.0cm}, width=0.45\textwidth]{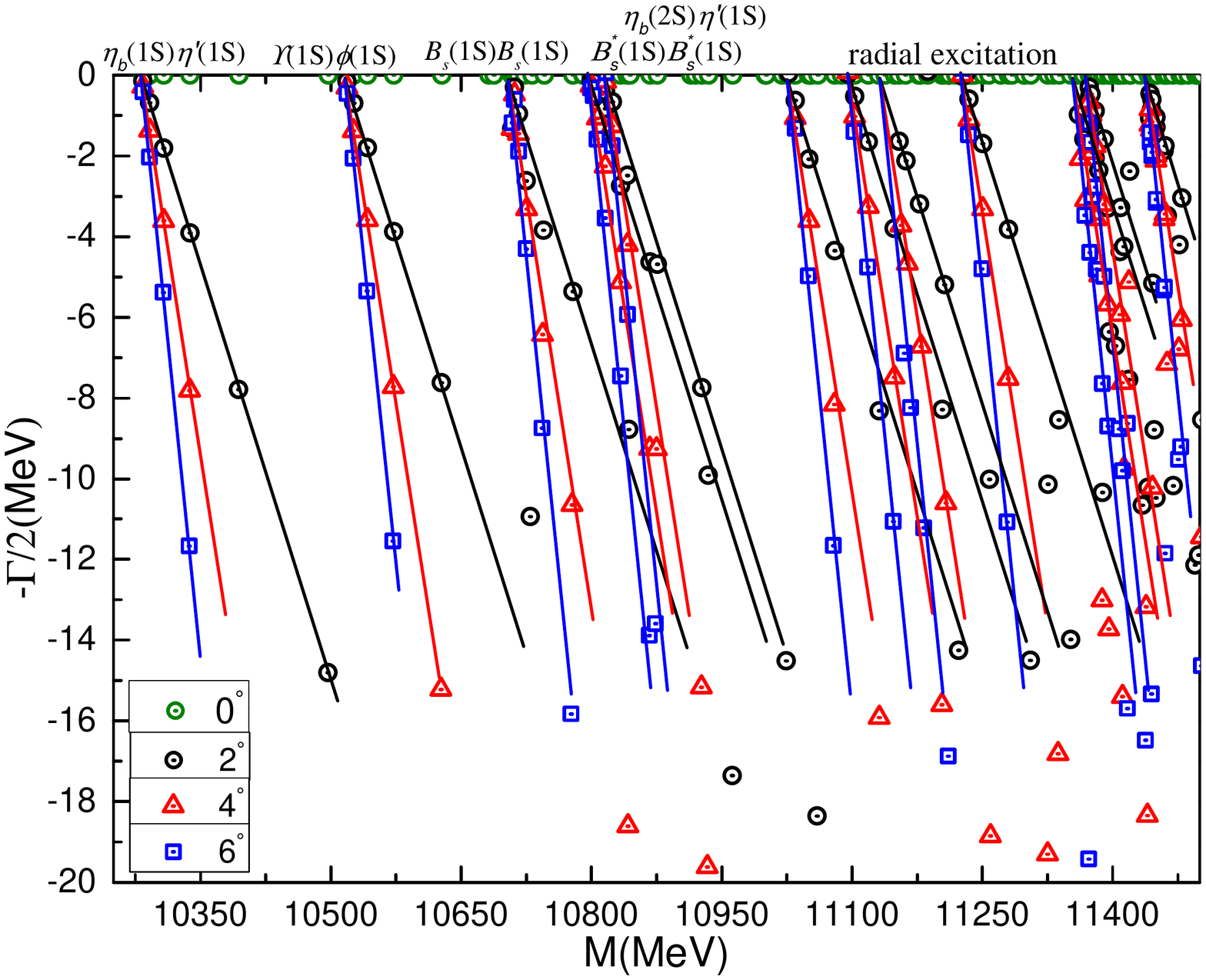} \\
\includegraphics[clip, trim={3.0cm 2.0cm 3.0cm 1.0cm}, width=0.45\textwidth]{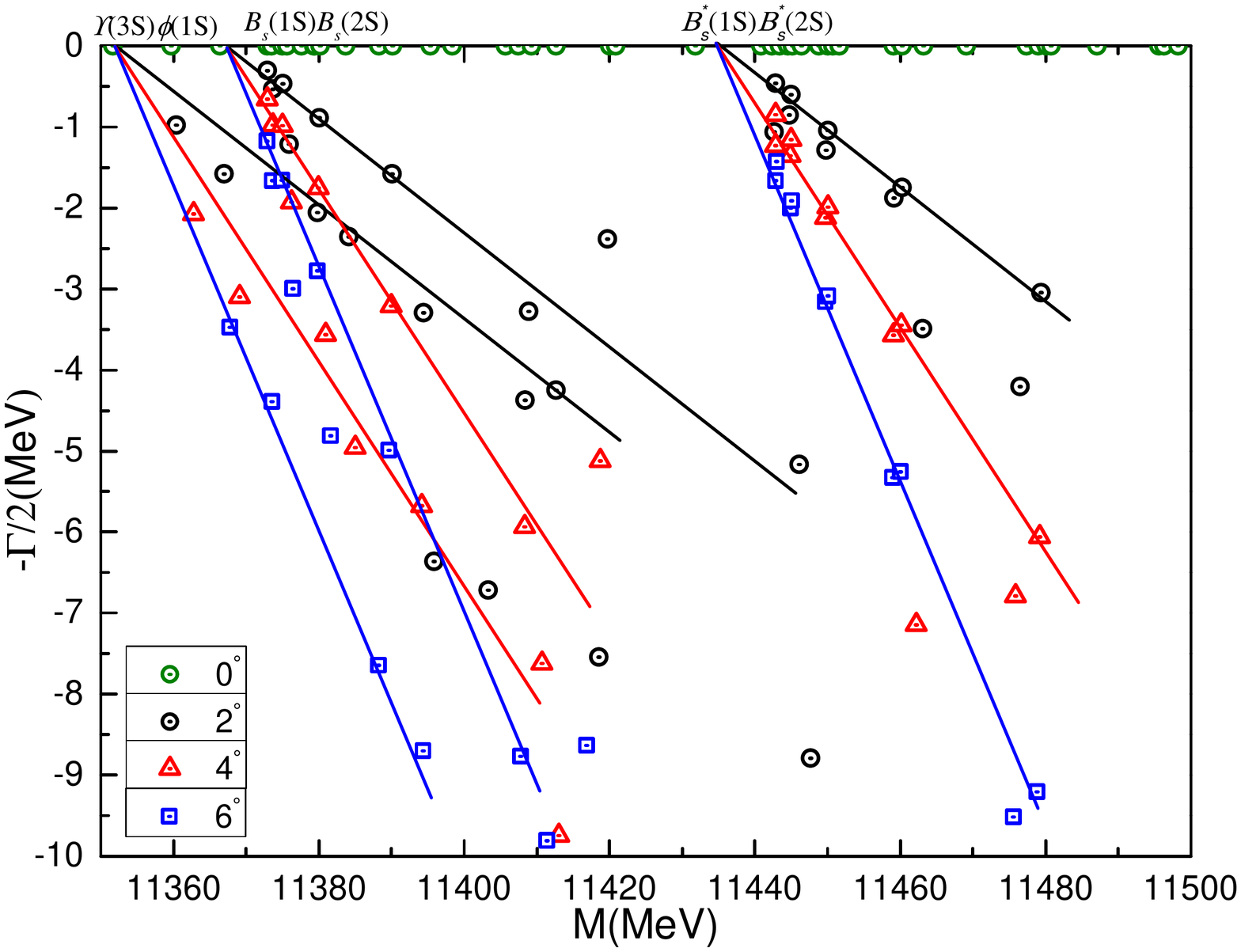}
\caption{\label{PP10} {\it Top panel:} The complete coupled-channels calculation of the $b\bar{b}s\bar{s}$ tetraquark system with $I(J^P)=0(0^+)$ quantum numbers. {\it Bottom panel:} Enlarged top panel, with real values of energy ranging from $11.35\,\text{GeV}$ to $11.50\,\text{GeV}$. We use the complex-scaling method of the chiral quark model varying $\theta$ from $0^\circ$ to $6^\circ$.}
\end{figure}

{\bf The $\bm{I(J^P)=0(0^+)}$ sector:} Table~\ref{GresultCC10} lists the calculated masses of the $0(0^+)$ $b\bar{b}s\bar{s}$ tetraquark system. Firstly, no bound state is found in each single channel, and also in the different variants of the coupled-channels calculations. The lowest mass obtained is $10282$ MeV, which is just the theoretical value of $\eta_b \eta'$ threshold. The remaining channels are generally located within an energy interval $10.5-11.1$ GeV.

When a fully coupled-channel calculation is performed using the complex-range method, the scattering nature of $\eta_b \eta'$, $\Upsilon \phi$ and $B^{(*)}_s \bar{B}^{(*)}_s$ are clearly shown in Fig.~\ref{PP10}. Generally, no stable pole is found in the top panel. Besides, an enlarged energy region from $11.35$ GeV to $11.50$ GeV is presented in the bottom panel, the radial excitations $\Upsilon(3S)\phi(1S)$, $B_s(1S) \bar{B}_s(2S)$ and $B^*_s(1S) \bar{B}^*_s(2S)$ are shown no resonance is obtained, too.

\begin{table}[!t]
\caption{\label{GresultCC11} Lowest-lying $b\bar{b}s\bar{s}$ tetraquark states with $I(J^P)=0(1^+)$ calculated within the real range formulation of the chiral quark model. The results are similarly organized as those in Table~\ref{GresultCC1}.
(unit: MeV).}
\begin{ruledtabular}
\begin{tabular}{lcccc}
~~Channel   & Index & $\chi_J^{\sigma_i}$;~$\chi_I^{f_j}$;~$\chi_k^c$ & $M$ & Mixed~~ \\
        &   &$[i; ~j; ~k]$ &  \\[2ex]
$(\eta_b \phi)^1 (10082)$          & 1  & [1;~3;~1]  & $10465$ & \\
$(\Upsilon \eta')^1 (10008)$  & 2  & [2;~3;~1]   & $10333$ &  \\
$(\Upsilon \phi)^1 (10242)$  & 3  & [3;~3;~1]   & $10516$ &  \\
$(B_s \bar{B}^*_s)^1 (10605)$          & 4  & [1;~3;~1]  & $10755$ & \\
$(B^*_s \bar{B}^*_s)^1 (10650)$  & 5  & [3;~3;~1]   & $10800$ & $10333$ \\[2ex]
$(\eta_b \phi)^8$  & 6  & [1;~3;~2]  & $11082$ & \\
$(\Upsilon \eta')^8$    & 7  & [2;~3;~2]   & $11102$ &  \\
$(\Upsilon \phi)^8$  & 8  & [3;~3;~2]   & $11075$ &  \\
$(B_s \bar{B}^*_s)^8$          & 9 & [1;~3;~2]  & $10985$ & \\
$(B^*_s \bar{B}^*_s)^8$      & 10  & [3;~3;~2]   & $10951$ & $10848$ \\[2ex]
$(bs)(\bar{s}\bar{b})^*$      & 11   & [4;~3;~3]  & $10986$ & \\
$(bs)(\bar{s}\bar{b})^*$      & 12   & [4;~3;~4]  & $10969$ & \\
$(bs)^*(\bar{s}\bar{b})^*$  & 13  & [6;~3;~3]   & $10949$ & \\
$(bs)^*(\bar{s}\bar{b})^*$  & 14  & [6;~3;~4]   & $10902$ & $10782$ \\[2ex]
$K_1$  & 15  & [7;~3;~5]   & $11090$ & \\
  & 16  & [8;~3;~5]   & $11081$ & \\
  & 17  & [9;~3;~5]   & $11080$ & \\
  & 18  & [7;~3;~6]   & $10782$ & \\
  & 19  & [8;~3;~6]   & $10799$ & \\
  & 20  & [9;~3;~6]   & $10764$ & $10763$ \\[2ex]
$K_3$  & 21  & [13;~3;~9]   & $10905$ & \\
  & 22  & [14;~3;~9]   & $10910$ & \\
  & 23  & [15;~3;~9]   & $10946$ & \\
  & 24  & [13;~3;~10]   & $10954$ & \\
  & 25  & [14;~3;~10]   & $10953$ & \\
  & 26  & [15;~3;~10]   & $10979$ & $10770$ \\[2ex]
\multicolumn{4}{c}{Complete coupled-channels:} & $10333$
\end{tabular}
\end{ruledtabular}
\end{table}

\begin{figure}[!t]
\includegraphics[width=0.49\textwidth, trim={2.3cm 2.0cm 2.0cm 1.0cm}]{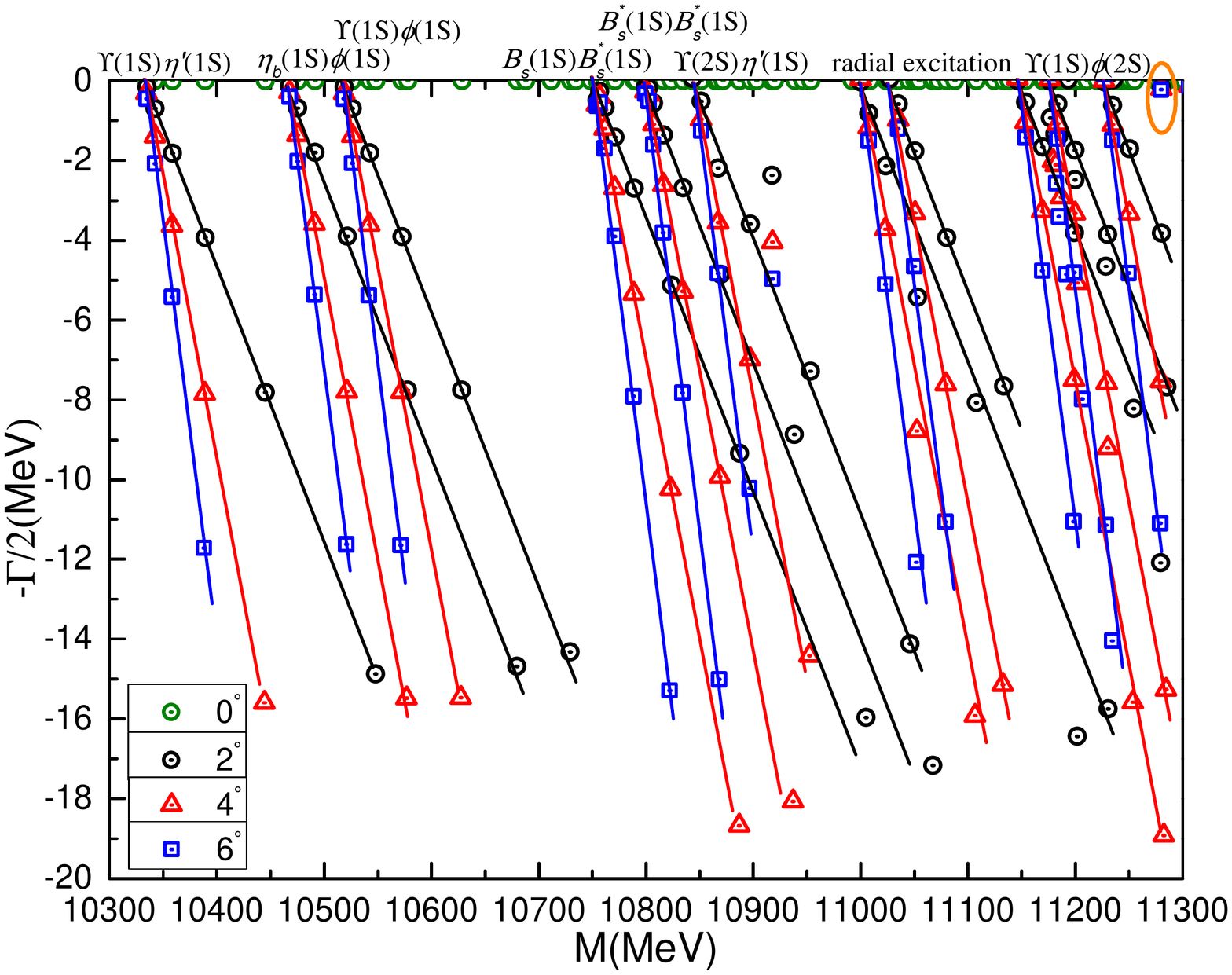}
\caption{\label{PP11} The complete coupled-channels calculation of the $b\bar{b}s\bar{s}$ tetraquark system with $I(J^P)=0(1^+)$ quantum numbers. We use the complex-scaling method of the chiral quark model varying $\theta$ from $0^\circ$ to $6^\circ$.}
\end{figure}

{\bf The $\bm{I(J^P)=0(1^+)}$ sector:} 26 channels contribute to the $b\bar{b}s\bar{s}$ tetraquark system with $0(1^+)$ quantum numbers.
The five color-singlet meson-meson channels, $\eta_b \phi$, $\Upsilon \eta'$, $\Upsilon \phi$ and $B^{(*)}_s \bar{B}^{(*)}_s$, are all scattering states and the lowest mass $10333$ GeV is just the value of non-interacting $\Upsilon \eta'$ threshold. The other exotic quark arrangements are generally located in $10.76-11.10$ GeV. Meanwhile, coupled-channel has little effect on these results.
  
Figure~\ref{PP11} presents distributions of $\eta_b \phi$, $\Upsilon \eta'$, $\Upsilon \phi$ and $B^{(*)}_s \bar{B}^{(*)}_s$ channels, in both ground and radial excitation states, when coupled-channel calculation is performed with CSM employed. One extremely narrow resonance is found circled above $\Upsilon(1S)\phi(2S)$ threshold. The calculated complex-energy is $11280+i0.4$ MeV. The quite narrow resonance width indicates that it should be stable against two-mesons strong decay processes.

\begin{table}[!t]
\caption{\label{GresultCC12} Lowest-lying $b\bar{b}s\bar{s}$ tetraquark states with $I(J^P)=0(2^+)$ calculated within the real range formulation of the chiral quark model. The results are similarly organized as those in Table~\ref{GresultCC1}.
(unit: MeV).}
\begin{ruledtabular}
\begin{tabular}{lcccc}
~~Channel   & Index & $\chi_J^{\sigma_i}$;~$\chi_I^{f_j}$;~$\chi_k^c$ & $M$ & Mixed~~ \\
        &   &$[i; ~j; ~k]$ &  \\[2ex]
$(\Upsilon \phi)^1 (10242)$  & 1  & [1;~3;~1]   & $10516$ &  \\
$(B^*_s \bar{B}^*_s)^1 (10650)$  & 2  & [1;~3;~1]   & $10800$ & $10516$ \\[2ex]
$(\Upsilon \phi)^8$  & 3  & [1;~3;~2]   & $11090$ &  \\
$(B^*_s \bar{B}^*_s)^8$  & 4  & [1;~3;~2]   & $10974$ & $10886$ \\[2ex]
$(bs)^*(\bar{s}\bar{b})^*$  & 5  & [1;~3;~3]   & $10972$ & \\
$(bs)^*(\bar{s}\bar{b})^*$  & 6  & [1;~3;~4]   & $10931$ & $10822$ \\[2ex]
$K_1$  & 7  & [1;~3;~5]   & $11088$ & \\
  & 8  & [1;~3;~6]   & $10815$ & $10815$ \\[2ex]
$K_3$  & 9  & [1;~3;~9]   & $10928$ & \\
  & 10  & [1;~3;~10]   & $10962$ & $10823$ \\[2ex]
\multicolumn{4}{c}{Complete coupled-channels:} & $10516$
\end{tabular}
\end{ruledtabular}
\end{table}

\begin{figure}[ht]
\includegraphics[width=0.49\textwidth, trim={2.3cm 2.0cm 2.0cm 1.0cm}]{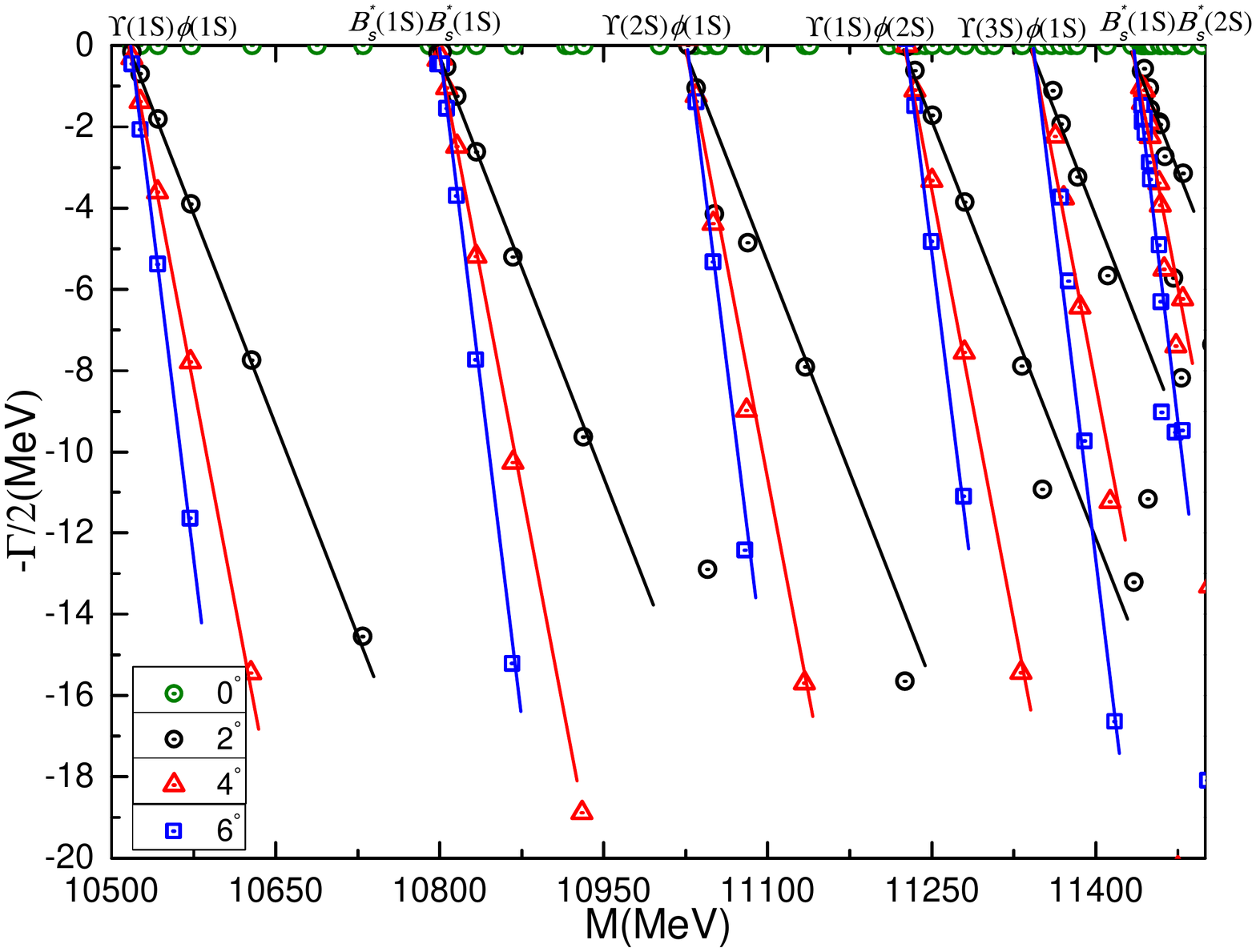}
\caption{\label{PP12} The complete coupled-channels calculation of the $bb\bar{s}\bar{s}$ tetraquark system with $I(J^P)=0(2^+)$ quantum numbers. We use the complex-scaling method of the chiral quark model varying $\theta$ from $0^\circ$ to $6^\circ$.}
\end{figure}

{\bf The $\bm{I(J^P)=0(2^+)}$ sector:} Table~\ref{GresultCC12} shows all the channels that can contribute to the mass of the highest spin state of $b\bar{b}s\bar{s}$ tetraquark. No bound state is still found in the single- and coupled-channel calculations, with the lowest mass indicating the value of $\Upsilon\phi$ threshold, $10516$ MeV. Our results of a complete coupled-channel calculation in complex-range are shown in Fig.~\ref{PP12}. Therein, one can find that $\Upsilon\phi$ and $B^*_s \bar{B}^*_s$ scattering states are well presented in $10.5-11.5$ GeV, and no stable resonance pole is acquired.

%%%%%%%%%%%%%%%%%%%%%%%%%%%%%%%%%%%%%%%%%%%%%%%%%%%%%%%%%%%%%%%%%%%%%%%%%%%%%%%%

\begin{table}[!t]
\caption{\label{GresultCCT} Summary of resonance structures found in the $b\bar{b}q\bar{q}$ $(q=u,\,d,\,s)$ tetraquark systems. The first column shows the isospin, total spin and parity quantum numbers of each singularity, the second column refers to the dominant channel, the obtained resonances are presented with the following notation: $E=M+i\Gamma$ in the last column. (unit: MeV).}
\begin{ruledtabular}
\begin{tabular}{llc}
~ $I(J^P)$ & Dominant Channel   & Complex Energy~~ \\
\hline
\multicolumn{3}{c}{$b\bar{b}q\bar{q}$ $(q=u,\,d)$ tetraquarks}\\
~~$0(0^+)$  & $B(1S)\bar{B}(2S)$   & $11322+i3.9$~~  \\
            & $B(1S)\bar{B}(2S)$   & $11328+i3.4$~~ \\[2ex]
~~$0(1^+)$  & $B(1S)\bar{B}^*(1S)$   & $10584+i1.6$~~ \\
            & $\Upsilon(3S)\omega(1S)$   & $11073+i1.8$~~ \\
            & $\Upsilon(3S)\omega(1S)$   & $11209+i2.8$~~ \\
            & $\Upsilon(3S)\omega(1S)$   & $11272+i5.0$~~ \\[2ex]
~~$0(2^+)$  & $B^*(1S\bar{B}^*(1S)$   & $10618+i1.0$~~ \\
                  & $B(1S)\bar{B}^*(2S)$   & $11416+i32.6$~~ \\[2ex]
~~$1(0^+)$  & $\Upsilon(2S)\rho(1S)$   & $10935+i1.4$~~  \\[2ex]
~~$1(1^+)$  & $B^*(1S)\bar{B}^*(1S)$   & $10750+i1.0$~~\\
                   & $\eta_b(1S)\rho(2S)$   & $10960+i0.7$~~ \\
                   & $\eta_b(1S)\rho(2S)$   & $10967+i2.6$~~\\[2ex]
~~$1(2^+)$  & $B^*(1S)\bar{B}^*(2S)$   & $11390+i7.8$~~ \\
\hline
\multicolumn{3}{c}{$b\bar{b}u\bar{s}$ tetraquarks}\\
~~$\frac{1}{2}(0^+)$  & $\Upsilon(1S)K^*(2S)$   & $11176+i0.8$~~  \\[2ex]
~~$\frac{1}{2}(1^+)$  & $\eta_b(1S)K^*(2S)$   & $11178+i0.8$~~ \\
                                   & $\eta_b(1S)K^*(2S)$   & $11086+i1.0$~~ \\
                                   & $\Upsilon(1S)K^*(2S)$   & $11092+i1.1$~~ \\[2ex]
~~$\frac{1}{2}(2^+)$  & $\Upsilon(2S)K^*(1S)$   & $11090+i1.1$~~ \\
\hline
\multicolumn{3}{c}{$b\bar{b}s\bar{s}$ tetraquarks}\\
~~$0(1^+)$  & $\Upsilon(1S)\phi(2S)$   & $11280+i0.4$~~ 
\end{tabular}
\end{ruledtabular}
\end{table}

\section{Summary}
\label{sec:summary}

The bottomonium-like tetraquarks $b\bar{b}q\bar{q}$ $(q=u,\,d,\,s)$ with spin-parity $J^P=0^+$, $1^+$ and $2^+$, and isospin $I=0,\,1$ or $\frac{1}{2}$, are systemically investigated by means of real- and complex-scaling range formulation of a chiral quark model, along with a high efficiency numerical approach for solving the 4-body Sch\"odinger equation, the Gaussian expansion method. The model contains one-gluon exchange, linear-screened color confining and Goldstone-boson exchanges (between light quarks) interactions, and it has been successfully applied to the description of hadron, hadron-hadron and multiquark phenomenology. We considered in our calculations all possible tetraquark arrangements allowed by quantum numbers: singlet- and hidden-color meson-meson configurations, diquark-antidiquark arrangements with their allowed color triplet-antitriplet and sextet-antisextet wave functions, and four K-type structures.

Several resonance structures are found in a complete coupled-channel calculation. They are summarized in Table~\ref{GresultCCT} which collects quantum numbers, dominant channel and pole position in the complex energy-plane. A brief review of them is as follows. On one hand, in the $b\bar{b}q\bar{q}$ $(q=u,\,d)$ tetraquark system, we found resonance states of $B^{(*)} \bar{B}^{(*)}$, $\Upsilon\omega$, $\Upsilon\rho$ and $\eta_b \rho$ nature with all possible $I(J^P)$ quantum numbers. Their masses are generally located in the energy range $11.0-11.3$ GeV and their widths are less than $10$ MeV. On the other hand, extremely narrow resonances, with two-meson strong decay widths less than $1.5$ MeV, are obtained in both $b\bar{b}u\bar{s}$ and $b\bar{b}s\bar{s}$ tetraquark systems. Particularly, four radial excitations of $\Upsilon K^*$ and $\eta_b K^*$ are found at $\sim11.1$ GeV in $J^P=0^+$, $1^+$ and $2^+$ channels of $b\bar{b}u\bar{s}$ system. One $\Upsilon(1S)\phi(2S)$ resonance state is obtained at $11.28$ GeV in the $J^P=1^+$ sector.

%%%%%%%%%%%%%%%%%%%%%%%%%%%%%%%%%%%%%%%%%%%%%%%%%%%%%%%%%%%%%%%%%%%%%%%%%%%%%%%%

% If you have acknowledgments, this puts in the proper section head.
\begin{acknowledgments}
Work partially financed by: the Zhejiang Provincial Natural Science Foundation under Grant No. LQ22A050004; National Natural Science Foundation of China under Grant Nos. 11535005 and 11775118; the Ministerio Espa\~nol de Ciencia e Innovaci\'on under grant No. PID2019-107844GB-C22; and Junta de Andaluc\'ia, contract nos. P18-FR-5057 and Operativo FEDER Andaluc\'ia 2014-2020 UHU-1264517.
\end{acknowledgments}

%%%%%%%%%%%%%%%%%%%%%%%%%%%%%%%%%%%%%%%%%%%%%%%%%%%%%%%%%%%%%%%%%%%%%%%%%%%%%%%%

% Create the reference section using BibTeX:
\bibliography{bbqq}

%merlin.mbs apsrev4-1.bst 2010-07-25 4.21a (PWD, AO, DPC) hacked
%Control: key (0)
%Control: author (8) initials jnrlst
%Control: editor formatted (1) identically to author
%Control: production of article title (-1) disabled
%Control: page (0) single
%Control: year (1) truncated
%Control: production of eprint (0) enabled
\begin{thebibliography}{70}%
\makeatletter
\providecommand \@ifxundefined [1]{%
 \@ifx{#1\undefined}
}%
\providecommand \@ifnum [1]{%
 \ifnum #1\expandafter \@firstoftwo
 \else \expandafter \@secondoftwo
 \fi
}%
\providecommand \@ifx [1]{%
 \ifx #1\expandafter \@firstoftwo
 \else \expandafter \@secondoftwo
 \fi
}%
\providecommand \natexlab [1]{#1}%
\providecommand \enquote  [1]{``#1''}%
\providecommand \bibnamefont  [1]{#1}%
\providecommand \bibfnamefont [1]{#1}%
\providecommand \citenamefont [1]{#1}%
\providecommand \href@noop [0]{\@secondoftwo}%
\providecommand \href [0]{\begingroup \@sanitize@url \@href}%
\providecommand \@href[1]{\@@startlink{#1}\@@href}%
\providecommand \@@href[1]{\endgroup#1\@@endlink}%
\providecommand \@sanitize@url [0]{\catcode `\\12\catcode `\$12\catcode
  `\&12\catcode `\#12\catcode `\^12\catcode `\_12\catcode `\%12\relax}%
\providecommand \@@startlink[1]{}%
\providecommand \@@endlink[0]{}%
\providecommand \url  [0]{\begingroup\@sanitize@url \@url }%
\providecommand \@url [1]{\endgroup\@href {#1}{\urlprefix }}%
\providecommand \urlprefix  [0]{URL }%
\providecommand \Eprint [0]{\href }%
\providecommand \doibase [0]{http://dx.doi.org/}%
\providecommand \selectlanguage [0]{\@gobble}%
\providecommand \bibinfo  [0]{\@secondoftwo}%
\providecommand \bibfield  [0]{\@secondoftwo}%
\providecommand \translation [1]{[#1]}%
\providecommand \BibitemOpen [0]{}%
\providecommand \bibitemStop [0]{}%
\providecommand \bibitemNoStop [0]{.\EOS\space}%
\providecommand \EOS [0]{\spacefactor3000\relax}%
\providecommand \BibitemShut  [1]{\csname bibitem#1\endcsname}%
\let\auto@bib@innerbib\@empty
%</preamble>
\bibitem [{\citenamefont {Aaij}\ \emph {et~al.}(2015)\citenamefont {Aaij} \emph
  {et~al.}}]{Aaij:2015tga}%
  \BibitemOpen
  \bibfield  {author} {\bibinfo {author} {\bibfnamefont {R.}~\bibnamefont
  {Aaij}} \emph {et~al.} (\bibinfo {collaboration} {LHCb}),\ }\href {\doibase
  10.1103/PhysRevLett.115.072001} {\bibfield  {journal} {\bibinfo  {journal}
  {Phys. Rev. Lett.}\ }\textbf {\bibinfo {volume} {115}},\ \bibinfo {pages}
  {072001} (\bibinfo {year} {2015})}\BibitemShut {NoStop}%
\bibitem [{\citenamefont {Aaij}\ \emph {et~al.}(2019)\citenamefont {Aaij} \emph
  {et~al.}}]{lhcb:2019pc}%
  \BibitemOpen
  \bibfield  {author} {\bibinfo {author} {\bibfnamefont {R.}~\bibnamefont
  {Aaij}} \emph {et~al.} (\bibinfo {collaboration} {LHCb}),\ }\href {\doibase
  10.1103/PhysRevLett.122.222001} {\bibfield  {journal} {\bibinfo  {journal}
  {Phys. Rev. Lett.}\ }\textbf {\bibinfo {volume} {122}},\ \bibinfo {pages}
  {222001} (\bibinfo {year} {2019})}\BibitemShut {NoStop}%
\bibitem [{\citenamefont {Aaij}\ \emph {et~al.}(2022)\citenamefont {Aaij} \emph
  {et~al.}}]{LHCb:2021chn}%
  \BibitemOpen
  \bibfield  {author} {\bibinfo {author} {\bibfnamefont {R.}~\bibnamefont
  {Aaij}} \emph {et~al.} (\bibinfo {collaboration} {LHCb}),\ }\href {\doibase
  10.1103/PhysRevLett.128.062001} {\bibfield  {journal} {\bibinfo  {journal}
  {Phys. Rev. Lett.}\ }\textbf {\bibinfo {volume} {128}},\ \bibinfo {pages}
  {062001} (\bibinfo {year} {2022})},\ \Eprint
  {http://arxiv.org/abs/2108.04720} {arXiv:2108.04720 [hep-ex]} \BibitemShut
  {NoStop}%
\bibitem [{\citenamefont {Aaij}\ \emph
  {et~al.}(2021{\natexlab{a}})\citenamefont {Aaij} \emph
  {et~al.}}]{LHCb:2020jpq}%
  \BibitemOpen
  \bibfield  {author} {\bibinfo {author} {\bibfnamefont {R.}~\bibnamefont
  {Aaij}} \emph {et~al.} (\bibinfo {collaboration} {LHCb}),\ }\href {\doibase
  10.1016/j.scib.2021.02.030} {\bibfield  {journal} {\bibinfo  {journal} {Sci.
  Bull.}\ }\textbf {\bibinfo {volume} {66}},\ \bibinfo {pages} {1391} (\bibinfo
  {year} {2021}{\natexlab{a}})},\ \Eprint {http://arxiv.org/abs/2012.10380}
  {arXiv:2012.10380 [hep-ex]} \BibitemShut {NoStop}%
\bibitem [{\citenamefont {Aaij}\ \emph
  {et~al.}(2020{\natexlab{a}})\citenamefont {Aaij} \emph
  {et~al.}}]{LHCb:2020pxc}%
  \BibitemOpen
  \bibfield  {author} {\bibinfo {author} {\bibfnamefont {R.}~\bibnamefont
  {Aaij}} \emph {et~al.} (\bibinfo {collaboration} {LHCb}),\ }\href {\doibase
  10.1103/PhysRevD.102.112003} {\bibfield  {journal} {\bibinfo  {journal}
  {Phys. Rev. D}\ }\textbf {\bibinfo {volume} {102}},\ \bibinfo {pages}
  {112003} (\bibinfo {year} {2020}{\natexlab{a}})},\ \Eprint
  {http://arxiv.org/abs/2009.00026} {arXiv:2009.00026 [hep-ex]} \BibitemShut
  {NoStop}%
\bibitem [{\citenamefont {Aaij}\ \emph
  {et~al.}(2020{\natexlab{b}})\citenamefont {Aaij} \emph
  {et~al.}}]{LHCb:2020bls}%
  \BibitemOpen
  \bibfield  {author} {\bibinfo {author} {\bibfnamefont {R.}~\bibnamefont
  {Aaij}} \emph {et~al.} (\bibinfo {collaboration} {LHCb}),\ }\href {\doibase
  10.1103/PhysRevLett.125.242001} {\bibfield  {journal} {\bibinfo  {journal}
  {Phys. Rev. Lett.}\ }\textbf {\bibinfo {volume} {125}},\ \bibinfo {pages}
  {242001} (\bibinfo {year} {2020}{\natexlab{b}})},\ \Eprint
  {http://arxiv.org/abs/2009.00025} {arXiv:2009.00025 [hep-ex]} \BibitemShut
  {NoStop}%
\bibitem [{\citenamefont {Aaij}\ \emph
  {et~al.}(2021{\natexlab{b}})\citenamefont {Aaij} \emph
  {et~al.}}]{LHCb:2021vvq}%
  \BibitemOpen
  \bibfield  {author} {\bibinfo {author} {\bibfnamefont {R.}~\bibnamefont
  {Aaij}} \emph {et~al.} (\bibinfo {collaboration} {LHCb}),\ }\href@noop {} {\
  (\bibinfo {year} {2021}{\natexlab{b}})},\ \Eprint
  {http://arxiv.org/abs/2109.01038} {arXiv:2109.01038 [hep-ex]} \BibitemShut
  {NoStop}%
\bibitem [{\citenamefont {Aaij}\ \emph
  {et~al.}(2021{\natexlab{c}})\citenamefont {Aaij} \emph
  {et~al.}}]{LHCb:2021auc}%
  \BibitemOpen
  \bibfield  {author} {\bibinfo {author} {\bibfnamefont {R.}~\bibnamefont
  {Aaij}} \emph {et~al.} (\bibinfo {collaboration} {LHCb}),\ }\href@noop {} {\
  (\bibinfo {year} {2021}{\natexlab{c}})},\ \Eprint
  {http://arxiv.org/abs/2109.01056} {arXiv:2109.01056 [hep-ex]} \BibitemShut
  {NoStop}%
\bibitem [{\citenamefont {Ablikim}\ \emph {et~al.}(2021)\citenamefont {Ablikim}
  \emph {et~al.}}]{BESIII:2020qkh}%
  \BibitemOpen
  \bibfield  {author} {\bibinfo {author} {\bibfnamefont {M.}~\bibnamefont
  {Ablikim}} \emph {et~al.} (\bibinfo {collaboration} {BESIII}),\ }\href
  {\doibase 10.1103/PhysRevLett.126.102001} {\bibfield  {journal} {\bibinfo
  {journal} {Phys. Rev. Lett.}\ }\textbf {\bibinfo {volume} {126}},\ \bibinfo
  {pages} {102001} (\bibinfo {year} {2021})},\ \Eprint
  {http://arxiv.org/abs/2011.07855} {arXiv:2011.07855 [hep-ex]} \BibitemShut
  {NoStop}%
\bibitem [{\citenamefont {Aaij}\ \emph
  {et~al.}(2021{\natexlab{d}})\citenamefont {Aaij} \emph
  {et~al.}}]{LHCb:2021uow}%
  \BibitemOpen
  \bibfield  {author} {\bibinfo {author} {\bibfnamefont {R.}~\bibnamefont
  {Aaij}} \emph {et~al.} (\bibinfo {collaboration} {LHCb}),\ }\href {\doibase
  10.1103/PhysRevLett.127.082001} {\bibfield  {journal} {\bibinfo  {journal}
  {Phys. Rev. Lett.}\ }\textbf {\bibinfo {volume} {127}},\ \bibinfo {pages}
  {082001} (\bibinfo {year} {2021}{\natexlab{d}})},\ \Eprint
  {http://arxiv.org/abs/2103.01803} {arXiv:2103.01803 [hep-ex]} \BibitemShut
  {NoStop}%
\bibitem [{\citenamefont {Aaij}\ \emph
  {et~al.}(2020{\natexlab{c}})\citenamefont {Aaij} \emph
  {et~al.}}]{LHCb:2020bwg}%
  \BibitemOpen
  \bibfield  {author} {\bibinfo {author} {\bibfnamefont {R.}~\bibnamefont
  {Aaij}} \emph {et~al.} (\bibinfo {collaboration} {LHCb}),\ }\href {\doibase
  10.1016/j.scib.2020.08.032} {\bibfield  {journal} {\bibinfo  {journal} {Sci.
  Bull.}\ }\textbf {\bibinfo {volume} {65}},\ \bibinfo {pages} {1983} (\bibinfo
  {year} {2020}{\natexlab{c}})},\ \Eprint {http://arxiv.org/abs/2006.16957}
  {arXiv:2006.16957 [hep-ex]} \BibitemShut {NoStop}%
\bibitem [{\citenamefont {Bondar}\ \emph {et~al.}(2012)\citenamefont {Bondar}
  \emph {et~al.}}]{Belle:2011aa}%
  \BibitemOpen
  \bibfield  {author} {\bibinfo {author} {\bibfnamefont {A.}~\bibnamefont
  {Bondar}} \emph {et~al.} (\bibinfo {collaboration} {Belle}),\ }\href
  {\doibase 10.1103/PhysRevLett.108.122001} {\bibfield  {journal} {\bibinfo
  {journal} {Phys. Rev. Lett.}\ }\textbf {\bibinfo {volume} {108}},\ \bibinfo
  {pages} {122001} (\bibinfo {year} {2012})},\ \Eprint
  {http://arxiv.org/abs/1110.2251} {arXiv:1110.2251 [hep-ex]} \BibitemShut
  {NoStop}%
\bibitem [{\citenamefont {Garmash}\ \emph {et~al.}(2015)\citenamefont {Garmash}
  \emph {et~al.}}]{Belle:2014vzn}%
  \BibitemOpen
  \bibfield  {author} {\bibinfo {author} {\bibfnamefont {A.}~\bibnamefont
  {Garmash}} \emph {et~al.} (\bibinfo {collaboration} {Belle}),\ }\href
  {\doibase 10.1103/PhysRevD.91.072003} {\bibfield  {journal} {\bibinfo
  {journal} {Phys. Rev. D}\ }\textbf {\bibinfo {volume} {91}},\ \bibinfo
  {pages} {072003} (\bibinfo {year} {2015})},\ \Eprint
  {http://arxiv.org/abs/1403.0992} {arXiv:1403.0992 [hep-ex]} \BibitemShut
  {NoStop}%
\bibitem [{\citenamefont {Garmash}\ \emph {et~al.}(2016)\citenamefont {Garmash}
  \emph {et~al.}}]{Belle:2015upu}%
  \BibitemOpen
  \bibfield  {author} {\bibinfo {author} {\bibfnamefont {A.}~\bibnamefont
  {Garmash}} \emph {et~al.} (\bibinfo {collaboration} {Belle}),\ }\href
  {\doibase 10.1103/PhysRevLett.116.212001} {\bibfield  {journal} {\bibinfo
  {journal} {Phys. Rev. Lett.}\ }\textbf {\bibinfo {volume} {116}},\ \bibinfo
  {pages} {212001} (\bibinfo {year} {2016})},\ \Eprint
  {http://arxiv.org/abs/1512.07419} {arXiv:1512.07419 [hep-ex]} \BibitemShut
  {NoStop}%
\bibitem [{\citenamefont {Cleven}\ \emph {et~al.}(2011)\citenamefont {Cleven},
  \citenamefont {Guo}, \citenamefont {Hanhart},\ and\ \citenamefont
  {Meissner}}]{Cleven:2011gp}%
  \BibitemOpen
  \bibfield  {author} {\bibinfo {author} {\bibfnamefont {M.}~\bibnamefont
  {Cleven}}, \bibinfo {author} {\bibfnamefont {F.-K.}\ \bibnamefont {Guo}},
  \bibinfo {author} {\bibfnamefont {C.}~\bibnamefont {Hanhart}}, \ and\
  \bibinfo {author} {\bibfnamefont {U.-G.}\ \bibnamefont {Meissner}},\ }\href
  {\doibase 10.1140/epja/i2011-11120-6} {\bibfield  {journal} {\bibinfo
  {journal} {Eur. Phys. J. A}\ }\textbf {\bibinfo {volume} {47}},\ \bibinfo
  {pages} {120} (\bibinfo {year} {2011})},\ \Eprint
  {http://arxiv.org/abs/1107.0254} {arXiv:1107.0254 [hep-ph]} \BibitemShut
  {NoStop}%
\bibitem [{\citenamefont {He}(2014)}]{He:2014nya}%
  \BibitemOpen
  \bibfield  {author} {\bibinfo {author} {\bibfnamefont {J.}~\bibnamefont
  {He}},\ }\href {\doibase 10.1103/PhysRevD.90.076008} {\bibfield  {journal}
  {\bibinfo  {journal} {Phys. Rev. D}\ }\textbf {\bibinfo {volume} {90}},\
  \bibinfo {pages} {076008} (\bibinfo {year} {2014})},\ \Eprint
  {http://arxiv.org/abs/1409.8506} {arXiv:1409.8506 [hep-ph]} \BibitemShut
  {NoStop}%
\bibitem [{\citenamefont {Sun}\ \emph {et~al.}(2011)\citenamefont {Sun},
  \citenamefont {He}, \citenamefont {Liu}, \citenamefont {Luo},\ and\
  \citenamefont {Zhu}}]{Sun:2011uh}%
  \BibitemOpen
  \bibfield  {author} {\bibinfo {author} {\bibfnamefont {Z.-F.}\ \bibnamefont
  {Sun}}, \bibinfo {author} {\bibfnamefont {J.}~\bibnamefont {He}}, \bibinfo
  {author} {\bibfnamefont {X.}~\bibnamefont {Liu}}, \bibinfo {author}
  {\bibfnamefont {Z.-G.}\ \bibnamefont {Luo}}, \ and\ \bibinfo {author}
  {\bibfnamefont {S.-L.}\ \bibnamefont {Zhu}},\ }\href {\doibase
  10.1103/PhysRevD.84.054002} {\bibfield  {journal} {\bibinfo  {journal} {Phys.
  Rev. D}\ }\textbf {\bibinfo {volume} {84}},\ \bibinfo {pages} {054002}
  (\bibinfo {year} {2011})},\ \Eprint {http://arxiv.org/abs/1106.2968}
  {arXiv:1106.2968 [hep-ph]} \BibitemShut {NoStop}%
\bibitem [{\citenamefont {Yang}\ \emph
  {et~al.}(2019{\natexlab{a}})\citenamefont {Yang}, \citenamefont {Tan},
  \citenamefont {Zong},\ and\ \citenamefont {Ping}}]{Yang:2017rmm}%
  \BibitemOpen
  \bibfield  {author} {\bibinfo {author} {\bibfnamefont {Y.-C.}\ \bibnamefont
  {Yang}}, \bibinfo {author} {\bibfnamefont {Z.-Y.}\ \bibnamefont {Tan}},
  \bibinfo {author} {\bibfnamefont {H.-S.}\ \bibnamefont {Zong}}, \ and\
  \bibinfo {author} {\bibfnamefont {J.}~\bibnamefont {Ping}},\ }\href {\doibase
  10.1007/s00601-018-1477-5} {\bibfield  {journal} {\bibinfo  {journal} {Few
  Body Syst.}\ }\textbf {\bibinfo {volume} {60}},\ \bibinfo {pages} {9}
  (\bibinfo {year} {2019}{\natexlab{a}})},\ \Eprint
  {http://arxiv.org/abs/1712.09285} {arXiv:1712.09285 [hep-ph]} \BibitemShut
  {NoStop}%
\bibitem [{\citenamefont {Zhang}\ \emph {et~al.}(2022)\citenamefont {Zhang},
  \citenamefont {Kang},\ and\ \citenamefont {Guo}}]{Zhang:2022hfa}%
  \BibitemOpen
  \bibfield  {author} {\bibinfo {author} {\bibfnamefont {L.}~\bibnamefont
  {Zhang}}, \bibinfo {author} {\bibfnamefont {X.-W.}\ \bibnamefont {Kang}}, \
  and\ \bibinfo {author} {\bibfnamefont {X.-H.}\ \bibnamefont {Guo}},\
  }\href@noop {} {\  (\bibinfo {year} {2022})},\ \Eprint
  {http://arxiv.org/abs/2203.02301} {arXiv:2203.02301 [hep-ph]} \BibitemShut
  {NoStop}%
\bibitem [{\citenamefont {Nieves}\ and\ \citenamefont
  {Valderrama}(2011)}]{Nieves:2011zz}%
  \BibitemOpen
  \bibfield  {author} {\bibinfo {author} {\bibfnamefont {J.}~\bibnamefont
  {Nieves}}\ and\ \bibinfo {author} {\bibfnamefont {M.~P.}\ \bibnamefont
  {Valderrama}},\ }\href {\doibase 10.1103/PhysRevD.84.056015} {\bibfield
  {journal} {\bibinfo  {journal} {Phys. Rev. D}\ }\textbf {\bibinfo {volume}
  {84}},\ \bibinfo {pages} {056015} (\bibinfo {year} {2011})},\ \Eprint
  {http://arxiv.org/abs/1106.0600} {arXiv:1106.0600 [hep-ph]} \BibitemShut
  {NoStop}%
\bibitem [{\citenamefont {Dias}\ \emph {et~al.}(2015)\citenamefont {Dias},
  \citenamefont {Aceti},\ and\ \citenamefont {Oset}}]{Dias:2014pva}%
  \BibitemOpen
  \bibfield  {author} {\bibinfo {author} {\bibfnamefont {J.~M.}\ \bibnamefont
  {Dias}}, \bibinfo {author} {\bibfnamefont {F.}~\bibnamefont {Aceti}}, \ and\
  \bibinfo {author} {\bibfnamefont {E.}~\bibnamefont {Oset}},\ }\href {\doibase
  10.1103/PhysRevD.91.076001} {\bibfield  {journal} {\bibinfo  {journal} {Phys.
  Rev. D}\ }\textbf {\bibinfo {volume} {91}},\ \bibinfo {pages} {076001}
  (\bibinfo {year} {2015})},\ \Eprint {http://arxiv.org/abs/1410.1785}
  {arXiv:1410.1785 [hep-ph]} \BibitemShut {NoStop}%
\bibitem [{\citenamefont {Guo}\ \emph {et~al.}(2016)\citenamefont {Guo},
  \citenamefont {Hanhart}, \citenamefont {Kalashnikova}, \citenamefont
  {Matuschek}, \citenamefont {Mizuk}, \citenamefont {Nefediev}, \citenamefont
  {Wang},\ and\ \citenamefont {Wynen}}]{Guo:2016bjq}%
  \BibitemOpen
  \bibfield  {author} {\bibinfo {author} {\bibfnamefont {F.~K.}\ \bibnamefont
  {Guo}}, \bibinfo {author} {\bibfnamefont {C.}~\bibnamefont {Hanhart}},
  \bibinfo {author} {\bibfnamefont {Y.~S.}\ \bibnamefont {Kalashnikova}},
  \bibinfo {author} {\bibfnamefont {P.}~\bibnamefont {Matuschek}}, \bibinfo
  {author} {\bibfnamefont {R.~V.}\ \bibnamefont {Mizuk}}, \bibinfo {author}
  {\bibfnamefont {A.~V.}\ \bibnamefont {Nefediev}}, \bibinfo {author}
  {\bibfnamefont {Q.}~\bibnamefont {Wang}}, \ and\ \bibinfo {author}
  {\bibfnamefont {J.~L.}\ \bibnamefont {Wynen}},\ }\href {\doibase
  10.1103/PhysRevD.93.074031} {\bibfield  {journal} {\bibinfo  {journal} {Phys.
  Rev. D}\ }\textbf {\bibinfo {volume} {93}},\ \bibinfo {pages} {074031}
  (\bibinfo {year} {2016})},\ \Eprint {http://arxiv.org/abs/1602.00940}
  {arXiv:1602.00940 [hep-ph]} \BibitemShut {NoStop}%
\bibitem [{\citenamefont {Wang}\ \emph {et~al.}(2018)\citenamefont {Wang},
  \citenamefont {Baru}, \citenamefont {Filin}, \citenamefont {Hanhart},
  \citenamefont {Nefediev},\ and\ \citenamefont {Wynen}}]{Wang:2018jlv}%
  \BibitemOpen
  \bibfield  {author} {\bibinfo {author} {\bibfnamefont {Q.}~\bibnamefont
  {Wang}}, \bibinfo {author} {\bibfnamefont {V.}~\bibnamefont {Baru}}, \bibinfo
  {author} {\bibfnamefont {A.~A.}\ \bibnamefont {Filin}}, \bibinfo {author}
  {\bibfnamefont {C.}~\bibnamefont {Hanhart}}, \bibinfo {author} {\bibfnamefont
  {A.~V.}\ \bibnamefont {Nefediev}}, \ and\ \bibinfo {author} {\bibfnamefont
  {J.~L.}\ \bibnamefont {Wynen}},\ }\href {\doibase 10.1103/PhysRevD.98.074023}
  {\bibfield  {journal} {\bibinfo  {journal} {Phys. Rev. D}\ }\textbf {\bibinfo
  {volume} {98}},\ \bibinfo {pages} {074023} (\bibinfo {year} {2018})},\
  \Eprint {http://arxiv.org/abs/1805.07453} {arXiv:1805.07453 [hep-ph]}
  \BibitemShut {NoStop}%
\bibitem [{\citenamefont {Ortega}\ \emph {et~al.}(2021)\citenamefont {Ortega},
  \citenamefont {Segovia},\ and\ \citenamefont {Fernandez}}]{Ortega:2021xst}%
  \BibitemOpen
  \bibfield  {author} {\bibinfo {author} {\bibfnamefont {P.~G.}\ \bibnamefont
  {Ortega}}, \bibinfo {author} {\bibfnamefont {J.}~\bibnamefont {Segovia}}, \
  and\ \bibinfo {author} {\bibfnamefont {F.}~\bibnamefont {Fernandez}},\ }\href
  {\doibase 10.1103/PhysRevD.104.094004} {\bibfield  {journal} {\bibinfo
  {journal} {Phys. Rev. D}\ }\textbf {\bibinfo {volume} {104}},\ \bibinfo
  {pages} {094004} (\bibinfo {year} {2021})},\ \Eprint
  {http://arxiv.org/abs/2107.02544} {arXiv:2107.02544 [hep-ph]} \BibitemShut
  {NoStop}%
\bibitem [{\citenamefont {Sadl}\ and\ \citenamefont
  {Prelovsek}(2021)}]{Sadl:2021bme}%
  \BibitemOpen
  \bibfield  {author} {\bibinfo {author} {\bibfnamefont {M.}~\bibnamefont
  {Sadl}}\ and\ \bibinfo {author} {\bibfnamefont {S.}~\bibnamefont
  {Prelovsek}},\ }\href {\doibase 10.1103/PhysRevD.104.114503} {\bibfield
  {journal} {\bibinfo  {journal} {Phys. Rev. D}\ }\textbf {\bibinfo {volume}
  {104}},\ \bibinfo {pages} {114503} (\bibinfo {year} {2021})},\ \Eprint
  {http://arxiv.org/abs/2109.08560} {arXiv:2109.08560 [hep-lat]} \BibitemShut
  {NoStop}%
\bibitem [{\citenamefont {Dai}\ \emph {et~al.}(2022)\citenamefont {Dai},
  \citenamefont {Oset}, \citenamefont {Feijoo}, \citenamefont {Molina},
  \citenamefont {Roca}, \citenamefont {Torres},\ and\ \citenamefont
  {Khemchandani}}]{Dai:2022ulk}%
  \BibitemOpen
  \bibfield  {author} {\bibinfo {author} {\bibfnamefont {L.~R.}\ \bibnamefont
  {Dai}}, \bibinfo {author} {\bibfnamefont {E.}~\bibnamefont {Oset}}, \bibinfo
  {author} {\bibfnamefont {A.}~\bibnamefont {Feijoo}}, \bibinfo {author}
  {\bibfnamefont {R.}~\bibnamefont {Molina}}, \bibinfo {author} {\bibfnamefont
  {L.}~\bibnamefont {Roca}}, \bibinfo {author} {\bibfnamefont {A.~M.}\
  \bibnamefont {Torres}}, \ and\ \bibinfo {author} {\bibfnamefont {K.~P.}\
  \bibnamefont {Khemchandani}},\ }\href@noop {} {\  (\bibinfo {year} {2022})},\
  \Eprint {http://arxiv.org/abs/2201.04840} {arXiv:2201.04840 [hep-ph]}
  \BibitemShut {NoStop}%
\bibitem [{\citenamefont {Goerke}\ \emph {et~al.}(2017)\citenamefont {Goerke},
  \citenamefont {Gutsche}, \citenamefont {Ivanov}, \citenamefont {K\"orner},\
  and\ \citenamefont {Lyubovitskij}}]{Goerke:2017svb}%
  \BibitemOpen
  \bibfield  {author} {\bibinfo {author} {\bibfnamefont {F.}~\bibnamefont
  {Goerke}}, \bibinfo {author} {\bibfnamefont {T.}~\bibnamefont {Gutsche}},
  \bibinfo {author} {\bibfnamefont {M.~A.}\ \bibnamefont {Ivanov}}, \bibinfo
  {author} {\bibfnamefont {J.~G.}\ \bibnamefont {K\"orner}}, \ and\ \bibinfo
  {author} {\bibfnamefont {V.~E.}\ \bibnamefont {Lyubovitskij}},\ }\href
  {\doibase 10.1103/PhysRevD.96.054028} {\bibfield  {journal} {\bibinfo
  {journal} {Phys. Rev. D}\ }\textbf {\bibinfo {volume} {96}},\ \bibinfo
  {pages} {054028} (\bibinfo {year} {2017})},\ \Eprint
  {http://arxiv.org/abs/1707.00539} {arXiv:1707.00539 [hep-ph]} \BibitemShut
  {NoStop}%
\bibitem [{\citenamefont {Wang}\ \emph {et~al.}(2019)\citenamefont {Wang},
  \citenamefont {Liu}, \citenamefont {Ma}, \citenamefont {Liu}, \citenamefont
  {Chen}, \citenamefont {Deng},\ and\ \citenamefont {Zhu}}]{Wang:2018pwi}%
  \BibitemOpen
  \bibfield  {author} {\bibinfo {author} {\bibfnamefont {G.-J.}\ \bibnamefont
  {Wang}}, \bibinfo {author} {\bibfnamefont {X.-H.}\ \bibnamefont {Liu}},
  \bibinfo {author} {\bibfnamefont {L.}~\bibnamefont {Ma}}, \bibinfo {author}
  {\bibfnamefont {X.}~\bibnamefont {Liu}}, \bibinfo {author} {\bibfnamefont
  {X.-L.}\ \bibnamefont {Chen}}, \bibinfo {author} {\bibfnamefont {W.-Z.}\
  \bibnamefont {Deng}}, \ and\ \bibinfo {author} {\bibfnamefont {S.-L.}\
  \bibnamefont {Zhu}},\ }\href {\doibase 10.1140/epjc/s10052-019-7059-y}
  {\bibfield  {journal} {\bibinfo  {journal} {Eur. Phys. J. C}\ }\textbf
  {\bibinfo {volume} {79}},\ \bibinfo {pages} {567} (\bibinfo {year} {2019})},\
  \Eprint {http://arxiv.org/abs/1811.10339} {arXiv:1811.10339 [hep-ph]}
  \BibitemShut {NoStop}%
\bibitem [{\citenamefont {Dong}\ \emph
  {et~al.}(2021{\natexlab{a}})\citenamefont {Dong}, \citenamefont {Guo},\ and\
  \citenamefont {Zou}}]{Dong:2020hxe}%
  \BibitemOpen
  \bibfield  {author} {\bibinfo {author} {\bibfnamefont {X.-K.}\ \bibnamefont
  {Dong}}, \bibinfo {author} {\bibfnamefont {F.-K.}\ \bibnamefont {Guo}}, \
  and\ \bibinfo {author} {\bibfnamefont {B.-S.}\ \bibnamefont {Zou}},\ }\href
  {\doibase 10.1103/PhysRevLett.126.152001} {\bibfield  {journal} {\bibinfo
  {journal} {Phys. Rev. Lett.}\ }\textbf {\bibinfo {volume} {126}},\ \bibinfo
  {pages} {152001} (\bibinfo {year} {2021}{\natexlab{a}})},\ \Eprint
  {http://arxiv.org/abs/2011.14517} {arXiv:2011.14517 [hep-ph]} \BibitemShut
  {NoStop}%
\bibitem [{\citenamefont {Chen}\ \emph {et~al.}(2016)\citenamefont {Chen},
  \citenamefont {Chen}, \citenamefont {Liu},\ and\ \citenamefont
  {Zhu}}]{Chen:2016qju}%
  \BibitemOpen
  \bibfield  {author} {\bibinfo {author} {\bibfnamefont {H.-X.}\ \bibnamefont
  {Chen}}, \bibinfo {author} {\bibfnamefont {W.}~\bibnamefont {Chen}}, \bibinfo
  {author} {\bibfnamefont {X.}~\bibnamefont {Liu}}, \ and\ \bibinfo {author}
  {\bibfnamefont {S.-L.}\ \bibnamefont {Zhu}},\ }\href {\doibase
  10.1016/j.physrep.2016.05.004} {\bibfield  {journal} {\bibinfo  {journal}
  {Phys. Rept.}\ }\textbf {\bibinfo {volume} {639}},\ \bibinfo {pages} {1}
  (\bibinfo {year} {2016})},\ \Eprint {http://arxiv.org/abs/1601.02092}
  {arXiv:1601.02092 [hep-ph]} \BibitemShut {NoStop}%
\bibitem [{\citenamefont {Chen}\ \emph {et~al.}(2017)\citenamefont {Chen},
  \citenamefont {Chen}, \citenamefont {Liu}, \citenamefont {Liu},\ and\
  \citenamefont {Zhu}}]{Chen:2016spr}%
  \BibitemOpen
  \bibfield  {author} {\bibinfo {author} {\bibfnamefont {H.-X.}\ \bibnamefont
  {Chen}}, \bibinfo {author} {\bibfnamefont {W.}~\bibnamefont {Chen}}, \bibinfo
  {author} {\bibfnamefont {X.}~\bibnamefont {Liu}}, \bibinfo {author}
  {\bibfnamefont {Y.-R.}\ \bibnamefont {Liu}}, \ and\ \bibinfo {author}
  {\bibfnamefont {S.-L.}\ \bibnamefont {Zhu}},\ }\href {\doibase
  10.1088/1361-6633/aa6420} {\bibfield  {journal} {\bibinfo  {journal} {Rept.
  Prog. Phys.}\ }\textbf {\bibinfo {volume} {80}},\ \bibinfo {pages} {076201}
  (\bibinfo {year} {2017})},\ \Eprint {http://arxiv.org/abs/1609.08928}
  {arXiv:1609.08928 [hep-ph]} \BibitemShut {NoStop}%
\bibitem [{\citenamefont {Guo}\ \emph {et~al.}(2018)\citenamefont {Guo},
  \citenamefont {Hanhart}, \citenamefont {Mei\ss{}ner}, \citenamefont {Wang},
  \citenamefont {Zhao},\ and\ \citenamefont {Zou}}]{Guo:2017jvc}%
  \BibitemOpen
  \bibfield  {author} {\bibinfo {author} {\bibfnamefont {F.-K.}\ \bibnamefont
  {Guo}}, \bibinfo {author} {\bibfnamefont {C.}~\bibnamefont {Hanhart}},
  \bibinfo {author} {\bibfnamefont {U.-G.}\ \bibnamefont {Mei\ss{}ner}},
  \bibinfo {author} {\bibfnamefont {Q.}~\bibnamefont {Wang}}, \bibinfo {author}
  {\bibfnamefont {Q.}~\bibnamefont {Zhao}}, \ and\ \bibinfo {author}
  {\bibfnamefont {B.-S.}\ \bibnamefont {Zou}},\ }\href {\doibase
  10.1103/RevModPhys.90.015004} {\bibfield  {journal} {\bibinfo  {journal}
  {Rev. Mod. Phys.}\ }\textbf {\bibinfo {volume} {90}},\ \bibinfo {pages}
  {015004} (\bibinfo {year} {2018})},\ \Eprint
  {http://arxiv.org/abs/1705.00141} {arXiv:1705.00141 [hep-ph]} \BibitemShut
  {NoStop}%
\bibitem [{\citenamefont {Liu}\ \emph {et~al.}(2019)\citenamefont {Liu},
  \citenamefont {Chen}, \citenamefont {Chen}, \citenamefont {Liu},\ and\
  \citenamefont {Zhu}}]{Liu:2019zoy}%
  \BibitemOpen
  \bibfield  {author} {\bibinfo {author} {\bibfnamefont {Y.-R.}\ \bibnamefont
  {Liu}}, \bibinfo {author} {\bibfnamefont {H.-X.}\ \bibnamefont {Chen}},
  \bibinfo {author} {\bibfnamefont {W.}~\bibnamefont {Chen}}, \bibinfo {author}
  {\bibfnamefont {X.}~\bibnamefont {Liu}}, \ and\ \bibinfo {author}
  {\bibfnamefont {S.-L.}\ \bibnamefont {Zhu}},\ }\href {\doibase
  10.1016/j.ppnp.2019.04.003} {\bibfield  {journal} {\bibinfo  {journal} {Prog.
  Part. Nucl. Phys.}\ }\textbf {\bibinfo {volume} {107}},\ \bibinfo {pages}
  {237} (\bibinfo {year} {2019})},\ \Eprint {http://arxiv.org/abs/1903.11976}
  {arXiv:1903.11976 [hep-ph]} \BibitemShut {NoStop}%
\bibitem [{\citenamefont {Yang}\ \emph
  {et~al.}(2020{\natexlab{a}})\citenamefont {Yang}, \citenamefont {Ping},\ and\
  \citenamefont {Segovia}}]{Yang:2020atz}%
  \BibitemOpen
  \bibfield  {author} {\bibinfo {author} {\bibfnamefont {G.}~\bibnamefont
  {Yang}}, \bibinfo {author} {\bibfnamefont {J.}~\bibnamefont {Ping}}, \ and\
  \bibinfo {author} {\bibfnamefont {J.}~\bibnamefont {Segovia}},\ }\href
  {\doibase 10.3390/sym12111869} {\bibfield  {journal} {\bibinfo  {journal}
  {Symmetry}\ }\textbf {\bibinfo {volume} {12}},\ \bibinfo {pages} {1869}
  (\bibinfo {year} {2020}{\natexlab{a}})},\ \Eprint
  {http://arxiv.org/abs/2009.00238} {arXiv:2009.00238 [hep-ph]} \BibitemShut
  {NoStop}%
\bibitem [{\citenamefont {Dong}\ \emph
  {et~al.}(2021{\natexlab{b}})\citenamefont {Dong}, \citenamefont {Guo},\ and\
  \citenamefont {Zou}}]{Dong:2021bvy}%
  \BibitemOpen
  \bibfield  {author} {\bibinfo {author} {\bibfnamefont {X.-K.}\ \bibnamefont
  {Dong}}, \bibinfo {author} {\bibfnamefont {F.-K.}\ \bibnamefont {Guo}}, \
  and\ \bibinfo {author} {\bibfnamefont {B.-S.}\ \bibnamefont {Zou}},\ }\href
  {\doibase 10.1088/1572-9494/ac27a2} {\bibfield  {journal} {\bibinfo
  {journal} {Commun. Theor. Phys.}\ }\textbf {\bibinfo {volume} {73}},\
  \bibinfo {pages} {125201} (\bibinfo {year} {2021}{\natexlab{b}})},\ \Eprint
  {http://arxiv.org/abs/2108.02673} {arXiv:2108.02673 [hep-ph]} \BibitemShut
  {NoStop}%
\bibitem [{\citenamefont {Chen}(2021)}]{Chen:2021erj}%
  \BibitemOpen
  \bibfield  {author} {\bibinfo {author} {\bibfnamefont {H.-X.}\ \bibnamefont
  {Chen}},\ }\href@noop {} {\  (\bibinfo {year} {2021})},\ \Eprint
  {http://arxiv.org/abs/2103.08586} {arXiv:2103.08586 [hep-ph]} \BibitemShut
  {NoStop}%
\bibitem [{\citenamefont {Yang}\ \emph
  {et~al.}(2021{\natexlab{a}})\citenamefont {Yang}, \citenamefont {Ping},\ and\
  \citenamefont {Segovia}}]{Yang:2021zhe}%
  \BibitemOpen
  \bibfield  {author} {\bibinfo {author} {\bibfnamefont {G.}~\bibnamefont
  {Yang}}, \bibinfo {author} {\bibfnamefont {J.}~\bibnamefont {Ping}}, \ and\
  \bibinfo {author} {\bibfnamefont {J.}~\bibnamefont {Segovia}},\ }\href
  {\doibase 10.1103/PhysRevD.104.094035} {\bibfield  {journal} {\bibinfo
  {journal} {Phys. Rev. D}\ }\textbf {\bibinfo {volume} {104}},\ \bibinfo
  {pages} {094035} (\bibinfo {year} {2021}{\natexlab{a}})},\ \Eprint
  {http://arxiv.org/abs/2109.04311} {arXiv:2109.04311 [hep-ph]} \BibitemShut
  {NoStop}%
\bibitem [{\citenamefont {Yang}\ \emph {et~al.}(2017)\citenamefont {Yang},
  \citenamefont {Ping},\ and\ \citenamefont {Wang}}]{Yang:2015bmv}%
  \BibitemOpen
  \bibfield  {author} {\bibinfo {author} {\bibfnamefont {G.}~\bibnamefont
  {Yang}}, \bibinfo {author} {\bibfnamefont {J.}~\bibnamefont {Ping}}, \ and\
  \bibinfo {author} {\bibfnamefont {F.}~\bibnamefont {Wang}},\ }\href {\doibase
  10.1103/PhysRevD.95.014010} {\bibfield  {journal} {\bibinfo  {journal} {Phys.
  Rev.}\ }\textbf {\bibinfo {volume} {D95}},\ \bibinfo {pages} {014010}
  (\bibinfo {year} {2017})}\BibitemShut {NoStop}%
\bibitem [{\citenamefont {Yang}\ \emph
  {et~al.}(2020{\natexlab{b}})\citenamefont {Yang}, \citenamefont {Ping},\ and\
  \citenamefont {Segovia}}]{gy:2020dcp}%
  \BibitemOpen
  \bibfield  {author} {\bibinfo {author} {\bibfnamefont {G.}~\bibnamefont
  {Yang}}, \bibinfo {author} {\bibfnamefont {J.~L.}\ \bibnamefont {Ping}}, \
  and\ \bibinfo {author} {\bibfnamefont {J.}~\bibnamefont {Segovia}},\ }\href
  {\doibase 10.1103/PhysRevD.101.074030} {\bibfield  {journal} {\bibinfo
  {journal} {Phys. Rev. D}\ }\textbf {\bibinfo {volume} {101}},\ \bibinfo
  {pages} {074030} (\bibinfo {year} {2020}{\natexlab{b}})}\BibitemShut
  {NoStop}%
\bibitem [{\citenamefont {Yang}\ \emph
  {et~al.}(2019{\natexlab{b}})\citenamefont {Yang}, \citenamefont {Ping},\ and\
  \citenamefont {Segovia}}]{Yang:2018oqd}%
  \BibitemOpen
  \bibfield  {author} {\bibinfo {author} {\bibfnamefont {G.}~\bibnamefont
  {Yang}}, \bibinfo {author} {\bibfnamefont {J.}~\bibnamefont {Ping}}, \ and\
  \bibinfo {author} {\bibfnamefont {J.}~\bibnamefont {Segovia}},\ }\href
  {\doibase 10.1103/PhysRevD.99.014035} {\bibfield  {journal} {\bibinfo
  {journal} {Phys. Rev.}\ }\textbf {\bibinfo {volume} {D99}},\ \bibinfo {pages}
  {014035} (\bibinfo {year} {2019}{\natexlab{b}})}\BibitemShut {NoStop}%
\bibitem [{\citenamefont {Yang}\ \emph
  {et~al.}(2020{\natexlab{c}})\citenamefont {Yang}, \citenamefont {Ping},\ and\
  \citenamefont {Segovia}}]{gy:2020dht}%
  \BibitemOpen
  \bibfield  {author} {\bibinfo {author} {\bibfnamefont {G.}~\bibnamefont
  {Yang}}, \bibinfo {author} {\bibfnamefont {J.~L.}\ \bibnamefont {Ping}}, \
  and\ \bibinfo {author} {\bibfnamefont {J.}~\bibnamefont {Segovia}},\ }\href
  {\doibase 10.1103/PhysRevD.101.014001} {\bibfield  {journal} {\bibinfo
  {journal} {Phys. Rev. D}\ }\textbf {\bibinfo {volume} {101}},\ \bibinfo
  {pages} {014001} (\bibinfo {year} {2020}{\natexlab{c}})}\BibitemShut
  {NoStop}%
\bibitem [{\citenamefont {Yang}\ \emph
  {et~al.}(2020{\natexlab{d}})\citenamefont {Yang}, \citenamefont {Ping},\ and\
  \citenamefont {Segovia}}]{gy:2020dhts}%
  \BibitemOpen
  \bibfield  {author} {\bibinfo {author} {\bibfnamefont {G.}~\bibnamefont
  {Yang}}, \bibinfo {author} {\bibfnamefont {J.}~\bibnamefont {Ping}}, \ and\
  \bibinfo {author} {\bibfnamefont {J.}~\bibnamefont {Segovia}},\ }\href
  {\doibase 10.1103/PhysRevD.102.054023} {\bibfield  {journal} {\bibinfo
  {journal} {Phys. Rev. D}\ }\textbf {\bibinfo {volume} {102}},\ \bibinfo
  {pages} {054023} (\bibinfo {year} {2020}{\natexlab{d}})}\BibitemShut
  {NoStop}%
\bibitem [{\citenamefont {Yang}\ \emph
  {et~al.}(2021{\natexlab{b}})\citenamefont {Yang}, \citenamefont {Ping},\ and\
  \citenamefont {Segovia}}]{Yang:2021hrb}%
  \BibitemOpen
  \bibfield  {author} {\bibinfo {author} {\bibfnamefont {G.}~\bibnamefont
  {Yang}}, \bibinfo {author} {\bibfnamefont {J.}~\bibnamefont {Ping}}, \ and\
  \bibinfo {author} {\bibfnamefont {J.}~\bibnamefont {Segovia}},\ }\href
  {\doibase 10.1103/PhysRevD.104.014006} {\bibfield  {journal} {\bibinfo
  {journal} {Phys. Rev. D}\ }\textbf {\bibinfo {volume} {104}},\ \bibinfo
  {pages} {014006} (\bibinfo {year} {2021}{\natexlab{b}})},\ \Eprint
  {http://arxiv.org/abs/2104.08814} {arXiv:2104.08814 [hep-ph]} \BibitemShut
  {NoStop}%
\bibitem [{\citenamefont {Yang}\ \emph
  {et~al.}(2021{\natexlab{c}})\citenamefont {Yang}, \citenamefont {Ping},\ and\
  \citenamefont {Segovia}}]{Yang:2021izl}%
  \BibitemOpen
  \bibfield  {author} {\bibinfo {author} {\bibfnamefont {G.}~\bibnamefont
  {Yang}}, \bibinfo {author} {\bibfnamefont {J.}~\bibnamefont {Ping}}, \ and\
  \bibinfo {author} {\bibfnamefont {J.}~\bibnamefont {Segovia}},\ }\href
  {\doibase 10.1103/PhysRevD.103.074011} {\bibfield  {journal} {\bibinfo
  {journal} {Phys. Rev. D}\ }\textbf {\bibinfo {volume} {103}},\ \bibinfo
  {pages} {074011} (\bibinfo {year} {2021}{\natexlab{c}})},\ \Eprint
  {http://arxiv.org/abs/2101.04933} {arXiv:2101.04933 [hep-ph]} \BibitemShut
  {NoStop}%
\bibitem [{\citenamefont {Aguilar}\ and\ \citenamefont
  {Combes}(1971)}]{JA22269}%
  \BibitemOpen
  \bibfield  {author} {\bibinfo {author} {\bibfnamefont {J.}~\bibnamefont
  {Aguilar}}\ and\ \bibinfo {author} {\bibfnamefont {J.~M.}\ \bibnamefont
  {Combes}},\ }\href {\doibase 10.1007/BF01877510} {\bibfield  {journal}
  {\bibinfo  {journal} {Commun. Math. Phys.}\ }\textbf {\bibinfo {volume}
  {22}},\ \bibinfo {pages} {269} (\bibinfo {year} {1971})}\BibitemShut
  {NoStop}%
%%CITATION = ARXIV:1809.06193;%%
\bibitem [{\citenamefont {Balslev}\ and\ \citenamefont
  {Combes}(1971)}]{EB22280}%
  \BibitemOpen
  \bibfield  {author} {\bibinfo {author} {\bibfnamefont {E.}~\bibnamefont
  {Balslev}}\ and\ \bibinfo {author} {\bibfnamefont {J.~M.}\ \bibnamefont
  {Combes}},\ }\href {\doibase 10.1007/BF01877511} {\bibfield  {journal}
  {\bibinfo  {journal} {Commun. Math. Phys.}\ }\textbf {\bibinfo {volume}
  {22}},\ \bibinfo {pages} {280} (\bibinfo {year} {1971})}\BibitemShut
  {NoStop}%
%%CITATION = ARXIV:1809.06193;%%
\bibitem [{\citenamefont {Vijande}\ \emph {et~al.}(2005)\citenamefont
  {Vijande}, \citenamefont {Fernandez},\ and\ \citenamefont
  {Valcarce}}]{Vijande:2004he}%
  \BibitemOpen
  \bibfield  {author} {\bibinfo {author} {\bibfnamefont {J.}~\bibnamefont
  {Vijande}}, \bibinfo {author} {\bibfnamefont {F.}~\bibnamefont {Fernandez}},
  \ and\ \bibinfo {author} {\bibfnamefont {A.}~\bibnamefont {Valcarce}},\
  }\href {\doibase 10.1088/0954-3899/31/5/017} {\bibfield  {journal} {\bibinfo
  {journal} {J. Phys.}\ }\textbf {\bibinfo {volume} {G31}},\ \bibinfo {pages}
  {481} (\bibinfo {year} {2005})},\ \Eprint
  {http://arxiv.org/abs/hep-ph/0411299} {arXiv:hep-ph/0411299 [hep-ph]}
  \BibitemShut {NoStop}%
%%CITATION = HEP-PH/0411299;%%
\bibitem [{\citenamefont {Scadron}(1982)}]{Scadron:1982eg}%
  \BibitemOpen
  \bibfield  {author} {\bibinfo {author} {\bibfnamefont {M.~D.}\ \bibnamefont
  {Scadron}},\ }\href {\doibase 10.1103/PhysRevD.26.239} {\bibfield  {journal}
  {\bibinfo  {journal} {Phys. Rev.}\ }\textbf {\bibinfo {volume} {D26}},\
  \bibinfo {pages} {239} (\bibinfo {year} {1982})}\BibitemShut {NoStop}%
%%CITATION = PHRVA,D26,239;%%
\bibitem [{\citenamefont {Garcia-Martin}\ \emph {et~al.}(2011)\citenamefont
  {Garcia-Martin}, \citenamefont {Kaminski}, \citenamefont {Pelaez},\ and\
  \citenamefont {Ruiz~de Elvira}}]{Garcia-Martin:2011nna}%
  \BibitemOpen
  \bibfield  {author} {\bibinfo {author} {\bibfnamefont {R.}~\bibnamefont
  {Garcia-Martin}}, \bibinfo {author} {\bibfnamefont {R.}~\bibnamefont
  {Kaminski}}, \bibinfo {author} {\bibfnamefont {J.~R.}\ \bibnamefont
  {Pelaez}}, \ and\ \bibinfo {author} {\bibfnamefont {J.}~\bibnamefont {Ruiz~de
  Elvira}},\ }\href {\doibase 10.1103/PhysRevLett.107.072001} {\bibfield
  {journal} {\bibinfo  {journal} {Phys. Rev. Lett.}\ }\textbf {\bibinfo
  {volume} {107}},\ \bibinfo {pages} {072001} (\bibinfo {year} {2011})},\
  \Eprint {http://arxiv.org/abs/1107.1635} {arXiv:1107.1635 [hep-ph]}
  \BibitemShut {NoStop}%
\bibitem [{\citenamefont {Albaladejo}\ and\ \citenamefont
  {Oller}(2012)}]{Albaladejo:2012te}%
  \BibitemOpen
  \bibfield  {author} {\bibinfo {author} {\bibfnamefont {M.}~\bibnamefont
  {Albaladejo}}\ and\ \bibinfo {author} {\bibfnamefont {J.~A.}\ \bibnamefont
  {Oller}},\ }\href {\doibase 10.1103/PhysRevD.86.034003} {\bibfield  {journal}
  {\bibinfo  {journal} {Phys. Rev. D}\ }\textbf {\bibinfo {volume} {86}},\
  \bibinfo {pages} {034003} (\bibinfo {year} {2012})},\ \Eprint
  {http://arxiv.org/abs/1205.6606} {arXiv:1205.6606 [hep-ph]} \BibitemShut
  {NoStop}%
\bibitem [{\citenamefont {Pelaez}(2016)}]{Pelaez:2015qba}%
  \BibitemOpen
  \bibfield  {author} {\bibinfo {author} {\bibfnamefont {J.~R.}\ \bibnamefont
  {Pelaez}},\ }\href {\doibase 10.1016/j.physrep.2016.09.001} {\bibfield
  {journal} {\bibinfo  {journal} {Phys. Rept.}\ }\textbf {\bibinfo {volume}
  {658}},\ \bibinfo {pages} {1} (\bibinfo {year} {2016})},\ \Eprint
  {http://arxiv.org/abs/1510.00653} {arXiv:1510.00653 [hep-ph]} \BibitemShut
  {NoStop}%
\bibitem [{\citenamefont {Bali}\ \emph {et~al.}(2005)\citenamefont {Bali},
  \citenamefont {Neff}, \citenamefont {Duessel}, \citenamefont {Lippert},\ and\
  \citenamefont {Schilling}}]{Bali:2005fu}%
  \BibitemOpen
  \bibfield  {author} {\bibinfo {author} {\bibfnamefont {G.~S.}\ \bibnamefont
  {Bali}}, \bibinfo {author} {\bibfnamefont {H.}~\bibnamefont {Neff}}, \bibinfo
  {author} {\bibfnamefont {T.}~\bibnamefont {Duessel}}, \bibinfo {author}
  {\bibfnamefont {T.}~\bibnamefont {Lippert}}, \ and\ \bibinfo {author}
  {\bibfnamefont {K.}~\bibnamefont {Schilling}} (\bibinfo {collaboration}
  {SESAM}),\ }\href {\doibase 10.1103/PhysRevD.71.114513} {\bibfield  {journal}
  {\bibinfo  {journal} {Phys. Rev.}\ }\textbf {\bibinfo {volume} {D71}},\
  \bibinfo {pages} {114513} (\bibinfo {year} {2005})},\ \Eprint
  {http://arxiv.org/abs/hep-lat/0505012} {arXiv:hep-lat/0505012 [hep-lat]}
  \BibitemShut {NoStop}%
%%CITATION = HEP-LAT/0505012;%%
\bibitem [{\citenamefont {Segovia}\ \emph {et~al.}(2008)\citenamefont
  {Segovia}, \citenamefont {Entem},\ and\ \citenamefont
  {Fernandez}}]{Segovia:2008zza}%
  \BibitemOpen
  \bibfield  {author} {\bibinfo {author} {\bibfnamefont {J.}~\bibnamefont
  {Segovia}}, \bibinfo {author} {\bibfnamefont {D.~R.}\ \bibnamefont {Entem}},
  \ and\ \bibinfo {author} {\bibfnamefont {F.}~\bibnamefont {Fernandez}},\
  }\href {\doibase 10.1016/j.physletb.2008.02.051} {\bibfield  {journal}
  {\bibinfo  {journal} {Phys. Lett. B}\ }\textbf {\bibinfo {volume} {662}},\
  \bibinfo {pages} {33} (\bibinfo {year} {2008})}\BibitemShut {NoStop}%
\bibitem [{\citenamefont {Segovia}\ \emph {et~al.}(2013)\citenamefont
  {Segovia}, \citenamefont {Entem}, \citenamefont {Fernandez},\ and\
  \citenamefont {Hernandez}}]{Segovia:2013wma}%
  \BibitemOpen
  \bibfield  {author} {\bibinfo {author} {\bibfnamefont {J.}~\bibnamefont
  {Segovia}}, \bibinfo {author} {\bibfnamefont {D.~R.}\ \bibnamefont {Entem}},
  \bibinfo {author} {\bibfnamefont {F.}~\bibnamefont {Fernandez}}, \ and\
  \bibinfo {author} {\bibfnamefont {E.}~\bibnamefont {Hernandez}},\ }\href
  {\doibase 10.1142/S0218301313300269} {\bibfield  {journal} {\bibinfo
  {journal} {Int. J. Mod. Phys.}\ }\textbf {\bibinfo {volume} {E22}},\ \bibinfo
  {pages} {1330026} (\bibinfo {year} {2013})},\ \Eprint
  {http://arxiv.org/abs/1309.6926} {arXiv:1309.6926 [hep-ph]} \BibitemShut
  {NoStop}%
%%CITATION = ARXIV:1309.6926;%%
\bibitem [{\citenamefont {Segovia}\ \emph {et~al.}(2009)\citenamefont
  {Segovia}, \citenamefont {Yasser}, \citenamefont {Entem},\ and\ \citenamefont
  {Fernandez}}]{Segovia:2009zz}%
  \BibitemOpen
  \bibfield  {author} {\bibinfo {author} {\bibfnamefont {J.}~\bibnamefont
  {Segovia}}, \bibinfo {author} {\bibfnamefont {A.}~\bibnamefont {Yasser}},
  \bibinfo {author} {\bibfnamefont {D.}~\bibnamefont {Entem}}, \ and\ \bibinfo
  {author} {\bibfnamefont {F.}~\bibnamefont {Fernandez}},\ }\href {\doibase
  10.1103/PhysRevD.80.054017} {\bibfield  {journal} {\bibinfo  {journal} {Phys.
  Rev. D}\ }\textbf {\bibinfo {volume} {80}},\ \bibinfo {pages} {054017}
  (\bibinfo {year} {2009})}\BibitemShut {NoStop}%
\bibitem [{\citenamefont {Segovia}\ \emph {et~al.}(2011)\citenamefont
  {Segovia}, \citenamefont {Entem},\ and\ \citenamefont
  {Fernandez}}]{Segovia:2011zza}%
  \BibitemOpen
  \bibfield  {author} {\bibinfo {author} {\bibfnamefont {J.}~\bibnamefont
  {Segovia}}, \bibinfo {author} {\bibfnamefont {D.}~\bibnamefont {Entem}}, \
  and\ \bibinfo {author} {\bibfnamefont {F.}~\bibnamefont {Fernandez}},\ }\href
  {\doibase 10.1103/PhysRevD.83.114018} {\bibfield  {journal} {\bibinfo
  {journal} {Phys. Rev. D}\ }\textbf {\bibinfo {volume} {83}},\ \bibinfo
  {pages} {114018} (\bibinfo {year} {2011})}\BibitemShut {NoStop}%
\bibitem [{\citenamefont {Segovia}\ \emph {et~al.}(2015)\citenamefont
  {Segovia}, \citenamefont {Entem},\ and\ \citenamefont
  {Fernandez}}]{Segovia:2015dia}%
  \BibitemOpen
  \bibfield  {author} {\bibinfo {author} {\bibfnamefont {J.}~\bibnamefont
  {Segovia}}, \bibinfo {author} {\bibfnamefont {D.~R.}\ \bibnamefont {Entem}},
  \ and\ \bibinfo {author} {\bibfnamefont {F.}~\bibnamefont {Fernandez}},\
  }\href {\doibase 10.1103/PhysRevD.91.094020} {\bibfield  {journal} {\bibinfo
  {journal} {Phys. Rev. D}\ }\textbf {\bibinfo {volume} {91}},\ \bibinfo
  {pages} {094020} (\bibinfo {year} {2015})},\ \Eprint
  {http://arxiv.org/abs/1502.03827} {arXiv:1502.03827 [hep-ph]} \BibitemShut
  {NoStop}%
\bibitem [{\citenamefont {Segovia}\ \emph {et~al.}(2016)\citenamefont
  {Segovia}, \citenamefont {Ortega}, \citenamefont {Entem},\ and\ \citenamefont
  {Fern\'andez}}]{Segovia:2016xqb}%
  \BibitemOpen
  \bibfield  {author} {\bibinfo {author} {\bibfnamefont {J.}~\bibnamefont
  {Segovia}}, \bibinfo {author} {\bibfnamefont {P.~G.}\ \bibnamefont {Ortega}},
  \bibinfo {author} {\bibfnamefont {D.~R.}\ \bibnamefont {Entem}}, \ and\
  \bibinfo {author} {\bibfnamefont {F.}~\bibnamefont {Fern\'andez}},\ }\href
  {\doibase 10.1103/PhysRevD.93.074027} {\bibfield  {journal} {\bibinfo
  {journal} {Phys. Rev. D}\ }\textbf {\bibinfo {volume} {93}},\ \bibinfo
  {pages} {074027} (\bibinfo {year} {2016})},\ \Eprint
  {http://arxiv.org/abs/1601.05093} {arXiv:1601.05093 [hep-ph]} \BibitemShut
  {NoStop}%
\bibitem [{\citenamefont {Yang}\ \emph {et~al.}(2018)\citenamefont {Yang},
  \citenamefont {Ping},\ and\ \citenamefont {Segovia}}]{Yang:2017xpp}%
  \BibitemOpen
  \bibfield  {author} {\bibinfo {author} {\bibfnamefont {G.}~\bibnamefont
  {Yang}}, \bibinfo {author} {\bibfnamefont {J.}~\bibnamefont {Ping}}, \ and\
  \bibinfo {author} {\bibfnamefont {J.}~\bibnamefont {Segovia}},\ }\href
  {\doibase 10.1007/s00601-018-1433-4} {\bibfield  {journal} {\bibinfo
  {journal} {Few-Body Syst.}\ }\textbf {\bibinfo {volume} {59}},\ \bibinfo
  {pages} {113} (\bibinfo {year} {2018})},\ \Eprint
  {http://arxiv.org/abs/1709.09315} {arXiv:1709.09315 [hep-ph]} \BibitemShut
  {NoStop}%
%%CITATION = ARXIV:1709.09315;%%
\bibitem [{\citenamefont {Yang}\ \emph
  {et~al.}(2020{\natexlab{e}})\citenamefont {Yang}, \citenamefont {Ping},
  \citenamefont {Ortega},\ and\ \citenamefont {Segovia}}]{Yang:2019lsg}%
  \BibitemOpen
  \bibfield  {author} {\bibinfo {author} {\bibfnamefont {G.}~\bibnamefont
  {Yang}}, \bibinfo {author} {\bibfnamefont {J.}~\bibnamefont {Ping}}, \bibinfo
  {author} {\bibfnamefont {P.~G.}\ \bibnamefont {Ortega}}, \ and\ \bibinfo
  {author} {\bibfnamefont {J.}~\bibnamefont {Segovia}},\ }\href {\doibase
  10.1088/1674-1137/44/2/023102} {\bibfield  {journal} {\bibinfo  {journal}
  {Chin. Phys. C}\ }\textbf {\bibinfo {volume} {44}},\ \bibinfo {pages}
  {023102} (\bibinfo {year} {2020}{\natexlab{e}})},\ \Eprint
  {http://arxiv.org/abs/1904.10166} {arXiv:1904.10166 [hep-ph]} \BibitemShut
  {NoStop}%
\bibitem [{\citenamefont {Ortega}\ \emph
  {et~al.}(2016{\natexlab{a}})\citenamefont {Ortega}, \citenamefont {Segovia},
  \citenamefont {Entem},\ and\ \citenamefont {Fernandez}}]{Ortega:2016mms}%
  \BibitemOpen
  \bibfield  {author} {\bibinfo {author} {\bibfnamefont {P.~G.}\ \bibnamefont
  {Ortega}}, \bibinfo {author} {\bibfnamefont {J.}~\bibnamefont {Segovia}},
  \bibinfo {author} {\bibfnamefont {D.~R.}\ \bibnamefont {Entem}}, \ and\
  \bibinfo {author} {\bibfnamefont {F.}~\bibnamefont {Fernandez}},\ }\href
  {\doibase 10.1103/PhysRevD.94.074037} {\bibfield  {journal} {\bibinfo
  {journal} {Phys. Rev.}\ }\textbf {\bibinfo {volume} {D94}},\ \bibinfo {pages}
  {074037} (\bibinfo {year} {2016}{\natexlab{a}})},\ \Eprint
  {http://arxiv.org/abs/1603.07000} {arXiv:1603.07000 [hep-ph]} \BibitemShut
  {NoStop}%
%%CITATION = ARXIV:1603.07000;%%
\bibitem [{\citenamefont {Ortega}\ \emph {et~al.}(2017)\citenamefont {Ortega},
  \citenamefont {Segovia}, \citenamefont {Entem},\ and\ \citenamefont
  {Fernández}}]{Ortega:2016pgg}%
  \BibitemOpen
  \bibfield  {author} {\bibinfo {author} {\bibfnamefont {P.~G.}\ \bibnamefont
  {Ortega}}, \bibinfo {author} {\bibfnamefont {J.}~\bibnamefont {Segovia}},
  \bibinfo {author} {\bibfnamefont {D.~R.}\ \bibnamefont {Entem}}, \ and\
  \bibinfo {author} {\bibfnamefont {F.}~\bibnamefont {Fernández}},\ }\href
  {\doibase 10.1103/PhysRevD.95.034010} {\bibfield  {journal} {\bibinfo
  {journal} {Phys. Rev.}\ }\textbf {\bibinfo {volume} {D95}},\ \bibinfo {pages}
  {034010} (\bibinfo {year} {2017})},\ \Eprint
  {http://arxiv.org/abs/1612.04826} {arXiv:1612.04826 [hep-ph]} \BibitemShut
  {NoStop}%
%%CITATION = ARXIV:1612.04826;%%
\bibitem [{\citenamefont {Ortega}\ \emph
  {et~al.}(2016{\natexlab{b}})\citenamefont {Ortega}, \citenamefont {Segovia},
  \citenamefont {Entem},\ and\ \citenamefont {Fern\'andez}}]{Ortega:2016hde}%
  \BibitemOpen
  \bibfield  {author} {\bibinfo {author} {\bibfnamefont {P.~G.}\ \bibnamefont
  {Ortega}}, \bibinfo {author} {\bibfnamefont {J.}~\bibnamefont {Segovia}},
  \bibinfo {author} {\bibfnamefont {D.~R.}\ \bibnamefont {Entem}}, \ and\
  \bibinfo {author} {\bibfnamefont {F.}~\bibnamefont {Fern\'andez}},\ }\href
  {\doibase 10.1103/PhysRevD.94.114018} {\bibfield  {journal} {\bibinfo
  {journal} {Phys. Rev. D}\ }\textbf {\bibinfo {volume} {94}},\ \bibinfo
  {pages} {114018} (\bibinfo {year} {2016}{\natexlab{b}})},\ \Eprint
  {http://arxiv.org/abs/1608.01325} {arXiv:1608.01325 [hep-ph]} \BibitemShut
  {NoStop}%
\bibitem [{\citenamefont {Ortega}\ \emph {et~al.}(2018)\citenamefont {Ortega},
  \citenamefont {Segovia}, \citenamefont {Entem},\ and\ \citenamefont
  {Fern\'andez}}]{Ortega:2017qmg}%
  \BibitemOpen
  \bibfield  {author} {\bibinfo {author} {\bibfnamefont {P.~G.}\ \bibnamefont
  {Ortega}}, \bibinfo {author} {\bibfnamefont {J.}~\bibnamefont {Segovia}},
  \bibinfo {author} {\bibfnamefont {D.~R.}\ \bibnamefont {Entem}}, \ and\
  \bibinfo {author} {\bibfnamefont {F.}~\bibnamefont {Fern\'andez}},\ }\href
  {\doibase 10.1016/j.physletb.2018.01.005} {\bibfield  {journal} {\bibinfo
  {journal} {Phys. Lett. B}\ }\textbf {\bibinfo {volume} {778}},\ \bibinfo
  {pages} {1} (\bibinfo {year} {2018})},\ \Eprint
  {http://arxiv.org/abs/1706.02639} {arXiv:1706.02639 [hep-ph]} \BibitemShut
  {NoStop}%
\bibitem [{\citenamefont {Ortega}\ \emph {et~al.}(2019)\citenamefont {Ortega},
  \citenamefont {Segovia}, \citenamefont {Entem},\ and\ \citenamefont
  {Fern\'andez}}]{Ortega:2018cnm}%
  \BibitemOpen
  \bibfield  {author} {\bibinfo {author} {\bibfnamefont {P.~G.}\ \bibnamefont
  {Ortega}}, \bibinfo {author} {\bibfnamefont {J.}~\bibnamefont {Segovia}},
  \bibinfo {author} {\bibfnamefont {D.~R.}\ \bibnamefont {Entem}}, \ and\
  \bibinfo {author} {\bibfnamefont {F.}~\bibnamefont {Fern\'andez}},\ }\href
  {\doibase 10.1140/epjc/s10052-019-6552-7} {\bibfield  {journal} {\bibinfo
  {journal} {Eur. Phys. J. C}\ }\textbf {\bibinfo {volume} {79}},\ \bibinfo
  {pages} {78} (\bibinfo {year} {2019})},\ \Eprint
  {http://arxiv.org/abs/1808.00914} {arXiv:1808.00914 [hep-ph]} \BibitemShut
  {NoStop}%
\bibitem [{\citenamefont {Ortega}\ \emph {et~al.}(2020)\citenamefont {Ortega},
  \citenamefont {Segovia}, \citenamefont {Entem},\ and\ \citenamefont
  {Fernandez}}]{Ortega:2020uvc}%
  \BibitemOpen
  \bibfield  {author} {\bibinfo {author} {\bibfnamefont {P.~G.}\ \bibnamefont
  {Ortega}}, \bibinfo {author} {\bibfnamefont {J.}~\bibnamefont {Segovia}},
  \bibinfo {author} {\bibfnamefont {D.~R.}\ \bibnamefont {Entem}}, \ and\
  \bibinfo {author} {\bibfnamefont {F.}~\bibnamefont {Fernandez}},\ }\href
  {\doibase 10.1140/epjc/s10052-020-7764-6} {\bibfield  {journal} {\bibinfo
  {journal} {Eur. Phys. J. C}\ }\textbf {\bibinfo {volume} {80}},\ \bibinfo
  {pages} {223} (\bibinfo {year} {2020})},\ \Eprint
  {http://arxiv.org/abs/2001.08093} {arXiv:2001.08093 [hep-ph]} \BibitemShut
  {NoStop}%
\bibitem [{\citenamefont {Yang}\ and\ \citenamefont
  {Ping}(2018)}]{Yang:2017rpg}%
  \BibitemOpen
  \bibfield  {author} {\bibinfo {author} {\bibfnamefont {G.}~\bibnamefont
  {Yang}}\ and\ \bibinfo {author} {\bibfnamefont {J.}~\bibnamefont {Ping}},\
  }\href {\doibase 10.1103/PhysRevD.97.034023} {\bibfield  {journal} {\bibinfo
  {journal} {Phys. Rev.}\ }\textbf {\bibinfo {volume} {D97}},\ \bibinfo {pages}
  {034023} (\bibinfo {year} {2018})},\ \Eprint
  {http://arxiv.org/abs/1703.08845} {arXiv:1703.08845 [hep-ph]} \BibitemShut
  {NoStop}%
%%CITATION = ARXIV:1703.08845;%%
\bibitem [{\citenamefont {Yang}\ \emph
  {et~al.}(2020{\natexlab{f}})\citenamefont {Yang}, \citenamefont {Ping},\ and\
  \citenamefont {Segovia}}]{Yang:2020fou}%
  \BibitemOpen
  \bibfield  {author} {\bibinfo {author} {\bibfnamefont {G.}~\bibnamefont
  {Yang}}, \bibinfo {author} {\bibfnamefont {J.}~\bibnamefont {Ping}}, \ and\
  \bibinfo {author} {\bibfnamefont {J.}~\bibnamefont {Segovia}},\ }\href
  {\doibase 10.1103/PhysRevD.102.054023} {\bibfield  {journal} {\bibinfo
  {journal} {Phys. Rev. D}\ }\textbf {\bibinfo {volume} {102}},\ \bibinfo
  {pages} {054023} (\bibinfo {year} {2020}{\natexlab{f}})},\ \Eprint
  {http://arxiv.org/abs/2007.05190} {arXiv:2007.05190 [hep-ph]} \BibitemShut
  {NoStop}%
\bibitem [{\citenamefont {Yang}\ \emph
  {et~al.}(2020{\natexlab{g}})\citenamefont {Yang}, \citenamefont {Ping},\ and\
  \citenamefont {Segovia}}]{Yang:2020twg}%
  \BibitemOpen
  \bibfield  {author} {\bibinfo {author} {\bibfnamefont {G.}~\bibnamefont
  {Yang}}, \bibinfo {author} {\bibfnamefont {J.}~\bibnamefont {Ping}}, \ and\
  \bibinfo {author} {\bibfnamefont {J.}~\bibnamefont {Segovia}},\ }\href
  {\doibase 10.1103/PhysRevD.101.074030} {\bibfield  {journal} {\bibinfo
  {journal} {Phys. Rev. D}\ }\textbf {\bibinfo {volume} {101}},\ \bibinfo
  {pages} {074030} (\bibinfo {year} {2020}{\natexlab{g}})},\ \Eprint
  {http://arxiv.org/abs/2003.05253} {arXiv:2003.05253 [hep-ph]} \BibitemShut
  {NoStop}%
\bibitem [{\citenamefont {Hiyama}\ \emph {et~al.}(2003)\citenamefont {Hiyama},
  \citenamefont {Kino},\ and\ \citenamefont {Kamimura}}]{Hiyama:2003cu}%
  \BibitemOpen
  \bibfield  {author} {\bibinfo {author} {\bibfnamefont {E.}~\bibnamefont
  {Hiyama}}, \bibinfo {author} {\bibfnamefont {Y.}~\bibnamefont {Kino}}, \ and\
  \bibinfo {author} {\bibfnamefont {M.}~\bibnamefont {Kamimura}},\ }\href
  {\doibase 10.1016/S0146-6410(03)90015-9} {\bibfield  {journal} {\bibinfo
  {journal} {Prog. Part. Nucl. Phys.}\ }\textbf {\bibinfo {volume} {51}},\
  \bibinfo {pages} {223} (\bibinfo {year} {2003})}\BibitemShut {NoStop}%
%%CITATION = PPNPD,51,223;%%
\end{thebibliography}%

\end{document}